\documentclass[twocolumn,preprintnumbers,amsmath,amssymb]{revtex4}

\usepackage{graphicx}
\usepackage{dcolumn}
\usepackage{bm}
\usepackage[colorlinks=true,linkcolor=blue, citecolor=blue,urlcolor=blue]{hyperref}
\usepackage{setspace}
\usepackage{natbib}
\usepackage{hyperref}
\hypersetup{colorlinks=true,linkcolor=blue,citecolor=blue,urlcolor=blue}
\usepackage{epsfig, epstopdf}
\usepackage{titlesec}
\titlespacing{\section}{0pt}{10pt}{10pt}
\titlespacing{\subsection}{0pt}{10pt}{10pt}
\titlespacing{\subsubsection}{0pt}{10pt}{10pt}
\newcommand{\tc}{{T_{\rm{c}}}}
\newcommand{\cl}{\rm{Ca_{1-x}La_{x}FeAs{_2}}}
\newcommand{\cp}{\rm{Ca_{1-x}Pr_{x}FeAs{_2}}}
\newcommand{\cfa}{\rm{CaFeAs{_2}}}
\newcommand{\cn}{\rm{CeNi_{x}Bi{_2}}}
\newcommand{\cpb}{\rm{CePd_{x}Bi{_2}}}
\newcommand{\lnb}{\rm{LaNi_{x}Bi{_2}}}
\newcommand{\lns}{\rm{LaNi_{x}Sb{_2}}}
\newcommand{\lps}{\rm{LaPd{_x}Sb{_2}}}
\newcommand{\lpb}{\rm{LaPd{_x}Bi{_2}}}
\newcommand{\smb}{\rm{SrMnBi{_2}}}
\newcommand{\cmb}{\rm{CaMnBi{_2}}}
\newcommand{\emb}{\rm{EuMnBi{_2}}}
\newcommand{\lab}{\rm{LaAgBi{_2}}}
\newcommand{\szs}{\rm{SrZnSb{_2}}}
\newcommand{\p}{\rm{pnictide}}

\newcommand{\y}{\rm{superconductivity}}
\newcommand{\si}{\rm{superconducting}}
\newcommand{\fs}{\rm{Fermi~surface}}
\newcommand{\e}{et$~$al.}

\begin{document}

\title{Superconductivity and Dirac Fermions in 112-phase Pnictides}

\author{S. J. Ray}
\email[Corresponding author : ]{ray@iitp.ac.in, ray.sjr@gmail.com}

\affiliation{
	Department of Physics, 
	Indian Institute of Technology Patna, 
	Bihta 801103, India
	}
	
\affiliation{
	Institute of Materials Science, 
	Technische Universit\"{a}t Darmstadt, 
	Alarich-Weiss-Stra\ss e 2, Darmstadt 64287, Germany
	}

\author{L. Alff}
\affiliation{
	Institute of Materials Science, 
	Technische Universit\"{a}t Darmstadt, 
	Alarich-Weiss-Stra\ss e 2, Darmstadt 64287, Germany
	}

\begin{abstract}
This article reviews the status of current research on the 112-phase of pnictides. The 112-phase has gained augmented attention due to the recent discovery of high-temperature superconductivity in $\cl$ with a maximum critical temperature $\tc\sim$ 47\,K upon Sb substitution. The structural, magnetic, and electronic properties of $\cl$ bear some similarities with other superconducting pnictide phases, however, the different valence states of the pnictogen and the presence of a metallic spacer layer are unique features of the 112-system. Low-temperature superconductivity which coexists with antiferromagnetic order was observed in transition metal (Ni, Pd) deficient 112-compounds like $\cn$, $\lpb$, $\lps$, $\lns$. Besides superconductivity, the presence of naturally occurring anisotropic Dirac Fermionic states were observed in the layered 112-compounds $\smb$, $\cmb$, $\lab$ which are of significant interest for future nanoelectronics as an alternative to graphene. In these compounds, the linear energy dispersion resulted in a high magnetoresistance that stayed unsaturated even at the highest applied magnetic fields. Here, we describe various 112-type materials systems combining experimental results and theoretical predictions to stimulate further research on this less well-known member of the pnictide family.
\end{abstract}
\maketitle
{\footnotesize \tableofcontents}
\setcounter{figure}{0}

\section{Introduction}

The discovery of superconductivity in the pnictide family of superconductors LaOFeP\cite{Kamihara2006} has fuelled research on the Fe-based superconducting compounds after the observation of a critical temperature of $\tc \sim$ 26\,K in the isostructural compound La(O$_{1-x}$F$_x$)FeAs \cite{Kamihara2008}, which was soon increased up to 43\,K \cite{Takahasi2008} under the application of high pressure \cite{Hosono2015}. Very soon, an even higher $\tc$ of 54\,K was observed in SmFeAsO$_{1-x}$F$_x$ \cite{Chen2008, Liu2008}. With different levels of doping and elemental substitutions $\tc$ values of 55-58\,K \cite{Wang2008, Ren2008, Hanna2011, Fujioka2013} were observed in several other compounds in the 1111-family of $\p$s. In the next few years, superconductivity was found in various other Fe-based $\p$ systems like (Ba,K)Fe$_2$As$_2$ (122-type) \cite{Rotter2008}, FeSe (11-type) \cite{Hsu2008}, LiFeAs (111-type) \cite{Tapp2008, Wang2008a, Pitcher2008} and, most recently, $\cl$ (112-type) with a $\tc$ of 38\,K \cite{Katayama2013, Kudo2014a, Kudo2014b}. The fundamental interest in these materials lies in understanding the mechanism behind the coexistence of superconductivity and magnetism. The high critical fields and isotropic critical currents \cite{Gurevich2011, Moll2010, Chubukov2012} could be of interest for electrical power and magnetic applications. Apart from the presence of Fe in these compounds, which is believed to be harmful for conventional superconductivity, the uniqueness lies in the origin of superconductivity due to the presence of the Fe-$3d$ electrons\cite{Hosono2015}.

Although there are several structurally different phases of pnictides, they all contain a common Fe-pnictogen ($Pn$) layer in a tetrahedral arrangement that is separated by blocking layers. The composition of the blocking layer is believed to affect the superconducting properties \cite{Johnston2010}. The Fe-$Pn$ (or Fe-$Ch$) layers are tetrahedrally coordinated by the $Pn$ or $Ch$ (chalcogenide) atoms, and highest critical temperatures were observed for an ideal tetrahedral arrangement \cite{Lee2008} with the Fe-As-Fe bond angle closest to 109.47$^{\circ}$. Similar to the high-$T_c$ Cuprates, the pnictides are also quasi-two-dimensional with reduced electronic coupling along the $c$-axis, and the appearance of superconductivity is observed upon the suppression of the antiferromagnetic order \cite{Taylor2013}. In comparison to Cuprates, however, the range of superconducting materials is much larger in case of the pnictides offering a huge range of chemical substitution possibilities \cite{Aswathy2010}. Due to the limited availability of high-quality and larger sized single crystals \cite{Johnston2010}, research was focussed on the 1111- and 122-type compounds. Recently, significant interest has shifted to 11-type FeSe thin films\cite{Wang2012}, where evidence of strain induced interfacial superconductivity was observed up to a maximum $\tc \sim$ 109\,K \cite{Ge2015}.

A prediction of the existence of a 112-type $\p$ phase was made by Shim $\e$ \cite{Shim2009} in the hypothetical compounds BaFeAs$_2$ and BaFeSb$_2$ with metallic blocking layers unlike other $\p$ systems where these layers are insulating. While these compounds remain elusive, the related compound SrMnBi$_2$ was synthesized \cite{Wang2011, Wang2011b, Park2011} as single crystals with physical and structural similarities to BaFeAs$_2$. Despite having a large N\'{e}el temperature ($T_{\text{N}}\sim 290\,$K), no evidence of superconductivity was observed in SrMnBi$_2$. Band structure calculations suggested the presence of anisotropic Dirac fermions in the Bi square net layer which was experimentally confirmed later through the observation of quantum oscillations and angle-resolved photoelectron spectroscopy (ARPES) measurements \cite{Park2011}.

Superconductivity in the 112-type $\p$ was first reported in the Ce$TMPn_2$ ($TM =$ transition metal, $Pn =$ pnictogen) family of intermetallic compounds with $T_{\text{N}} \sim$ 5\,K although several heavy-fermion superconductors with layered tetragonal structures were known earlier \cite{Rosa2015b}. Superconductivity in $\cn$ with $\tc \sim$ 4.2 K \cite{Mizoguchi2011, Buckow2012} was claimed to originate from Ni deficiency, as no evidence of bulk superconductivity was observed in the parent compound CeNiBi$_2$ \cite{Jung2002, Kodama2011, Thamizhavel2003}. However, it has been suggested from the coexistence of light and heavy carriers that the superconducting charge carriers are hosted by the pnictogen square net layer \cite{Mizoguchi2011}. The presence of superconductivity with low superconducting volume fraction (SVF) were also observed in LaNi$_x$Bi$_2$ (SVF = 1-3\%) \cite{Mizoguchi2012, Lin2013}, NdNi$_x$Bi$_2$ (SVF = 14\%) \cite{Mizoguchi2011}, and YNi$_x$Bi$_2$ \cite{Mizoguchi2011} (SVF = 17\%) \cite{Mizoguchi2011}.

Interest in the 112-system has grown considerably after the discovery of a high $\tc$ of 34\,K in $\cl$ \cite{Katayama2013} and in $\cp$ with $\tc\sim$ 20\,K \cite{Yakita2014}. Subsequently, it was reported that adding a small amount of P (0.5\%) and Sb (1\%) substituting As in the parent compound leads to a drastic  enhancement of $\tc$  in $\cl$ to 41\,K and 43\,K, respectively \cite{Kudo2014a}. Later, it was revealed that a larger level of Sb doping can further enhance $\tc$ to 47\,K in Ca$_{1-x}$La$_x$Fe(As$_{1-y}$Sb$_y$) \cite{Kudo2014b} which is higher than the maximum $\tc$ observed in bulk 122, 11, 111-type $\p$s so far. One interesting fact about the superconductivity in the 112-phase is the existence of mixed-valance states of the pnictogen \cite{Mizoguchi2011, Mizoguchi2012} which has different valencies in the $Pn$ square net layer and the metal-pnictogen layers.

In the past two years, research interest again increased for $\cl$ \cite{Okada2014, Yakita2015, Katayama2015, Joseph2015a, Zhou2015, Hiroyuki2015, Zhou2014, Sala2014, Xing2015, Li2015, Kawasaki2015, Liu2015, Huang2015, Jiang2016} due to the availability of single crystals, high-$\tc$ in the bulk phase, and the possibility of a range of substitutions. Significant amount of research work was done earlier in other 112-type materials like LaPd$_x$Sb$_2$ \cite{Reiner2015}, LaPd$_x$Bi$_2$\cite{Han2013, Reiner2015}, CeNi$_x$Sb$_2$ \cite{Mizoguchi2011, Kodama2011, Buckow2012}, LaNi$_x$Sb$_2$ \cite{Buckow2013, Kurian2013} etc. with low $\tc\sim <5$\,K. Although primary research work in the 112-system started to look for possible high-$\tc$ materials, recently a large amount of work has also been performed with respect to anisotropic Dirac fermionic states\cite{Park2011, Wang2011, Wang2011b, Jo2014, Wang2012d, Lee2013, Jia2014, Guo2014, Feng2014} in (Ca/Sr)MnSb$_2$\cite{Wang2012f, He2012, Lee2013, Guo2014}, $\lab$\cite{Wang2012e, Wang2013, Wang2012d, Shi2015} and understanding their role in the observed large magnetoresistance\cite{May2014, He2012, Wang2011b, Wang2012f, Wang2012c, Wang2013} and magneto-thermopower generation\cite{Wang2012c, Wang2012d}. The purpose of this review article is to summarise the work done in the 112-systems so far, and to stimulate further research.

\begin{figure*}
\begin{center}
\includegraphics[width=17cm]{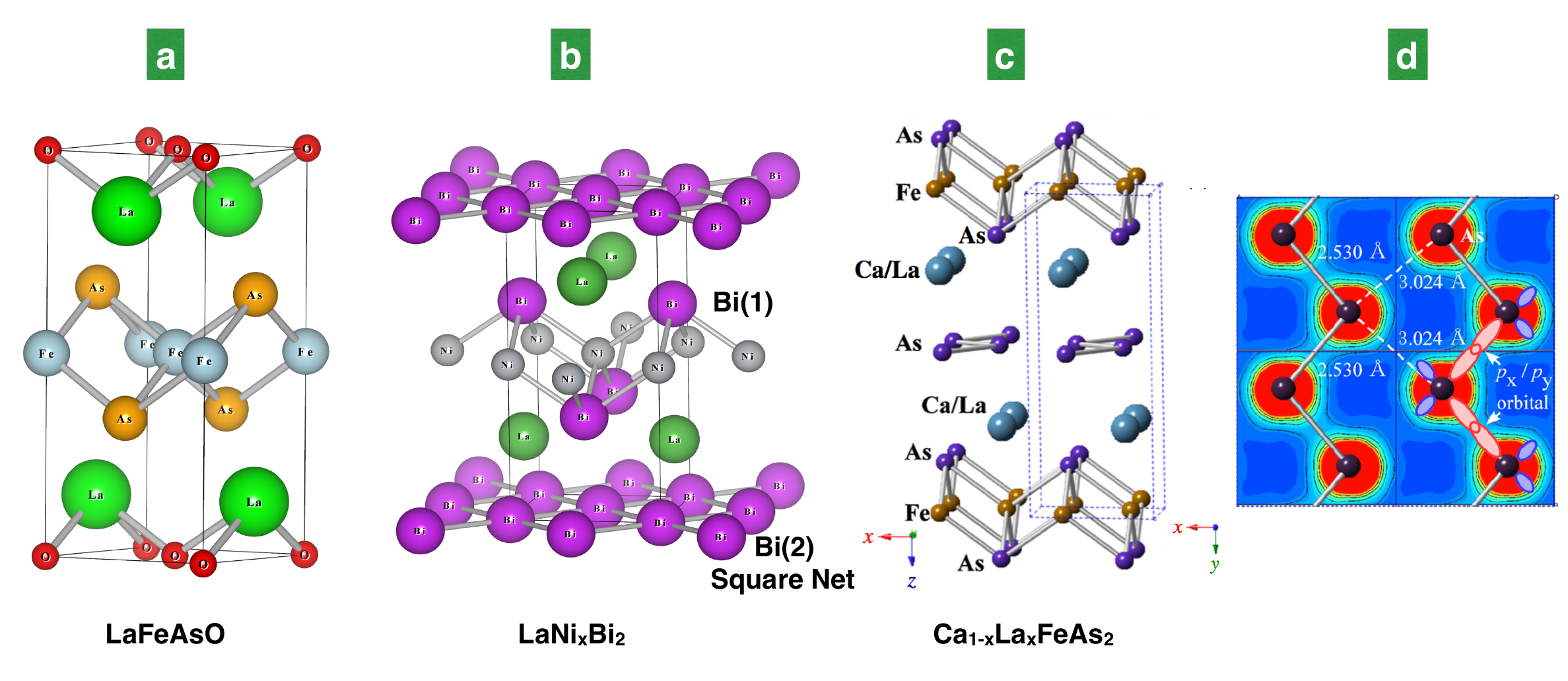}
\caption{{\small Schematic crystal structures of (a) 1111-type LaFeAsO, (b) 112-type $\lnb$ and (c) 112-type $\cl$\cite{Katayama2013}. (d) Top view of the As-zigzag chains in $\cl$. The colour map represents the contour of the charge distribution around As-atoms. Charge accumulation between neighbouring As-atoms are suggestive of the formation of the covalent bonds\cite{Katayama2013, Hosono2015}. [Fig.~\ref{fig.1}(c-d) : Reprinted with permission from Katayama $\e$\cite{Katayama2013}. Copyright 2013 by the Physics Society of Japan.]}}
\label{fig.1}
\end{center}
\end{figure*}

\section{Synthesis Techniques}

Similar to the other pnictide systems, primarily three techniques are used for bulk sample preparation of 112-pnictides\cite{Aswathy2010}: (a) solid state reaction, (b) high-pressure synthesis, and (c) self-flux method. The first two methods are mostly used for polycrystalline and powder samples, while the last one is convenient for single crystal growth. Growth of single crystalline thin films by molecular beam epitxy (MBE) is discussed in detail in Sec.~\ref{sec.thin_film_growth}.

\subsection{Bulk Synthesis}

The parent compound CaFeAs$_2$ has not yet been synthesised, but incorporation of a small amount of La in place of Ca\cite{Katayama2013, Kudo2014a} was found to be essential for stabilising the 112-phase and inducing $\y$. Single crystals of $\cl$ were grown using FeAs-self-flux \cite{Katayama2013, Kudo2014a, Kudo2014b, Xing2015, Katayama2015, Kawasaki2015} where all the constituents Ca, La, As, FeAs were mixed in appropriate stoichiometric ratio inside an aluminium crucible and sealed inside an evacuated quartz tube. In order to avoid contamination from the atmosphere, the whole process was carried out inside a glove box filled with argon gas. The sealed ampules were heated at appropriate temperatures typically at a maximum of 1100$^{\circ}$C and kept there for several hours. Finally the furnace was cooled slowly ($\sim 1.25^{\circ}$C/hour) to room temperature before taking the single crystals of maximum size of 2\,mm \cite{Zhou2015} out of the furnace. Zhou $\e$\cite{Zhou2014, Zhou2015} suggested that larger amounts of starting materials are necessary for the growth of large sized crystals and a small amount of CaO is helpful for crystallisation.

Polycrystalline samples were synthesised using solid state reaction inside a high-pressure cell \cite{Yakita2014, Sala2014}. For Ca$_{1-x}RE_x$FeAs$_2$ ($RE$ = rare earths from La $\rightarrow$ Gd) synthesis \cite{Sala2014}, a mixture of FeAs, $RE$As, Ca and As powders were mixed and pelletised which was later allowed to react inside a boron nitride crucible between 1000-1200$^{\circ}$C for 1\,h under 2\,GPa pressure \cite{Sala2014, Yakita2015, Yakita2015b}. Sala $\e$ \cite{Sala2014} pointed out that high pressure for synthesis is essential to incorporate smaller $RE$-ions and at much higher pressure ($>$ 2\,GPa) doping of Tb, Dy, Ho and Y into the 112-phase could be possible.

Single crystal synthesis of $\cn$\cite{Thamizhavel2003}, CePd$_{1-x}$Bi$_2$\cite{Han2015}, LaPd$_{1-x}$Bi$_2$ \cite{Han2013}, and SrMnBi$_2$ \cite{Wang2011} needed excess Bi flux and different temperature treatments while polycrystalline samples \cite{Mizoguchi2011, Kodama2011, Mizoguchi2012} were prepared inside an evacuated silica tube through standard solid state reaction at elevated temperatures for $RE$Ni$_x$Bi$_2$ ($RE$ = La, Ce, Nd, Y). It was observed that the Bi and Sb-based 112-systems decompose gradually when exposed to ambient atmosphere. Hence, storage of these materials in evacuated atmosphere is essential for achieving longer lifetime\cite{Mizoguchi2011, Kodama2011, Buckow2012, Buckow2013, Reiner2015}.

\section{Crystal Structure}

Initial reports on 112-type $\p$ systems were made on $RE$Ni$_x$Bi$_2$\cite{Mizoguchi2011} where $RE$ = La, Ce, Nd, Gd and other rare earth elements. These 112-compounds crystallise in the HfCuSi$_2$-type structure, which can be related to 1111-type compounds with the ZrCuSiAs-type structure. Structural similarity between the two phases can be found from Fig~\ref{fig.1}(a-b). The 1111-compound LaFeAsO\cite{Kamihara2008} contains two different anions (O/As) while the 112-compound $RE$Ni$_x$Bi$_2$ has only Bi as anion, but in two different valance states, namely Bi(1) and Bi(2). In the stoichiometrically deficient Ni$_x$Bi layer (analogous to the FeAs layer in LaFeAsO), Bi(1) forms a distorted tetrahedron in a trivalent charge state due to Coulomb attraction. This forces the other Bi(2) ion in a charge state of -1 to occupy a narrower space with a shorter Bi-Bi bond-length to form a square net layer as illustrated in Fig.~\ref{fig.1}(b). The presence of the two-dimensional square net layer is the most unique feature of the 112-phase which stabilises due to the Coulomb attraction driven relaxation between the $RE$-ions. The square net layer can be considered as the blocking layer of the 112-system, though it is metallic unlike the insulating blocking layers present in other $\p$ phases. The presence of the two different oxidation states of the pnictide was observed by XPS in $\lps$ \cite{Reiner2015} where two $3d$ photoelectron lines of Sb are separated by $\Delta E$ = 9.4 eV that correspond to -1 and -3 oxidation states of the Sb-atoms. A similar metallic square net layer was also observed in SrMnBi$_2$ (crystallizes in the SrZnBi$_2$-type structure with SG $I4/mmm$ (no. 139) confirmed via neutron scattering measurements\cite{Guo2014} as opposed to LaPd$_x$Pn$_2$ with $P4/nmm$ (no. 129) symmetry) which is metallic with a large $T_{\text{N}}\sim 290$\,K\cite{Wang2011}. Multiple Dirac cone like dispersions were observed close to the Fermi level. In the single crystalline $R$Ni$_x$Bi$_{2\pm y}$ (for $R =$ La, Ce-Nd, Sm, Gd-Dy), a monotonic decrease in the lattice parameters $a$ (2.1\%) and $c$ (5.3\%) were observed due to the lanthanide contraction\cite{Lin2013}.

Initial reports suggests that the 112-type $\cl$ crystallises in monoclinic structure with space group P$2_1$ (no.~4)\cite{Katayama2013} or P$2_{1}/m$ (no.~11)\cite{Yakita2014, Sala2014} which is different from the other $\p$ systems having tetragonal or orthorhombic space groups. However, recently Harter $\e$\cite{Harter2016} observed second harmonic generation (SHG) in $\cl$ that is a signature of noncentrosymmetric crystal structure, which suggets that the space group of $\cl$ should not be centrosymmetric P$2_{1}/m$, but noncentrosymmetric P$2_1$. The monoclinic structure stays stable up to 450\,K\cite{Katayama2013}. Alternatively stacked FeAs layers are present in $\cl$ (Fig.~\ref{fig.1}(c)) separated by zig-zag As bond layers with Ca/La placed in between them \cite{Hosono2015}. The distance between neighbouring FeAs layers is slightly larger than in the 1111-phase materials. The most interesting feature of this material is the presence of 2D As layers with two different bond lengths that were measured using synchrotron X-ray diffraction \cite{Katayama2013} as shown in Fig.~\ref{fig.1}(d). The shorter one ($\sim$ 2.53{\AA}) is identical to the As-As single bond length with As$^{-}$ ($4p^{4}$ configuration) forming a one-dimensional zig-zag chain along the $b$-axis, while the larger one ($\sim$ 3.02{\AA}) corresponds to the inter-chain distance between the zig-zag chains. In the FeAs layer, As is in a As$^{3-}$ valance state ($4p^{6}$ configuration). The chemical formula of $\cl$ can be written as (Ca$^{2+}_{1-x}RE^{3+}_{x}$)(Fe$^{2+}$As$^{3-}$)As$^{-}\cdot xe^{-}$ with excess charge injected inside the FeAs layer \cite{Hosono2015}. This can be compared to the 1111-type CaFeAsF whose structure can be written similarly as (Ca$^{2+}_{1-x}$RE$^{3+}_{x}$)(Fe$^{2+}$As$^{3-}$)F$^{-}\cdot xe^{-}$ where F$^{-}$ forms a square network. In this way, the 112-$\cl$ structure can be related to 1111-CaFeAsF, although the chemical bondings in the spacer layers are not the same in both cases. For CaFeAsF, the CaF layer is made of strong ionic bonds while CaAs layers in $\cl$ consists of zig-zag As chains of covalent bonds weakly coupled to the adjacent Ca layers. For this reason, the interlayer distance between FeAs planes in 112-$\cl$ ($\sim10.35$\AA) \cite{Kudo2014b} is higher than in CaFeAsF ($\sim8.6$\AA) \cite{Matsuishi2008}. This layered structure was confirmed by  high-angle annular dark field-scanning transmission electron microscopy (HAADF-STEM) measurements of (Ca,Pr)FeAs$_2$ \cite{Yakita2014} with an interlayer distance of 10.4\,{\AA}. X-ray scattering data\cite{Joseph2015b} suggested that microscopic manipulation of the electronically active FeAs layer is more effective compared to the larger structural tuning for controlling the superconducting properties of $\cl$.

Recently Joseph $\e$ \cite{Joseph2015a} reported the structural evolution of $\cl$ ($x = 0.18$) with temperature (110\,K - 300\,K) from powdered X-ray diffraction measurements. The expansion of the $c$-axis lattice constant goes through a distinct change around 150\,K which can be correlated with a change in the resistivity slope. In this temperature range, the in-plane lattice constants go through a constant thermal expansion which is an order of magnitude smaller ($\sim 0.3\times10^{-4}${\AA}/K) than for the $c$-lattice constant. The anisotropic thermal expansion suggests a change in the inter-layer interaction with temperature, although its relation to the resistivity change with temperature is not clear yet.

\begin{figure}
\begin{center}
\includegraphics[width=7cm]{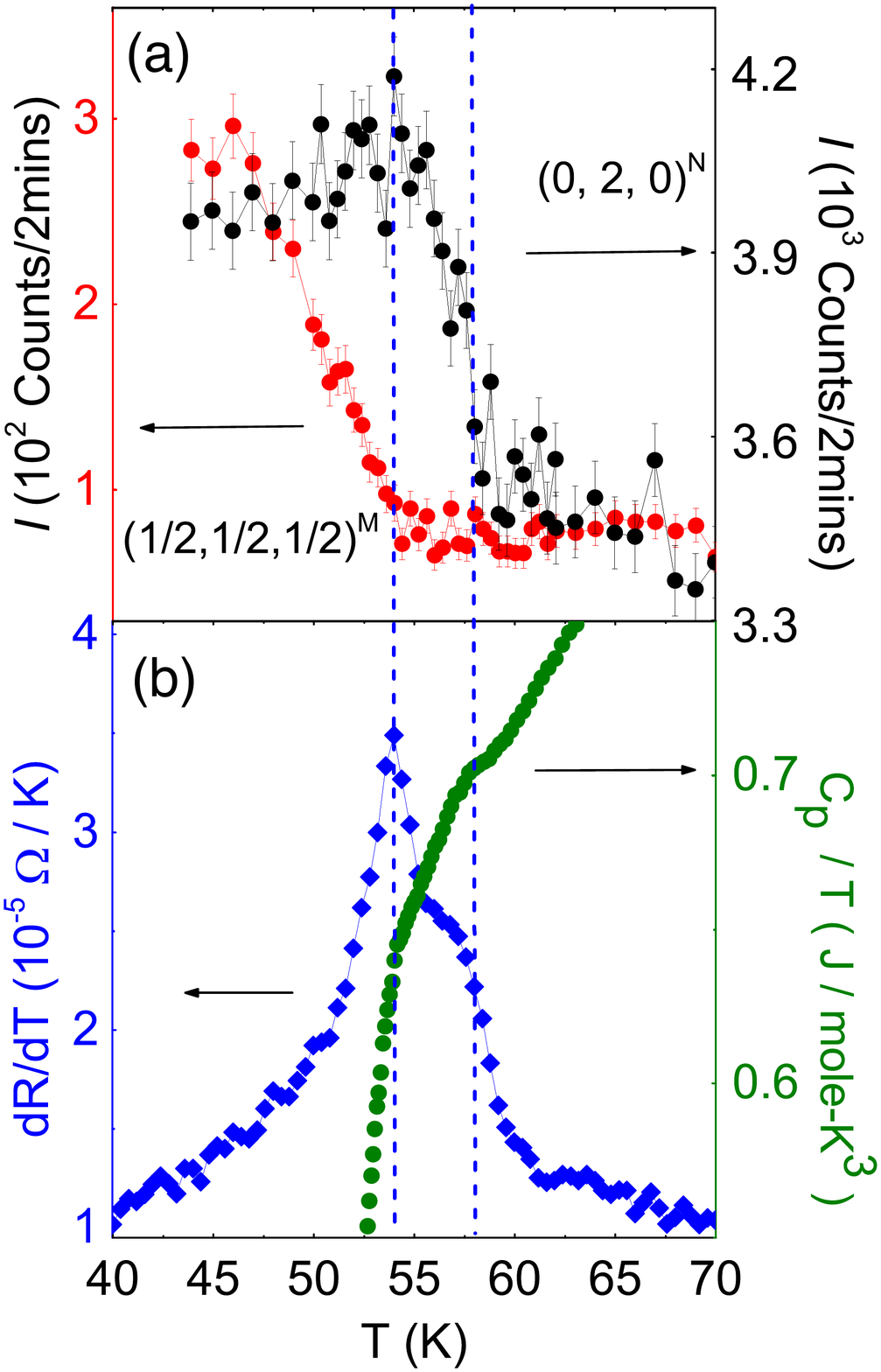}
\caption{{\small Temperature dependence of the (a) neutron intensity of nuclear (0 2 0)$^{N}$ and magnetic (1/2 1/2 1/2)$^{M}$ peaks, (b) Specific heat ($C_p$) and derivative of the in-plane resistivity ($\rho\parallel ab$) in $\cl$. The dotted lines are the temperatures at which the structural resp. magnetic phase transitions were observed\cite{Jiang2016}. Reprinted with permission from Jiang $\e$\cite{Jiang2016}. Copyright 2015 by American Physical Society.}}
\label{fig.2}
\end{center}
\end{figure}

Alternative claims have been made that instead of $\cfa$\cite{Shim2009}, $\cl$ ($x=0.27$)\cite{Jiang2016, Harter2016} is the non-superconducting parent compound of this family which is naturally structurally detwinned at ambient pressure and becomes superconducting on electron or hole-doping. Neutron diffraction and muon spin rotation ($\mu$SR) measurements suggested a structural phase transition from a monoclinic to a triclinic phase at 58\,K and a para\-magnetic to stripe AFM phase transition at 54\,K (Fig.~\ref{fig.2}), both of which can be suppressed by Co-substitution on the Fe-sites\cite{Jiang2016}. Additionally, the presence of two different crystallographic phases in different temperature regions were further confirmed from an optical 2$^{nd}$ harmonic generation study\cite{Harter2016} where no significant modification of the electronic structure was observed as a result of the phase transitions. A similar structural phase transition was recently reported in superconducting $\cl$ ($x=0.15$) around 100\,K using temperature dependent X-ray measurements\cite{Caglieris2016} while Kawasaki $\e$\cite{Kawasaki2015} reported an AFM ordering around $T_{\text{N}}\sim62$\,K for a sample with bulk $\tc\sim$ 35\,K. This suggests a correlation between the structural and AFM phase transition in $\cl$ occurring in similar temperature window and a weak coupling between the structural and magnetic order. 

Upon Co-doping into $\cl$ ($x=0.2$) \cite{Xing2015}, a mixture of 112- and 122-phases were observed. For low Co-content, almost pure 112-phase was found with mono\-clinic structure, but with an enhancement of the Co-doping level a mixture of the 112 and 122-phases were observed which for a higher Co-doping ($> 6\%$) resulted even in the complete disappearance of the 112-phase. Owing to the slightly smaller ionic radius of Co$^{2+}$ (74\,pm) compared to Fe$^{2+}$ (76\,pm), the successful substitution could be confirmed from the shift of the out-of-plane X-ray reflection. This suggests that the structural stability of the 112-phase only exists in a narrow window of Co-doping level putting strong constraints to the crystal growth conditions.

\section{Electronic Structure}

\subsection{Theoretical Investigations}

Density functional theory (DFT) band structure ana\-lysis of fully stoichiometric LaPdBi$_2$ revealed the almost equal contribution of all constituent atoms near the Fermi level, except for Pd which has a much higher dominance. Changing the Pd content in $\cpb$ affects the Fermi surface topology significantly and in the presence of significant Pd vacancies, Fermi surface nesting can be avoided. This suppresses any kind of charge density wave (CDW) in the Bi square net layer which allows the superconducting state to stabilise\cite{Han2013}.

\begin{figure}
\begin{center}
\includegraphics[width=7cm]{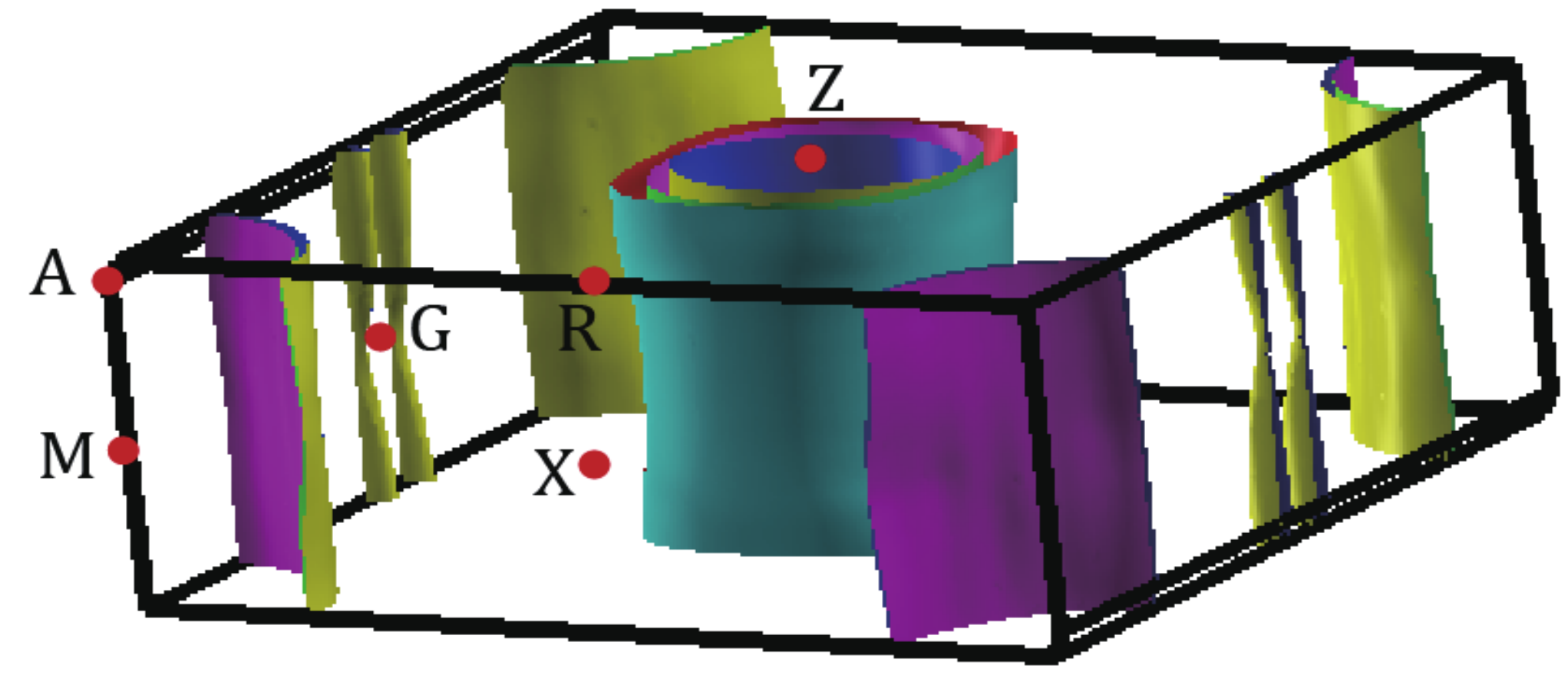}
\caption{{\small DFT calculated Fermi surface of CaFeAs$_2$ in the non-magnetic state\cite{Huang2015}. Reprinted with permission from Huang $\e$\cite{Huang2015}. Copyright 2015 by American Institute of Physics.}}
\label{fig.3}
\end{center}
\end{figure}

\begin{figure}
\begin{center}
\includegraphics[width=8cm]{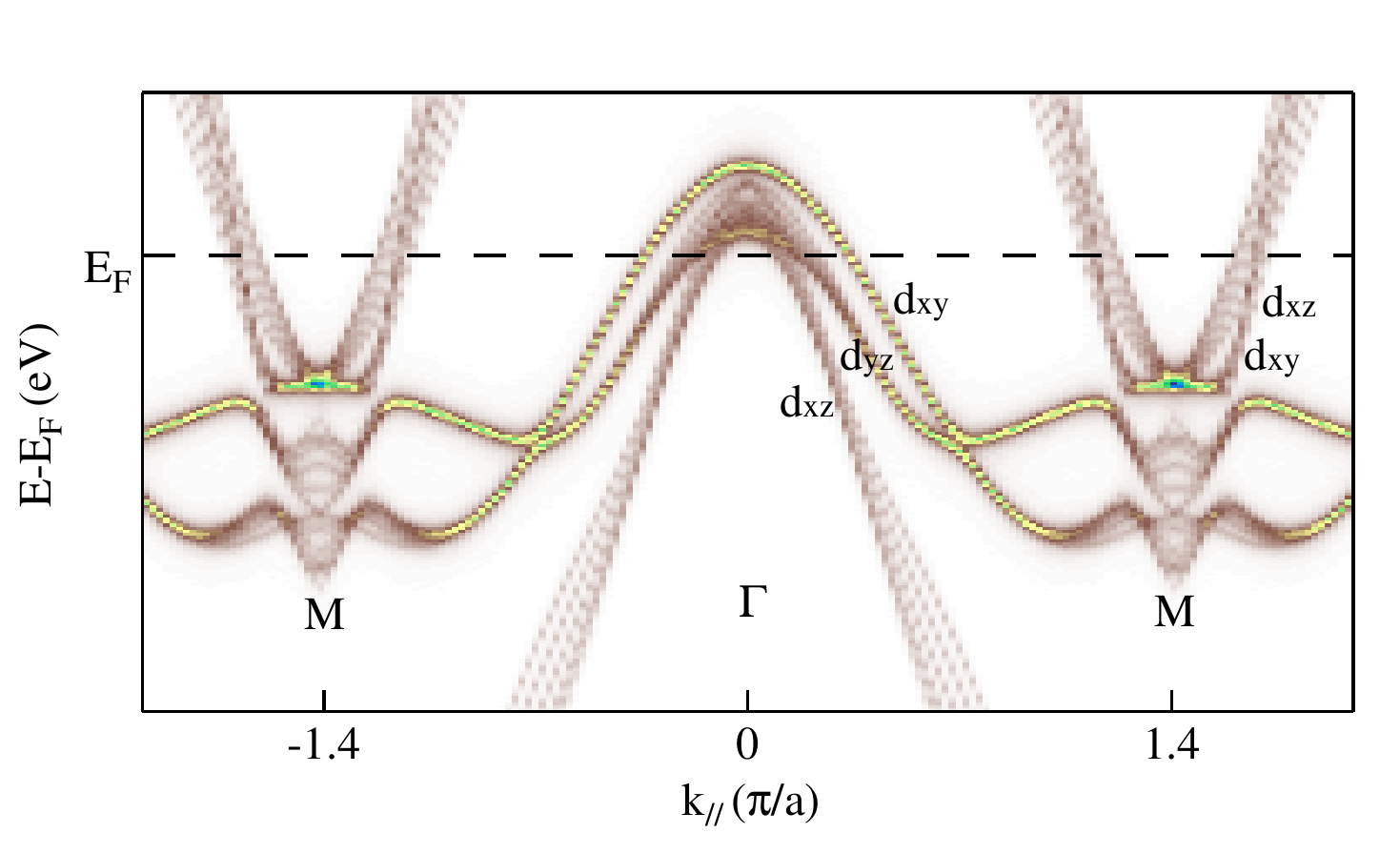}
\caption{{\small Band structure of $\cl$ projected to the in-plane BZ as obtained from ARPES measurements\cite{Liu2015}. Reprinted with permission from Liu $\e$\cite{Liu2015}. Copyright 2015 by American Institute of Physics.}}
\label{fig.4}
\end{center}
\end{figure}

The density of states of two hypothetical 112-structures BaFeAs$_2$ and BaFeSb$_2$\cite{Shim2009} showed a considerable amount of Fe-$3d$ states at $E_{\text{F}}$ with a small contribution from the spacer layer. This has been claimed to originate from the larger distance of separation between the As(1) and FeAs layers with minimal hybridisation between them. For CaFeAs$_2$, first-principles calculation also suggested the strong presence of Fe-3$d$ electrons in the density of states near the Fermi level \cite{Katayama2013, Huang2015}. The Fermi surface of CaFeAs$_2$ consists of two electron cylinders at the Brillouin zone (BZ) corner ($M$ point), four Dirac cone type electron cones at the BZ edge ($G$ point) and three hole cylinders with an additional three-dimensional (3D) hole pocket at the BZ centre ($\Gamma$ point)\cite{Katayama2013, Xu2013, Wang2014, Huang2015} as illustrated in Fig.~\ref{fig.3}. The presence of the additional hole pocket at (0,0,0) (likely originating from the hybridisation between the Fe $3d_{xz}/3d_{yz}$ and As(1) $4p$ orbitals from the FeAs layer) and four electron cones at the $G$ point (contributed by As(2) $p$-states) have not been found earlier in 1111\cite{Mazin2008, Singh2008} and 122\cite{Ding2008}-$\p$ phases. The nesting between the $\Gamma$ point hole pockets and $G$ point electron cones possibly results into a AFM spin density wave (SDW) phase which gets suppressed upon $RE$-doping favoring the superconducting state\cite{Wu2014, Huang2015}.

The band dispersion in $\cl$ showed considerable 2D character contributed mostly by the As layers. In the non-magnetic calculation, four hole-like bands (around $\Gamma(0,0)$ point) and two electron-like bands (around $M(\pi,\pi)$ point) were found from the band structure analysis (see Fig.~\ref{fig.4}) which has some similarity with BaFe$_2$As$_2$\cite{Xu2013}. It was predicted that for $T<\tc$, $\cl$ could work as a natural topological insulator/superconductor hybrid structure (FeAs layer to be responsible for the superconductivity and the As chain layer being the topological insulator) that could be ideal for the realisation of Majorana fermions\cite{Wu2015b}.

Nagai $\e$ \cite{Nagai2015} investigated the effect of Sb-substitution on the superconducting properties of CaFe(Sb$_x$As$_{1-x}$)$_2$. Sb-doping in the As zig-zag layer increased the lattice parameters $a,b$ with an overall increase in the unit cell volume leading to an overall stabilisation of the structure which is energetically more favourable than substitution in the FeAs layer. However, the calculated band structure with and without Sb-doping is very similar except a small shift along the $G-\Gamma$ direction. Primary investigations suggested the role of Sb-substitution in the enhancement of  $\tc$ to be possibly related to the stabilisation of the As-chains which have a crucial role in controlling $\tc$.

\subsection{Experimental Results}

ARPES measurements on $\cl$ indicated a band structure similar to the other $\p$ systems consisting of three hole like bands ($d_{xz}, d_{yz}, d_{xy}$ character) at the $\Gamma$ point and two electron like bands ($d_{xz}, d_{xy}$ types) at the $M$ point of the BZ\cite{Liu2015, Xu2013, Jiang2016, Li2015}, with reasonable nesting mostly originating from the Fe-$3d$ electrons (see Fig.~\ref{fig.4}). No evidence of the As-$p$ or Ca-$d$ states were found in the Fermi surface map scanned over a complete 3D momentum space which is similar to the absence of interstitial blocks on the Fermi surface in 1038-phase (CaFeAs)$_{10}$Pt$_{3.58}$As$_8$ and 1048-phase (CaFe$_{0.95}$Pt$_{0.05}$As)$_{10}$Pt$_3$As$_8$\cite{Thirupathaiah2013} and 1111-phase SmFe$_{1-x}$Co$_x$AsO\cite{Charnukha2015} systems, but in contrast to the predictions made from electronic structure calculations\cite{Wu2014}. However, Li $\e$\cite{Li2015} managed to resolve the presence of an additional 3D hole like band at the zone centre and a fast disappearing band near the X-point at the zone corner in $\cl$ (x=0.1) and also for $x=0.27$\cite{Jiang2016}. The presence of a considerable As-$4p_z$ orbital for the hole like band and As-$4p_{x,y}$ orbitals for the narrow band were resolved while considerable hybridisation between the As-$4p_z$ and Fe-$3d$ orbitals were found in the hole-like band.

\begin{figure}
\begin{center}
\includegraphics[width=8.5cm]{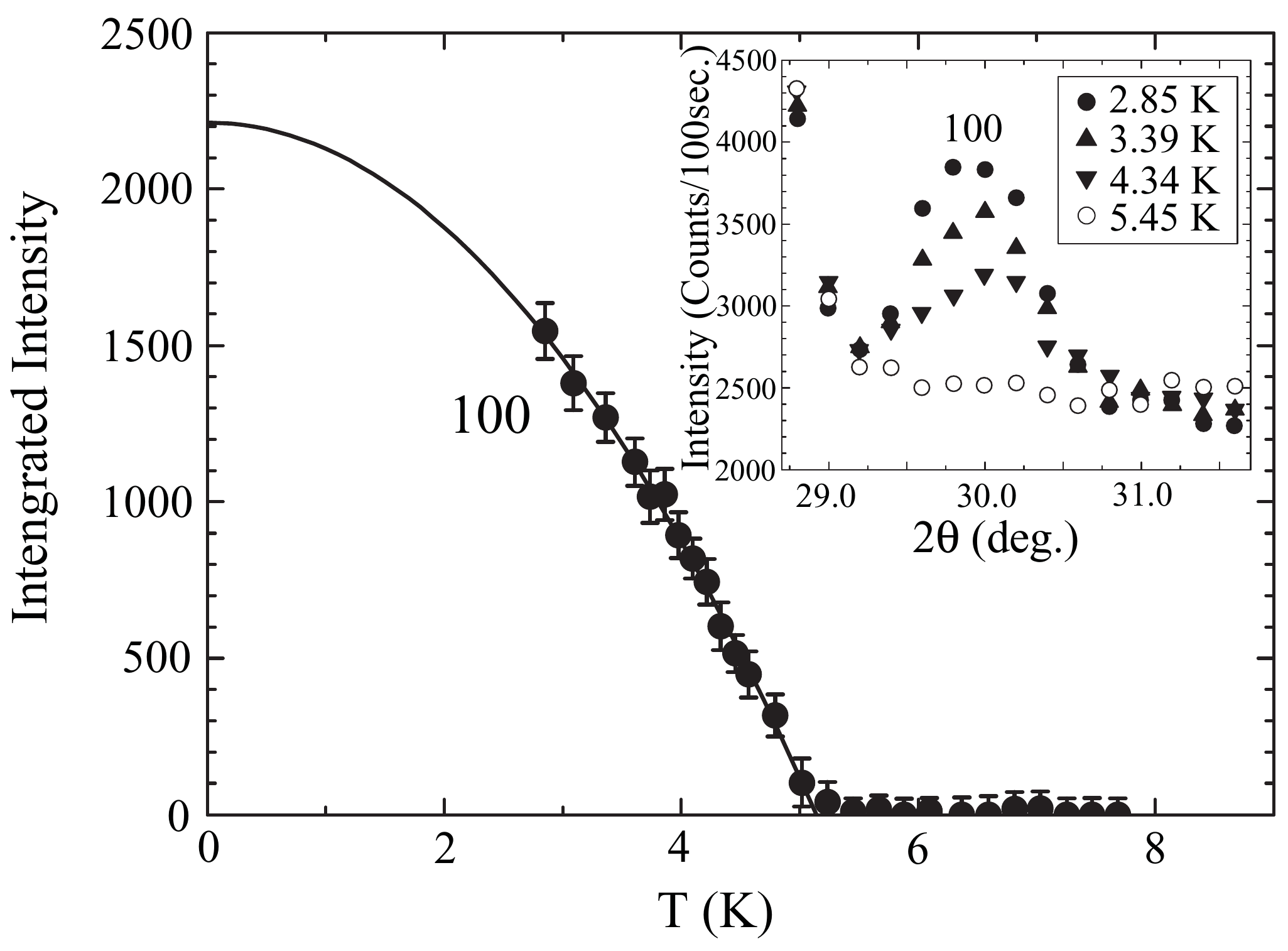}
\caption{{\small Temperature dependence of the integrated intensity of the (100) Bragg peak as obtained from neutron diffraction measurements of $\cn$. The solid line describes a $(1 - (T/T_N)^2)$ dependence of the intensity below 5\,K\cite{Kodama2011}. Reprinted with permission from Kodama $\e$\cite{Kodama2011}. Copyright 2011 by American Physical Society.}}
\label{fig.5}
\end{center}
\end{figure}

\section{Magnetic Structure}

The layered compound $\cpb$ is a Kondo-lattice AFM which is metallic above 75\,K, and exhibits below 75\,K a strong interplay between Kondo and crystal field effects (CEF)\cite{Han2015}. Magnetic susceptibility data suggested that the material is anisotropic due to the CEF effects with an AFM ordering temperature $T_{\text{N}} = 6$\,K\cite{Han2013}. The high-temperature susceptibility fitted using the Curie-Weiss law $\chi = C/(T-\theta_c)$ revealed a Curie Temperature $\theta_c = -1.5$\,K and an effective magnetic moment $\mu_{\rm{eff}}$ = 2.86$\mu_{_B}$/Ce atom which is of similar value as for a free Ce$^{3+}$ ion indicating the localised nature of the Ce-moments in $\cpb$. The negative $\theta_c$ supports the existence of AFM ordering of the Ce-atoms at higher temperatures\cite{Han2013}. The Hall-coefficient $R_{\text{H}}$ of $\cpb$ is negative with an average value of 1.7$\times10^{-4}$ cm$^{3}$/C which corresponds to a carrier concentration of 3.7$\times10^{22}$/cm$^{3}$. $R_{\text{H}}$ stays temperature independent suggesting a single-band character of the carriers \cite{Han2013}. The strong interplay between the Kondo and CEF interactions leads to a reconstruction of the Fermi surface topology which is most likely responsible for the absence of superconductivity in $\cpb$ as iso-structural $\lpb$ is superconducting and both have similar Fermi surfaces above 75\,K. Electronic structure calculation suggested that the Pd-vacancies induce strong scattering effects in the Pd$_x$Bi layer that enhances the CEF effect to quench the Ce moments at low temperature. The Fermi surface reconstruction and Ruderman-Kittel-Kasuya-Yoshida (RKKY) interaction induced magnetic ordering can induce superconductivity in the heavy-fermion system CeCu$_2$Si$_2$\cite{Steglich1979}, which is unlikely to occour in the case of $\cpb$ without external influences\cite{Han2015}.

\begin{figure}
\begin{center}
\includegraphics[width=8cm]{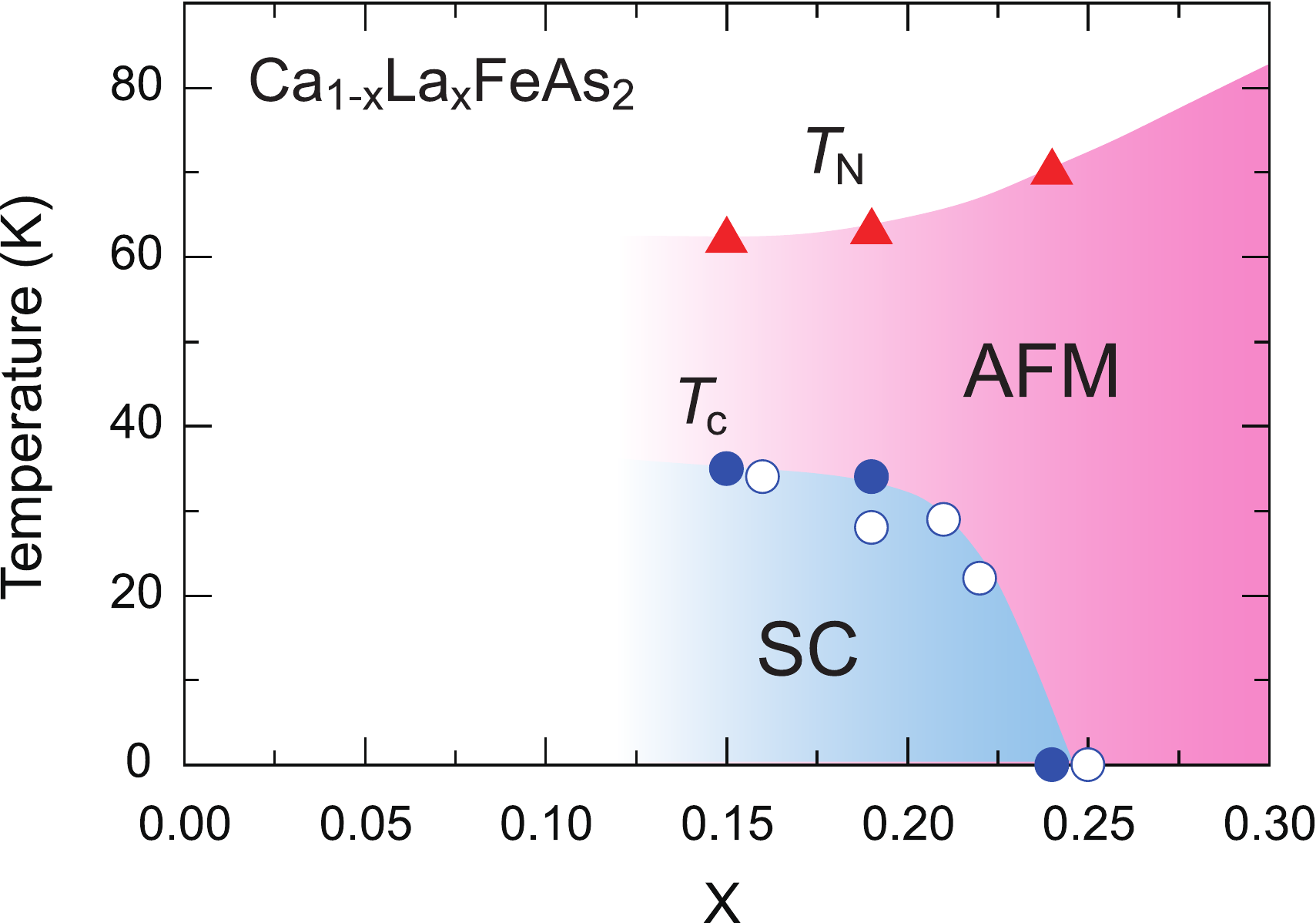}
\caption{{\small Phase diagram of $\cl$ for various doping levels of La\cite{Kawasaki2015}. Reprinted with permission from Kawasaki $\e$\cite{Kawasaki2015}. Copyright 2015 by American Physical Society.}}
\label{fig.6}
\end{center}
\end{figure}

The ground state of the Ce-based intermetallic compound CeNi$_x$Bi$_2$ is governed by the interplay between RKKY and Kondo interactions. The Kondo interaction strength is determined by the level of hybridisation between the Ce-$4f$ and conduction electrons favoring a non-magnetic ground state while the long range magnetically ordered state is preferred by the RKKY interaction\cite{Lin2013}. The parent compound of $\cn$ is CeNiBi$_2$ which is a moderately heavy-fermion antiferromagnet with a magnetic ordering temperature $T_{\text{N}}$ of 5\,K\cite{Kodama2011}. Rosa $\e$\cite{Rosa2015b} reported that $T_{\text{N}}$ in CeNi$_x$Bi$_{2-y}$ increases with an increase in $x$, which also enhances the magnetic anisotropy of Ce$^{3+}$ at low temperature.  Below 5\,K, the Ce moments order antiferromagnetically with the easy magnetisation axis along the $c$-direction with a saturated magnetic moment of $1.71\mu_{_B}$ as $T\rightarrow0$\,K.  Superconductivity is induced in CeNiBi$_2$ via Ni-deficiency similar to the oxygen deficiency in the 1111-system\cite{Kito2008}. From powder neutron diffraction measurements (see Fig.~\ref{fig.5}) clear Bragg peaks are observed at $q = (0,0,0)$ below the magnetic ordering temperature $T_{\text{N}}\sim 5.45$\,K and the intensity of the (100) Bragg peak increases down to the lowest temperature. However, no anomaly or jump was observed in the peak intensity around $\tc$ which is unlike the features observed in heavy-fermion superconductors where Bragg intensities are suppressed below $\tc$\cite{Nicklas2007}. This suggested that Ce-$4f$ electrons do not contribute to the superconductivity and the material cannot be considered to be a heavy-fermion superconductor. Although $\cn$ has a similar crystal structure like 1111-$\p$, the $\tc$ is much lower than in the 1111-system which can be due to the absence of magnetic fluctuations in $\cn$\cite{Kodama2011}. Anisotropic magnetic behavior was observed in the single crystalline $RE$Ni$_x$Bi$_{2\pm y}$ compounds ($RE =$ Ce-Nd, Sm, Gd-Dy) which order antiferromagnetically at low temperatures between 3.3\,K (Sm) and 10.2\,K (Tb)\cite{Lin2013}.

\begin{figure}
\begin{center}
\includegraphics[width=8cm]{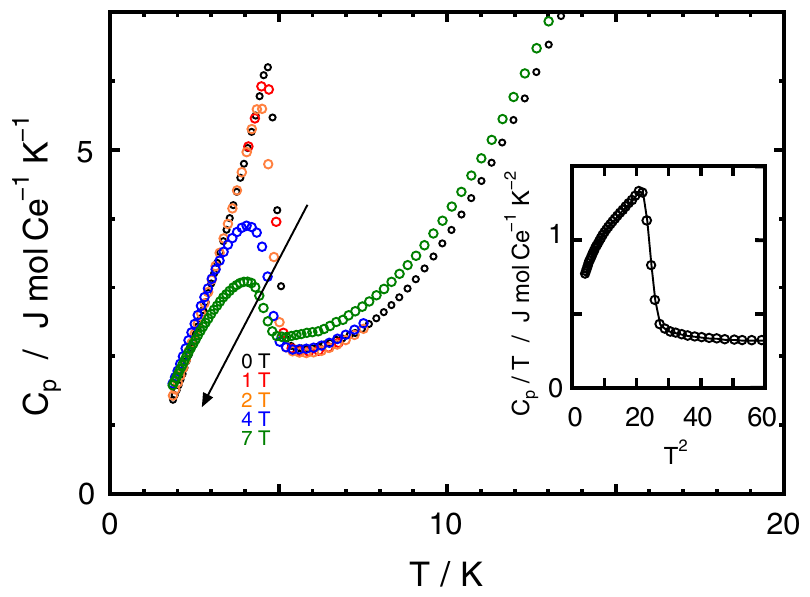}
\caption{{\small Temperature dependence of the specific heat ($C_p$) for different applied magnetic fields with distinct $\lambda$ peak around 5\,K in $\cn$. Inset: $C_p/T - T^2$ at 0\,Oe\cite{Mizoguchi2011}. Reprinted with permission from Mizoguchi $\e$\cite{Mizoguchi2011}. Copyright 2011 by American Physical Society.}}
\label{fig.7}
\end{center}
\end{figure}

From first-principles based investigation, the ground state of CaFeAs$_2$ was predicted to be a SDW type striped AFM phase driven by Fermi surface nesting with the magnetic moment of each Fe atom to be 2.1$\,\mu_{_{B}}$\cite{Huang2015}, significantly smaller in value than the LDA calculated value for 1111-LaFeAsO \cite{Katayama2013} and the hypothetical 112-compounds BaFeAs$_2$ and BaFeSb$_2$\cite{Shim2009}. Electron doping using rare-earth elements can help suppressing the SDW state and stabilising the superconducting state. Nuclear magnetic resonance (NMR) measurements revealed that the AFM ordering sets in below $T_{\text{N}} = 62$\,K for highly doped $\cl$ ($x = 0.15$) while superconductivity sets in at $\tc = 35$\,K \cite{Kawasaki2015}. With an increase in the doping concentration, AFM order gets enhanced as $T_{\text{N}}$ rising up to 70\,K\cite{Kawasaki2015} for $x = 0.24$ (see Fig.~\ref{fig.6}) which possibly originates from the nesting of the Fermi surfaces due to additional electron doping by La\cite{Xu2013, Li2015}. The AFM order in $\cl$ has been found to be robust against $RE$ doping. A similar phase diagram was also obtained for the 1111-system LaFeAsO$_{1-x}$H$_x$\cite{Fujiwara2013, Hiraishi2014} where heavy doping enhanced $T_{\text{N}}$ and, in both cases, a structural phase transition was observed above $T_{\text{N}}$\cite{Hiraishi2014, Jiang2016}. However, the superconducting and AFM orders coexist in $\cl$ microscopically\cite{Kawasaki2015}, while both orders are completely segregated in LaFeAsO$_{1-x}$H$_x$\cite{Sakurai2015}.

\section{Superconducting Properties}

\subsection{Low Critical Temperature Materials}

$\cn$  is a type-II superconductor ($\tc = 4.2$\,K)\cite{Mizoguchi2011} with a SVF of 96\%. Below 10\,K, the resistivity shows a $\rho(T)\sim T^2$ dependence and increases linearly with temperature up to 100\,K. For other $RE$-containing $RE$Ni$_x$Bi$_2$ systems with $RE =$ La, Nd, Y, also $\tc$ values around 4\,K were measured, but the SVF was significantly less than the Ce-containing phase. The specific heat data for $\cn$ (see Fig.~\ref{fig.7}) suggested the presence of two different types of carries with different effective masses. The AFM ordering around 5\,K was caused by the heavy carriers originating from the magnetic interaction between conduction electrons and Ce-$4f$ electrons in the Ni$_x$Bi plane, while the light carriers from the Bi square net layer were responsible for superconductivity. The entropy calculation suggested that the sharp jump in the specific heat around $\tc$ corresponds to the magnetic ordering of Ce-$4f$ moments, which are not involved in the occurrence of superconductivity. Lin $\e$\cite{Lin2013} claimed that the observed superconducting properties of $\lnb$ and $\cn$ are most likely due to the presence of Bi and Ni-Bi binary impurities as very little evidence of bulk superconductivity was found from their single crystalline samples. However, highly crystalline thin films also showed superconductivity around 4\,K\cite{Buckow2012,Reiner2015}. The zero-resistance state in LaPd$_x$Bi$_2$ \cite{Han2013} was observed around 2\,K with a bulk superconducting phase around the same temperature which is likely $s$-wave in nature. In the normal state, it is metallic with a residual resistivity ratio (RRR) of 2.45 indicating strong scattering in the conducting layers.

\subsection{High Critical Temperature Materials}

\subsubsection{Rare Earth Substitution}

\begin{figure}
\begin{center}
\includegraphics[width=6cm]{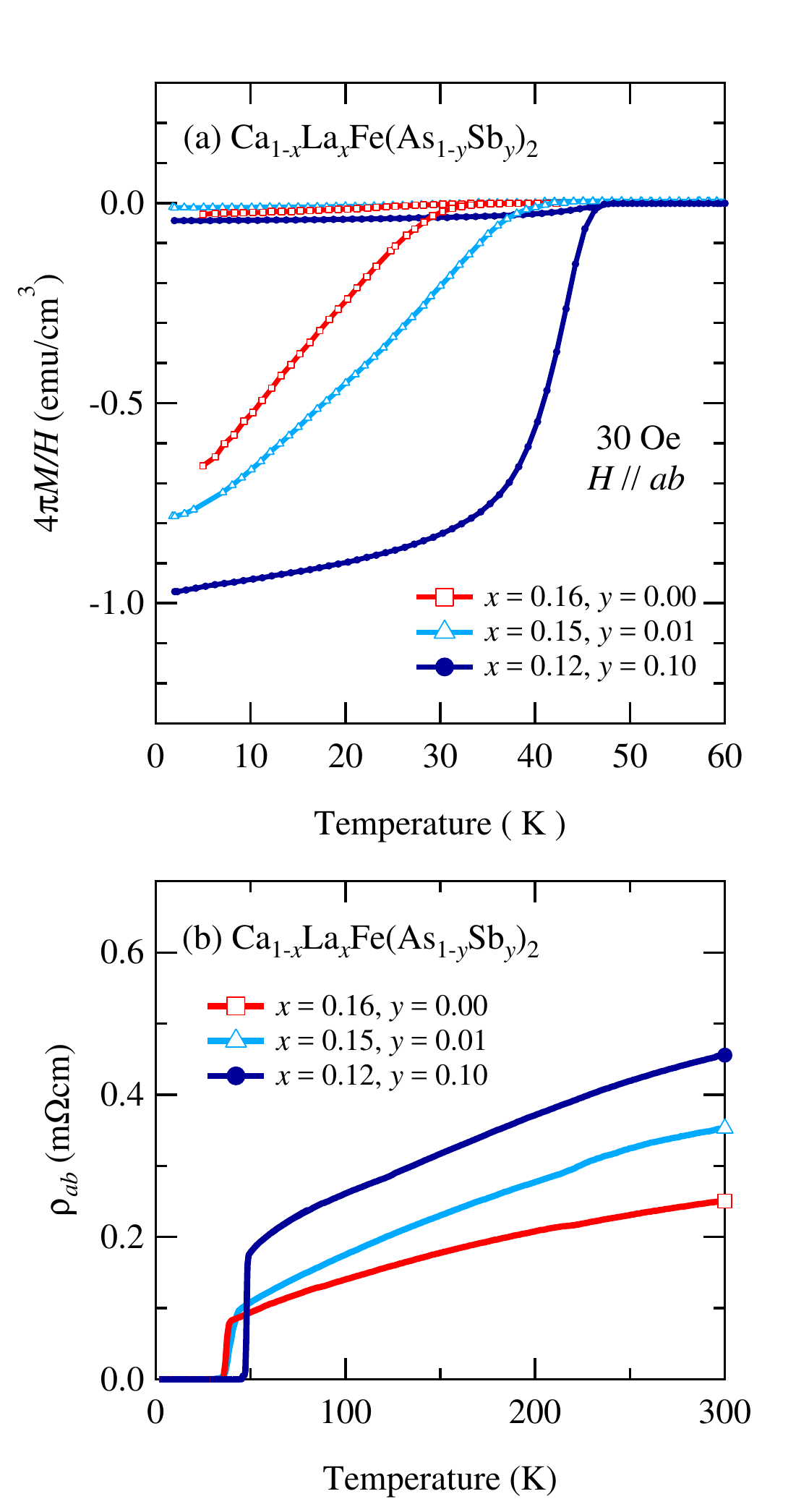}
\caption{{\small Temperature dependence of the (a) magnetisation at $H(\parallel ab) =$ 30\,G in ZFC and FC conditions  and (b) in-plane resistivity ($\rho\parallel ab$) for different La and Sb-doping in Ca$_{1-x}$La$_x$Fe(As$_{1-y}$Sb$_{y}$)$_2$\cite{Kudo2014b}. Reprinted with permission from Kudo $\e$\cite{Kudo2014b}. Copyright 2014 by Physics Society of Japan.}}
\label{fig.8}
\end{center}
\end{figure}

\begin{figure}
\begin{center}
\includegraphics[width=8.5cm]{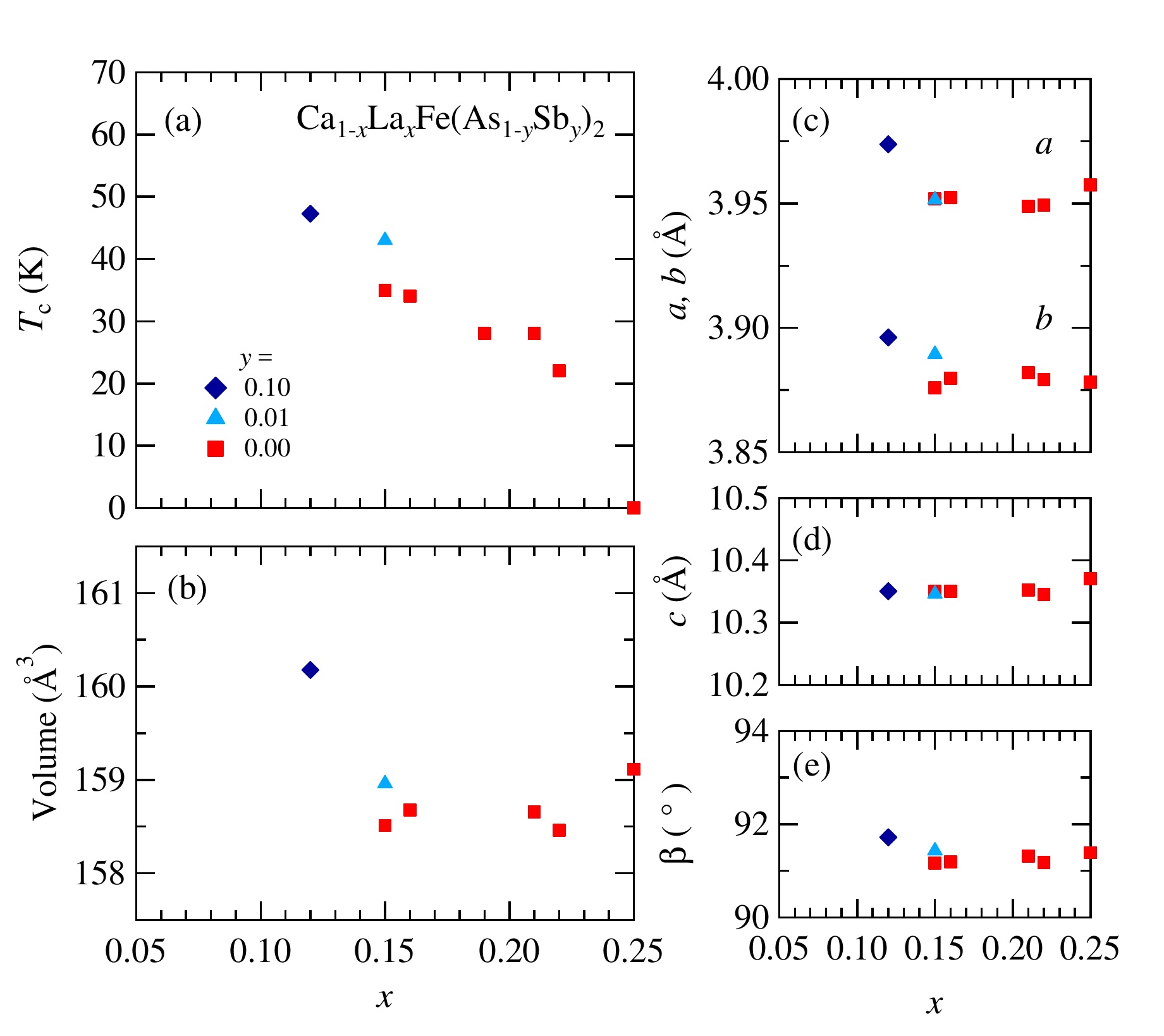}
\caption{{\small Dependence of (a) $\tc$, (b) unit cell volume, (c) $a$, $b$ lattice constants, (d) $c$-lattice constant, and (e) $\beta$ angle of Ca$_{1-x}$La$_x$Fe(As$_{1-y}$Sb$_{y}$)$_2$ for various La-doping levels $x$.\cite{Kudo2014b}. Reprinted with permission from Kudo $\e$\cite{Kudo2014b}. Copyright 2014 by Physics Society of Japan.}}
\vspace{-0.5cm}
\label{fig.9}
\end{center}
\end{figure}

The predicted mother compound of the (Ca,$RE$)-112 family CaFeAs$_2$ has not been synthesised yet and superconductivity was only observed in the $RE$-doped systems within a limited doping range. Doping of $RE$-elements also improved the $\tc$ value of 122-type (Ca,RE)Fe$_2$As$_2$ with $\tc^{onset}$ values : Ca$_{1-x}$Pr$_x$Fe$_2$As$_2$ (47K)\cite{Saha2012}, Ca$_{1-x}$La$_x$Fe$_2$(As$_{1-y}$P$_y$)$_2$ (45K)\cite{Kudo2013}, Ca$_{1-x}$Pr$_x$Fe$_2$As$_2$ (49K)\cite{Lv2011}. However, complete appearance of bulk superconductivity was not observed in some of these cases, possibly relating to the filamentary nature of the superconducting phase. 

The in-plane resistivity of $\cl$ with $x = 0.16$ goes through a superconducting transition at 36\,K \cite{Katayama2013} with a transition width of 2.4\,K and bulk superconductivity was observed from magnetisation measurements at 34\,K with a SVF of 66\%. For $x=0.21$, $T_c^{\text{onset}}$ increased to 45\,K \cite{Katayama2013}, but the zero-resistance state was only observed at 25\,K which is consistent with the $\tc$ determined from the magnetisation data. In (Ca$_{0.9}$Pr$_{0.1}$)Fe$_{1.3}$As$_{1.8}$O$_{0.2}$, a small level of O-doping helped enhancing the SVF  as bulk superconductivity was observed with a $\tc$ of 20\,K\cite{Yakita2014}.

Replacing As by a small amount of isovalent substitutional elements like P and Sb enhanced $\tc$ further \cite{Kudo2014a}. For 5\% P doping in Ca$_{1-x}$La$_x$Fe(As$_{0.995}$P$_{0.005}$)$_2$ ($x = 0.16$) bulk $\tc$ was enhanced to 41\,K with a SVF of 44\% at 5\,K. For $x=0.18$, $\tc$ got reduced to 39\,K, but the SVF increased to 77\%. For Sb-doped Ca$_{1-x}$La$_x$Fe(As$_{0.99}$Sb$_{0.01}$)$_2$ ($x = 0.16$), the bulk $\tc$ increased to 43\,K with a significant SVF of 78\%. From the trend in the $\tc$ vs.~$x$ phase diagram in Fig.~\ref{fig.9}(a), the highest $\tc$ was found for all materials at the lowest $x$ value. This suggests that a much higher $\tc$ could be expected if $x$ can be reduced further below $0.15$ which needs development of chemical synthesis techniques.

\begin{figure}
\begin{center}
\begin{tabular}{c}
\includegraphics[width=6cm]{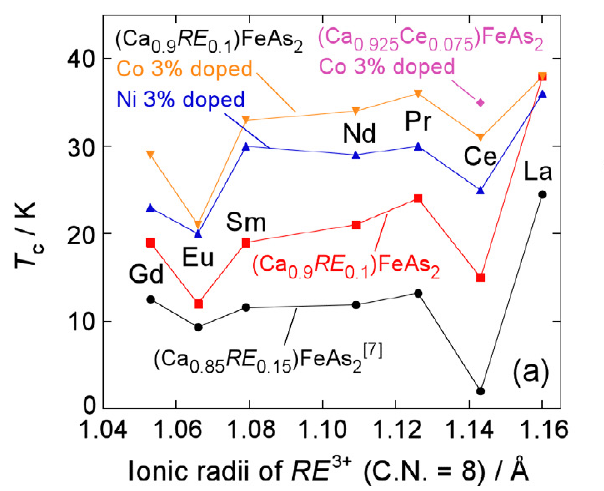} \\
\includegraphics[width=6cm]{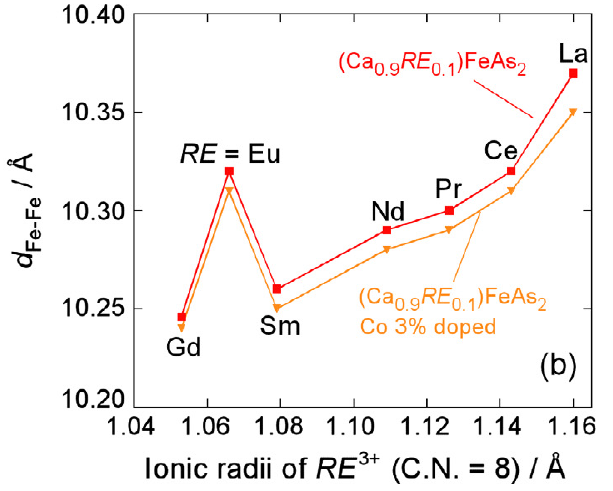}
\end{tabular}
\caption{{\small (a) $\tc$ and (b) $d_{Fe-Fe}$ of (Ca,$RE$)FeAs$_2$ as function of the ionic radii of various $RE^{3+}$ ions with coordination number eight\cite{Yakita2015}. Reprinted with permission from Yakita $\e$\cite{Yakita2015}. Copyright 2015 by Elsevier.}}
\vspace{-0.5cm}
\label{fig.10}
\end{center}
\end{figure}

Further enhancement of $\tc$ was possible with a higher level of Sb substitution in Ca$_{1-x}$La$_x$Fe(As$_{1-y}$Sb$_{y}$)$_2$ ($x = 0.12, y = 0.1$) to 47\,K \cite{Kudo2014b} with a SVF of 100\% at 2\,K (see Fig.~\ref{fig.8}) indicating the appearance of complete bulk superconductivity. This enhanced $\tc$ could originate from two different effects: (1) simultaneous Sb doping, (2) decrease in La content resulting in an increase in cell volume (Fig.~\ref{fig.9}(b)). The decrease in La content suggests the reduction of the number of charge carriers as the ionic radii of La$^{3+}$ and Ca$^{2+}$ are comparable. Secondly, additional Sb substitution can induce a negative chemical pressure due to an increase in cell volume as the ionic radius of Sb$^{3-}$(Sb$^{-}$) is larger than As$^{3-}$ (As$^{-}$) which was observed from the increase in the $a,b$ lattice constants (see Fig.~\ref{fig.9}(c-d)) and the higher level of localisation of the $d$-electrons with a larger sized $Pn$ atom. However, this mechanism is not valid for the $\tc$ enhancement in the P-doped system\cite{Kudo2014a} as neither the La content got reduced nor the lattice parameters changed in that case.

In Sb-substituted Ca$_{1-x}RE_x$Fe(As$_{1-y}$Sb$_{y}$)$_2$, there is a general trend of $\tc$ enhancement and improvement of the superconducting properties\cite{Kudo2014b}. In the absence of Sb ($y=0$), for the Ce-doped system, no evidence of bulk superconductivity was observed, while Pr- and Nd-doped systems exhibited $\tc$ at 10\,K and 11\,K respectively with a SVF of 5\%. For the Ce-doped system, the origin of the zero-resistance state was attributed to filamentary superconductivity as no clear diamagnetic signal was observed at $\tc$. Sb-doping improved $\y$ with a $\tc$ of 21\,K ($y = 0.01$) and 43\,K ($y = 0.1$) for the Ce-doped system, 26\,K ($y=0.01$) and 43\,K ($y=0.05$) for the Pr-doped system, 24\,K ($y = 0.01$) and 43\,K ($y = 0.05$) for the Nd-doped systems with a substantial increase in the SVF indicating the appearance of bulk superconductivity in all these phases. This suggests that irrespective of the $RE$ element, a $\tc$ above 40\,K can be obtained by tuning the Sb content.  Due to Sb substitution, the $b$-lattice parameter increased which helped enhancing $\tc$, but the $c$-value stayed almost unchanged for $x=0.15 - 0.25$ (see Fig.~\ref{fig.9}(c-e)). Kudo $\e$ \cite{Kudo2014b} suggested that $\tc$ can be increased beyond 50\,K if the $b$-lattice constant becomes equal to $a$ with a simultaneous reduction of the $c$-value to adjust the As-Fe-As bond angle.

\begin{figure}
\begin{center}
\includegraphics[width=8cm]{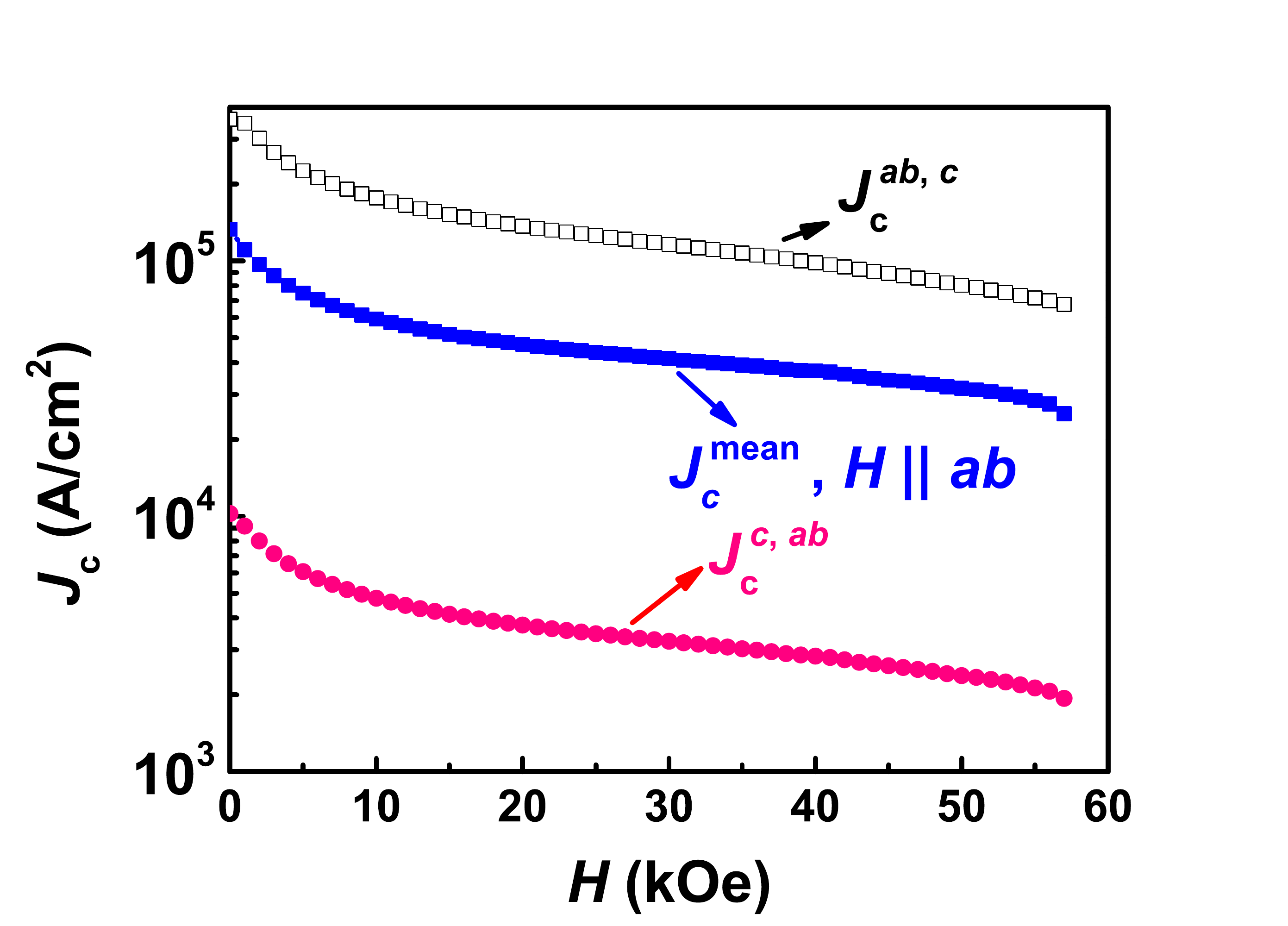}
\vspace{-0.3cm}
\caption{{\small Magnetic field dependence of critical current density $J_c$ of $\cl$ in the in-plane ($H\parallel ab$) and out-of-plane ($H\parallel c$) configuration\cite{Zhou2014}. Reprinted with permission from Zhou $\e$\cite{Zhou2014}. Copyright 2014 by Japanese Society of Applied Physics.}}
\vspace{-0.5cm}
\label{fig.11}
\end{center}
\end{figure}

In polycrystalline Ca$_{1-x}$$RE_{x}$FeAs$_2$, superconductivity was observed with a $\tc$ of 22.7\,K, 24.6\,K, 17.9\,K, 23.2\,K, 13.2\,K and 22.8\,K for La-, Pr-, Nd-, Sm-, Eu- and Gd-doped systems respectively, except for the Ce-doped system\cite{Sala2014}. Broad resistive transitions were observed due to the inhomogeneous distribution of the $RE$ atoms resulting in poor grain connectivity in the polycrystalline phases. There is an indication of a decrease of $\tc$ with decrease of the ionic radii of the $RE$ atoms, a similar behavior was observed for $d_{Fe-Fe}$ (interlayer distance between the neighbouring Fe planes). Note that none of these trends are clearly established due to the inhomogeneity of the phases. Compositional analysis revealed that the actual level of Ce in the Ca site is high due to the similar ionic radii \cite{Yakita2015} which could be a reason behind the absence of $\y$ in the Ce-doped system.

\subsubsection{Transition Metal Substitution}

In the $RE$-substituted material Ca$_{1-x}RE_x$FeAs$_2$, transition metal ($TM$)-co-doping indicates the direct substitution of the Fe atoms by Co-atoms which has been studied for $TM$ = Co, Ni, Mn\cite{Yakita2015, Yakita2015b, Xing2015}. Mn-co-doping suppressed superconductivity completely. However, it was found that a small level (3\%) of Co- and Ni-doping improved the superconducting properties of the (Ca,$RE$)112-systems\cite{Yakita2015, Xing2015}. Co-co-doped Ca$_{1-x}RE_x$(Fe$_{1-y}$Co$_y$)As$_2$ has an enhanced $\tc$ above 30\,K for $RE =$ La, Ce, Pr, Nd, Sm, in spite of the direct $TM$ substitution on the Fe-sites. This effect is similar for Ni-doped systems with slightly lower $\tc$ compared to the Co-doped phases suggesting an overdoped state of the Ni-doped systems.  On 3\% Co-co-doping, the $\tc$ of the (Ca,Pr)-112 system increased significantly from 23\,K ($TM$-free case) to 36\,K\cite{Yakita2015b}, while also the superconducting transition became sharper suggesting improved grain connectivity with significant enhancement in the $T_c^{\text{zero}}$ value (from 14\,K $\rightarrow$ 30\,K) and no change in the $\tc^{\text{onset}}$ value. With an increase in the Co-doping level, a linear suppression of $\tc^{\text{zero}}$ and $\tc^{\text{onset}}$ was observed at an average rate of 1.65\,K/Co\%\cite{Xing2015}.

Large diamagnetic screening was observed in the Co-co-doped (Ca, La)-112-system indicating the presence of bulk superconductivity\cite{Yakita2015}, although it was not clear why a direct substitution of Co for Fe would result in such behavior which is also contrary to doping effects in 122- and 1111-phases \cite{Sefat2008, Sefat2008b}. However, the improved $\si$ properties could arise from the decrease in La-content in the Co-doped phases, similar to Sb doping in (Ca,La)-112\cite{Kudo2014b} which resulted in an optimisation of the As-Fe-As bond angle.

In Ca$_{1-x}RE_x$(Fe$_{1-y}TM_y$)As$_2$ ($TM =$ Co, Ni), $\tc$ increased with increasing ionic radii for all $RE$ elements except for Ce and Eu. $d\rm{_{Fe-Fe}}$ went through a similar trend with ionic radii which is comparable to the trend observed earlier in the $TM$-free (Ca,RE)112-system\cite{Sala2014} (see Fig.~\ref{fig.10}). Co-co-doping resulted in slight reduction of $d\rm{_{Fe-Fe}}$ for all (Ca, RE)112-systems, keeping the $d\rm{_{Fe-Fe}}$ vs ionic radii trend the same for Co-free and Co-doped cases. For the (Ca,Ce)112-system, the reduction of the Ce-content increased the $\tc$ and with Co-co-doping followed the same trend. For the Eu-doped system, an exceptionally low $\tc$ and large $d\rm{_{Fe-Fe}}$ was measured which most likely indicates the coexistence of Eu$^{2+}$ and Eu$^{3+}$.

\begin{figure}
\begin{center}
\includegraphics[width=8cm]{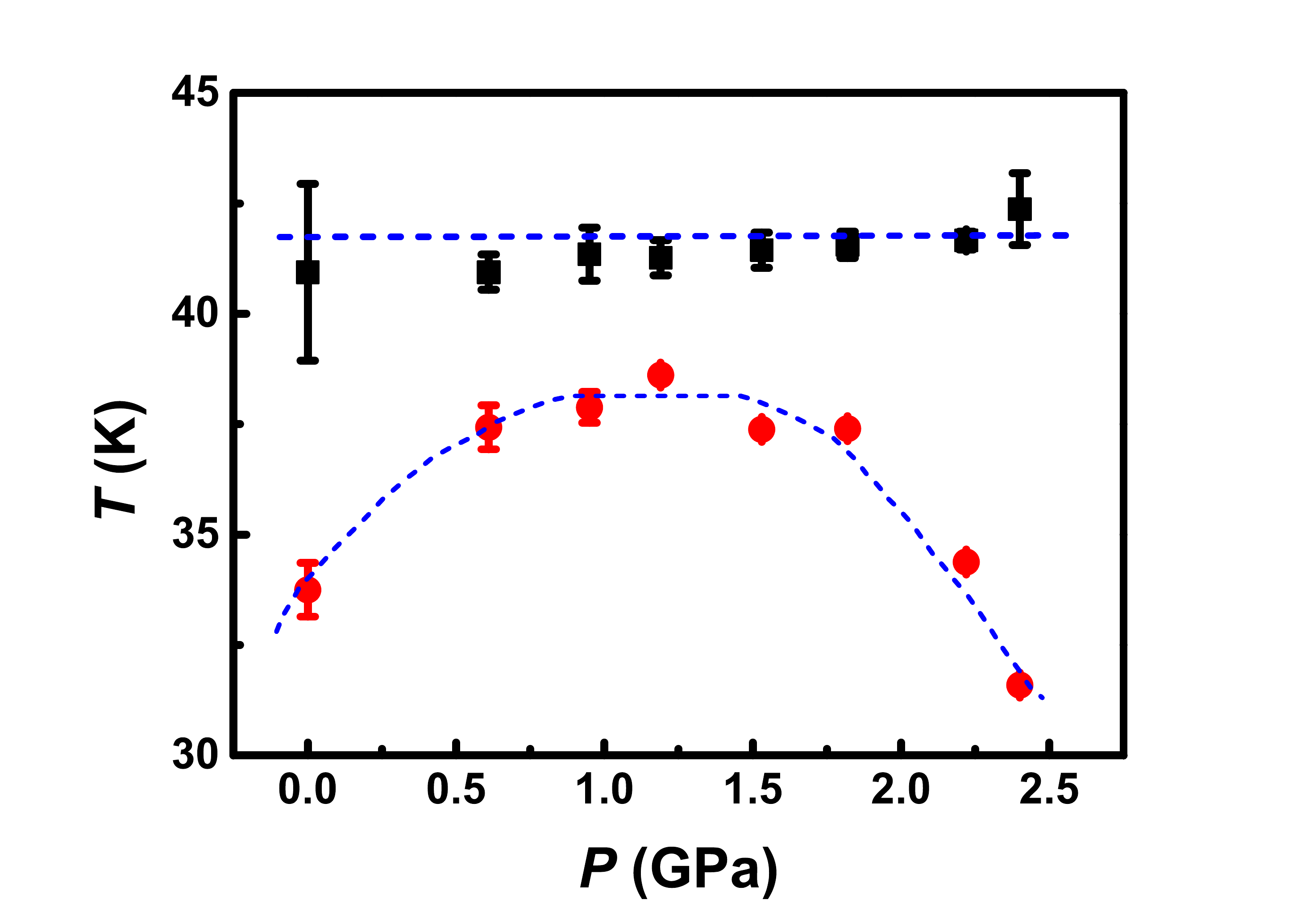}
\vspace{-0.3cm}
\caption{{\small Pressure dependence of $T_1$ and $\tc^{\text{zero}}$ of $\cl$ with $x=0.18$ single crystals\cite{Zhou2015}. Reprinted with permission from Zhou $\e$\cite{Zhou2015}. Copyright 2015 by Institute of  Physics.}}
\vspace{-0.5cm}
\label{fig.12}
\end{center}
\end{figure}

\subsubsection{Critical Current and Critical Field}

The lower critical fields ($H_{c1}$) for $\cn$, LaNi$_x$Bi$_2$, NdNi$_x$Bi$_2$, YNi$_x$Bi$_2$ are  65\,G, 90\,G, 55\,G and 67\,G respectively \cite{Mizoguchi2011}. The upper critical fields ($H_{c2}$) at zero-temperature for LaPd$_x$Bi$_2$ was 3\,T \cite{Han2013} which is relatively large for a $\tc\sim$ 2\,K suggesting type-II superconductivity. For $\cl$, critical fields near $T=0$ are as high as $H_{c2}^{c}(0) = 39.4$\,T (for $H\parallel c$) and $H_{c2}^{ab}(0) = 166.2$\,T (for $H\parallel ab$) which corresponds to coherence lengths\cite{Zhou2014, Caglieris2016} of $\xi_{c}(0) = 6.9$\,{\AA} and  $\xi_{ab}(0) = 28.9$\,{\AA}. The anisotropy $\gamma(0)$ near $\tc$ is 2.8 which lies in between 1111 ($5<\gamma<9.2$)\cite{Jaroszynski2008}- and 122 ($1<\gamma<2$)\cite{Yuan2009}-type $\p$s. $\gamma$ is a measure of the interlayer coupling strength between the FeAs and charge reservoir layers suggesting the presence of moderate anisotropy in 112-system. The anisotropic pinning potential in $\cl$ showed a field dependence for $H\parallel ab$ similar to the cuprates and the 1111-system\cite{Jaroszynski2008} suggesting a transition from single-vortex dominated pinning to a small bundle pinning. This indicates a relatively 2D nature of the superconducting state as that in the 122-system\cite{Zhou2014}. The critical current density $J_c$ for 112-$\cl$ is about $\sim10^5$A/cm$^2$ (see Fig.~\ref{fig.11}), only weakly dependent on the direction and strength of the applied magnetic field\cite{Zhou2014, Caglieris2016}. However, in Sb-doped Ca$_{0.85}$La$_{0.15}$Fe(As$_{0.92}$Sb$_{0.08}$)$_2$, significant improvement in the $J_c$ value\cite{Park2016} was measured ($\sim 2.2 \times 10^6$A/cm$^2$), which upon high energy proton irradiation can be enhaced upto a value of $6.2\times10^6$A/cm$^2$. The $J_c$ value is comparable to that of 11-type Fe$_{1+y}$(Te,Se)\cite{Sun2013} suggesting the importance of  strong bulk-dominated or artificial defect induced pinning in enhancing the value of $J_c$.

\begin{figure}
\begin{center}
\begin{tabular}{ll}
\includegraphics[width=4.5cm]{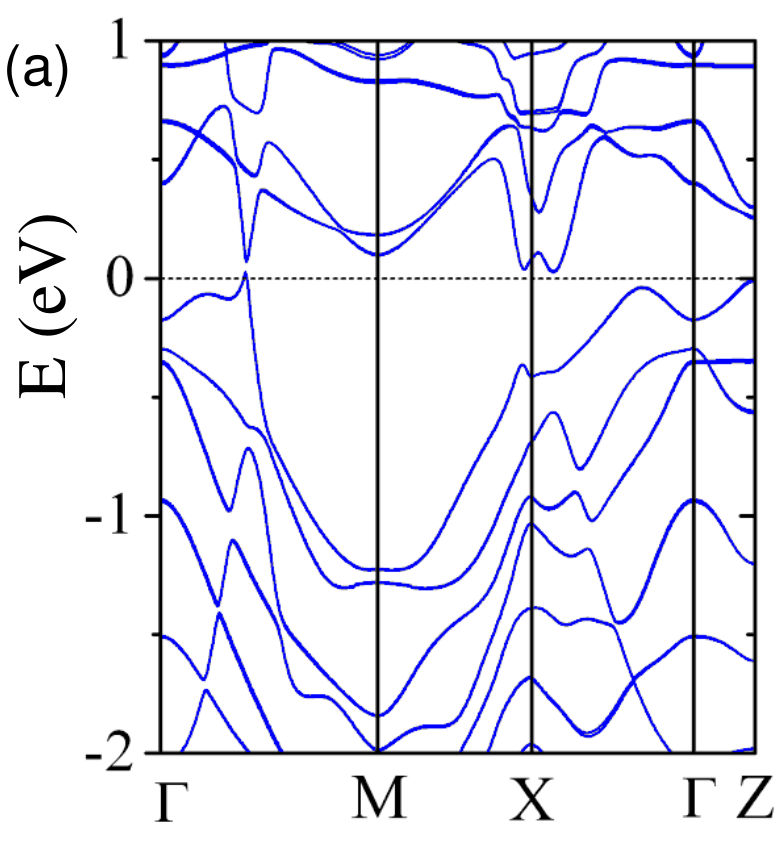} &
\includegraphics[width=4cm]{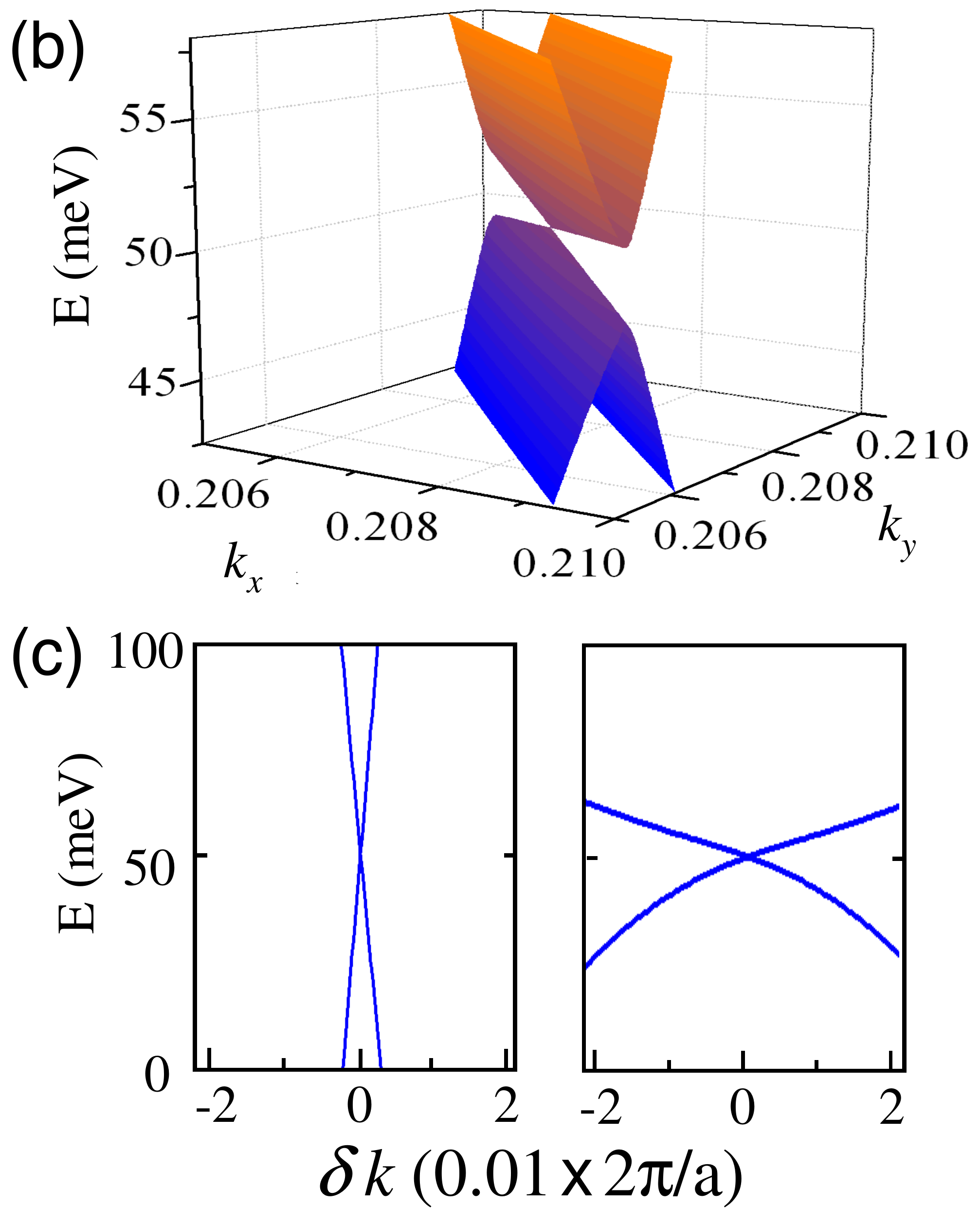}
\end{tabular}
\caption{{\small (a) Electronic band structure of $\smb$. (b) Anisotropic energy surfaces around the Dirac point $k_0 = (0.208, 0.208)$ for the (SrBi)$^+$ layer in $\smb$. (c) Energy dispersions near the Dirac point along parallel and perpendicular directions to the $\Gamma-M$ symmetry line in $\smb$\cite{Park2011}. Reprinted with permission from Park $\e$\cite{Park2011}. Copyright 2011 by American Physics Society.}}
\vspace{-0.5cm}
\label{fig.13}
\end{center}
\end{figure}

\subsubsection{Effect of Pressure}

Application of external pressure can be useful for the suppression of the AFM ground state and stabilisation of superconductivity in Fe-based superconductors without introducing substitutional elements or impurities. For $\cl$, Zhou $\e$ \cite{Zhou2015} observed that the resistivity went through a two-step decrease with temperature towards the superconducting phase while no effect of pressure was observed on the normal state resistivity. The temperature ($T_1$) at which the high-$T$ resistance decrease took place was found to be pressure independent. Both the $T_{c}^{\text{onset}}$ and $T_{c}^{\text{zero}}$ indicated a dome-shaped pressure dependence with its maximum at around 1.19\,GPa as illustrated in Fig.~\ref{fig.12}. The pressure coefficients were comparable to that of other $\p$ phases \cite{Ahilan2009, Wang2012b}. The maximum $T_{c}^{\text{zero}}$ was 38.5\,K at 1.19\,GPa which is much higher than the zero-pressure value of 34\,K, suggesting that the enhancement of $\tc$ is possible further in the doped 112-compounds via tuning of the As-Fe-As bond angle as result of the Sb substitution.

\section{Dirac Fermions}

\begin{figure}
\begin{center}
\vspace{-0.3cm}
\includegraphics[width=8cm]{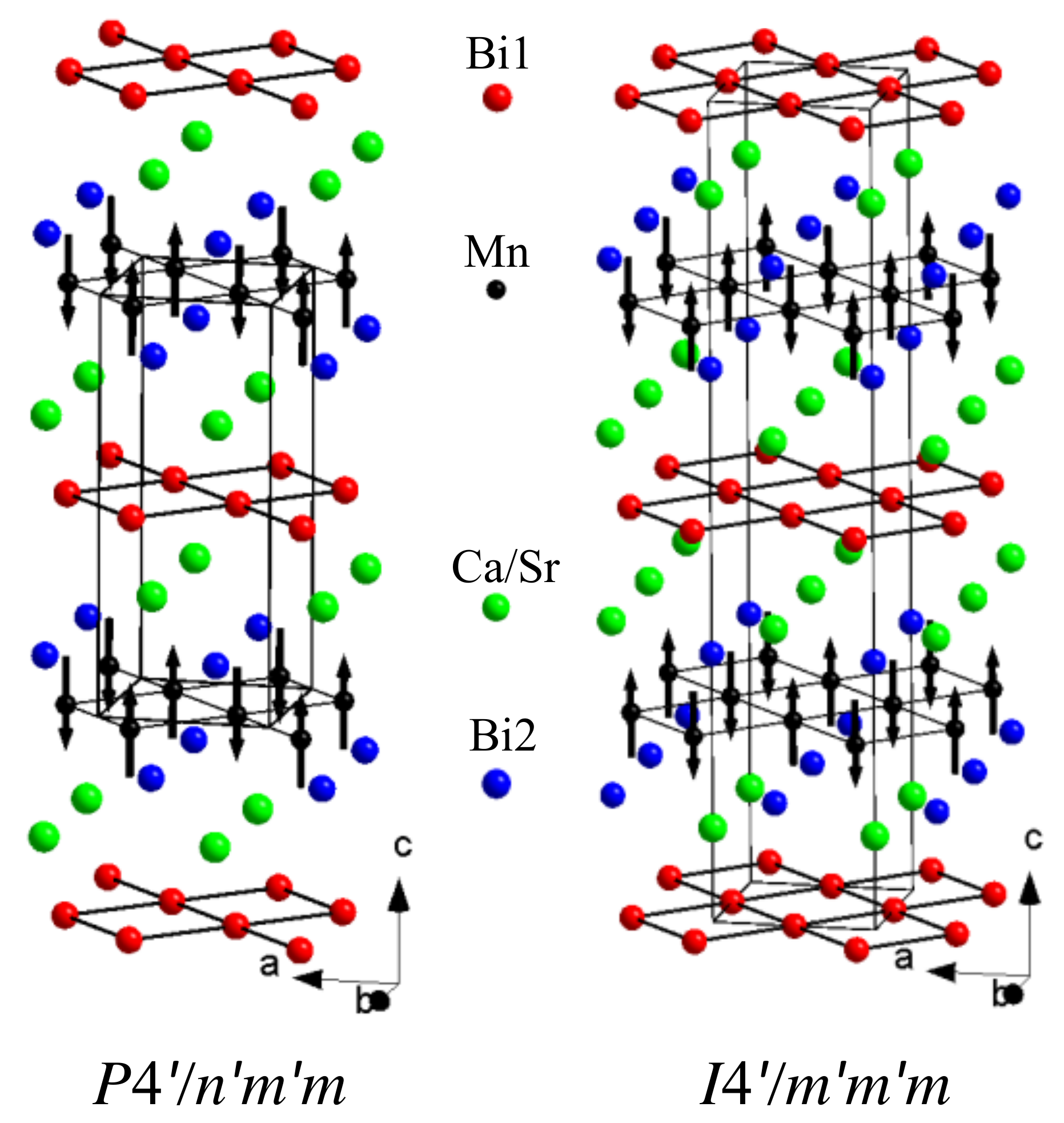}
\vspace{-0.3cm}
\caption{{\small Ground state magnetic configurations of $\cmb$ (left) and $\smb$ (right)\cite{Guo2014}. Reprinted with permission from Guo $\e$\cite{Guo2014}. Copyright 2014 by American Physics Society.}}
\vspace{-0.5cm}
\label{fig.14}
\end{center}
\end{figure}

Dirac materials like graphene\cite{Zhang2005} and topological insulators\cite{Hasan2010} have attracted significant research interest recently due to their novel quantum mechanical properties that could be useful for quantum computation, nanoelectronics and spintronics. The linear energy dispersion in these materials is governed by the relativistic Dirac equation and the crossing of the linearly dispersed bands at the Dirac point forming a Dirac cone\cite{Guo2014}. This configuration suppresses carrier backscattering and enhances electron mobility\cite{Park2011} resulting in novel quantum phenomena like anomalous quantum hall effect and a non-zero Berry phase\cite{Zhang2005, Hasan2010}. The linear energy dispersion also results in large magnetoresistance which increases linearly with magnetic field and does not saturate at higher field\cite{May2014, He2012, Wang2011b, Wang2012f, Wang2012c, Wang2013} as the lowest Landau level can be easily accessed by the Dirac fermions in the quantum limit at moderate applied field. Unlike graphene\cite{Zhang2005}, topological insulators\cite{Hasan2010} and $d$-wave superconductors\cite{Guo2014} with isotropic Dirac cone structures, the range of Dirac materials can be extended further by introducing anisotropy in the Dirac cone for making new electronic devices where electron propagation will be different in different directions from the Dirac point. While various approaches like coupled heterostructures\cite{Pardo2009}, application of strain\cite{Choi2010} etc. have been proposed to generate anisotropy\cite{Feng2014} in Dirac materials, layered intermetallic compounds like $\smb$\cite{Park2011, Wang2011, Wang2011b, Jo2014, Wang2012d, Lee2013, Jia2014, Guo2014, Feng2014}, $\cmb$\cite{Wang2012f, He2012, Lee2013, Guo2014}, $\lab$\cite{Wang2012e, Wang2013, Wang2012d, Shi2015} with 112-$\p$ structures naturally contain anisotropic Dirac cones (see Fig.~\ref{fig.13}). The Bi square net layers host the Dirac fermions in such compounds, and the linear energy dispersion originates from the crossing of the two Bi-6$p_{x,y}$ bands\cite{Park2011} which has also been supported by first-principles calculations\cite{Wang2013} and tight-binding analysis\cite{Lee2013}. The tetragonal unit cell of these materials (space group $I4/mmm$) is constituted by Mn-Bi$_{(2)}$ layers (analogous to the Fe$Pn$ layers) which are separated by Sr/Bi$_{(1)}$ layers (Fig.~\ref{fig.14}) on both sides with the $c$-axis length larger than in other $\p$ superconductors\cite{Shim2009, Wang2011}. Bi exists in two different valance states in the blocking and tetrahedral layers which is common for 112-phase $\p$s.

ARPES measurements on $\smb$\cite{Park2011} revealed the presence of a large circular $\fs$ at the zone centre ($\Gamma$-point) and a needle-like $\fs$ between the $\Gamma$ and $M$ point, as also predicted theoretically. Dirac type dispersion was observed from the needle-shaped $\fs$ along the $\Gamma-M$ line along which the estimated Fermi velocity ($v_F^{\parallel}) = 1.51\times10^{6}$\,m/s (comparable to that of graphene\cite{Zhang2005}), while the velocity perpendicular to the $\Gamma-M$ direction ($v_F^{\perp}$) is $\sim 1.91\times10^5$ m/s. The resulting anisotropy in $v_F$ along different directions of the Dirac cone is $v_F^{\parallel}/v_F^{\perp} \ge$ 5, in agreement with theoretical predictions\cite{Park2011, Jia2014}. Such anisotropy has been claimed to originate from the different levels of hybridisations in different directions, which along the $\Gamma-M$ line is determined by the overlap between neighbouring Bi atoms in the square net layer, and in the perpendicular direction is determined by the hybridisation strength between the Sr-$d_{xy, yz}$ and Bi-$p_{x,y}$ orbitals. Jo $\e$\cite{Jo2014} demonstrated that the interlayer conduction in $\smb$ can be valley-polarised under the presence of a tilted magnetic field which also enhances the anisotropy significantly ($\sim 100$) at high magnetic field.

\begin{figure}
\begin{center}
\includegraphics[width=9cm]{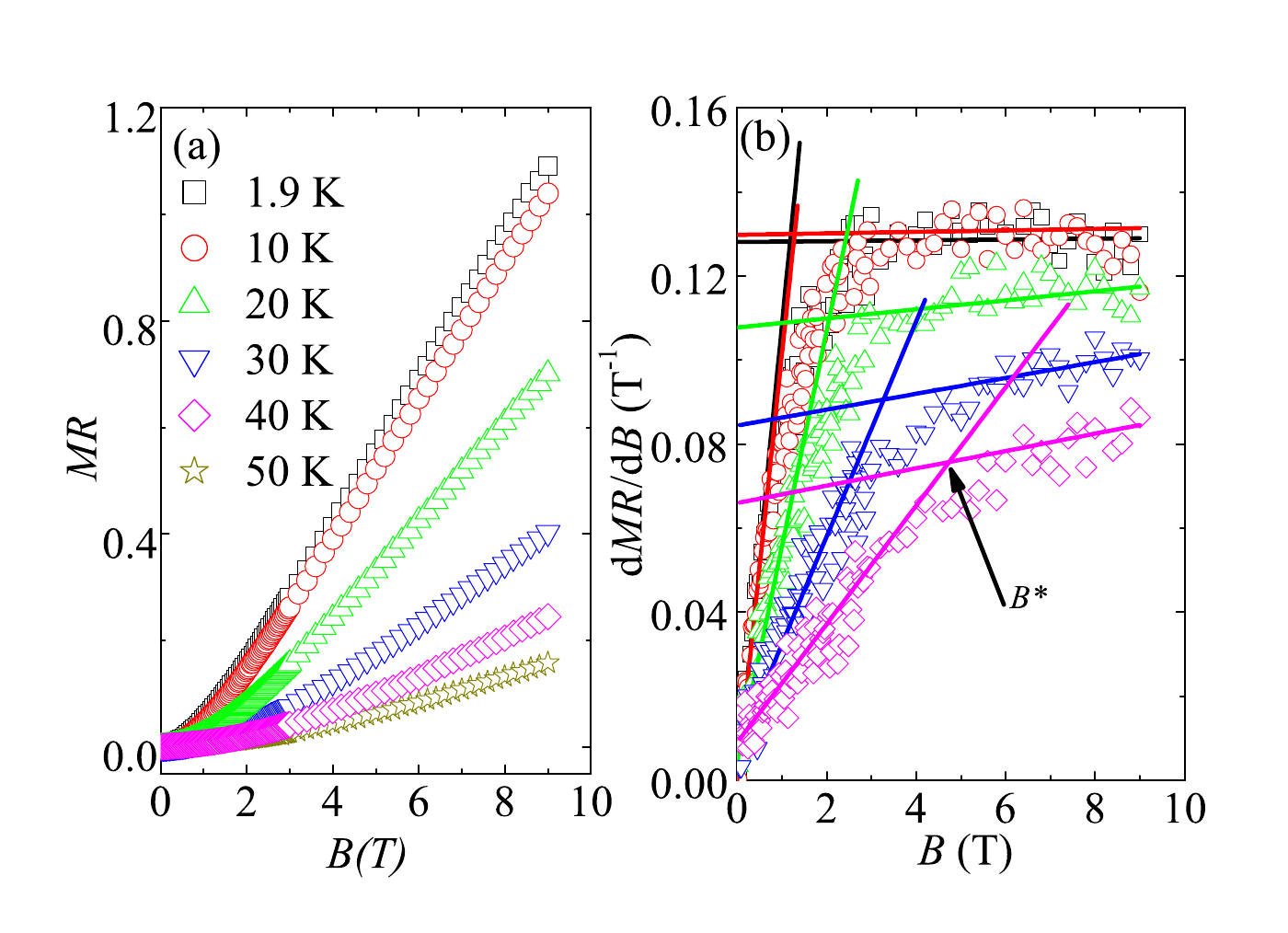}
\vspace{-0.7cm}
\caption{{\small (a) The magnetic field dependence of the in-plane $MR =\rho(B)-\rho(0))/\rho(0)$ at various temperatures. (b) First derivative of $MR$ (d($MR$)/d$B$) as function of applied magnetic field at different temperatures. Data at high field was fitted by $MR\sim f(B) + f(B^2)$ and at low field by $MR \sim f(B^2)$\cite{Wang2011b}. Reprinted with permission from Wang $\e$\cite{Wang2011b}. Copyright 2011 by American Physics Society.}}
\vspace{-0.5cm}
\label{fig.15}
\end{center}
\end{figure}

Unlike graphene with a negligible spin-orbit coupled (SOC) band gap\cite{Min2006}, the SOC gap for $\smb$ is $\sim$ 40\,meV at the Dirac point which can produce a large spin-hall effect\cite{Park2011}. The small effective mass ($0.29\,m_e$), relatively larger carrier mobility ($250$\,cm$^2$/Vs) and small Fermi surface volume support the existence of Dirac fermions in $\smb$. This is further confirmed by the observation of a non-zero Berry phase (=0.60(9)) from Shubnikov-de Haas (SdH) oscillations that is expected to be 0.5 for Dirac fermions for a graphene monolayer\cite{Zhang2005}. As the Bi square net layer hosts Dirac fermions, the SOC gap size could be engineered by replacing Bi with other pnictogens with lower atomic numbers\cite{Farhan2014, Wang2012c}. Feng $\e$\cite{Feng2014} observed differences in Dirac cone structures of $\cmb$ and $\smb$ originating from spin-orbit coupling and the arrangement of the Sr/Ca ions above and below the Bi square net layer which resulted in a larger gap size in $\smb$ than in $\cmb$.

It was thought earlier that the antiferromagnetic ordering of the Mn-atoms for the cases of (Sr/Ca/Eu)MnBi$_2$ is crucial for the anisotropic behavior, but this has been ruled out after the observation of a Dirac cone like structure close to $E_F$ along the $\Gamma-M$ direction in $\lab$ \cite{Wang2012e, Wang2013, Shi2015}. However, magnetism seems to be essential for Dirac materials as the long-range magnetic order couples the Dirac fermions which influences the transport behavior\cite{Guo2014} as observed from the differences in the magneto-transport behavior of $\smb$ and $\cmb$. Both materials posses N\'{e}el type in-plane AFM order, however, the neighboring MnBi$_4$ layers are coupled ferromagnetically in $\cmb$ and antiferromagnetically in $\smb$ which results in a $T_{\text{N}}$ anomaly in $\smb$, but not for $\smb$ as shown in Fig.~\ref{fig.14}. The opposite interlayer couplings were claimed to originate from the competition between the AFM super-exchange and FM double-exchange interactions\cite{Guo2014}.

\begin{figure}
\begin{center}
\includegraphics[width=8cm]{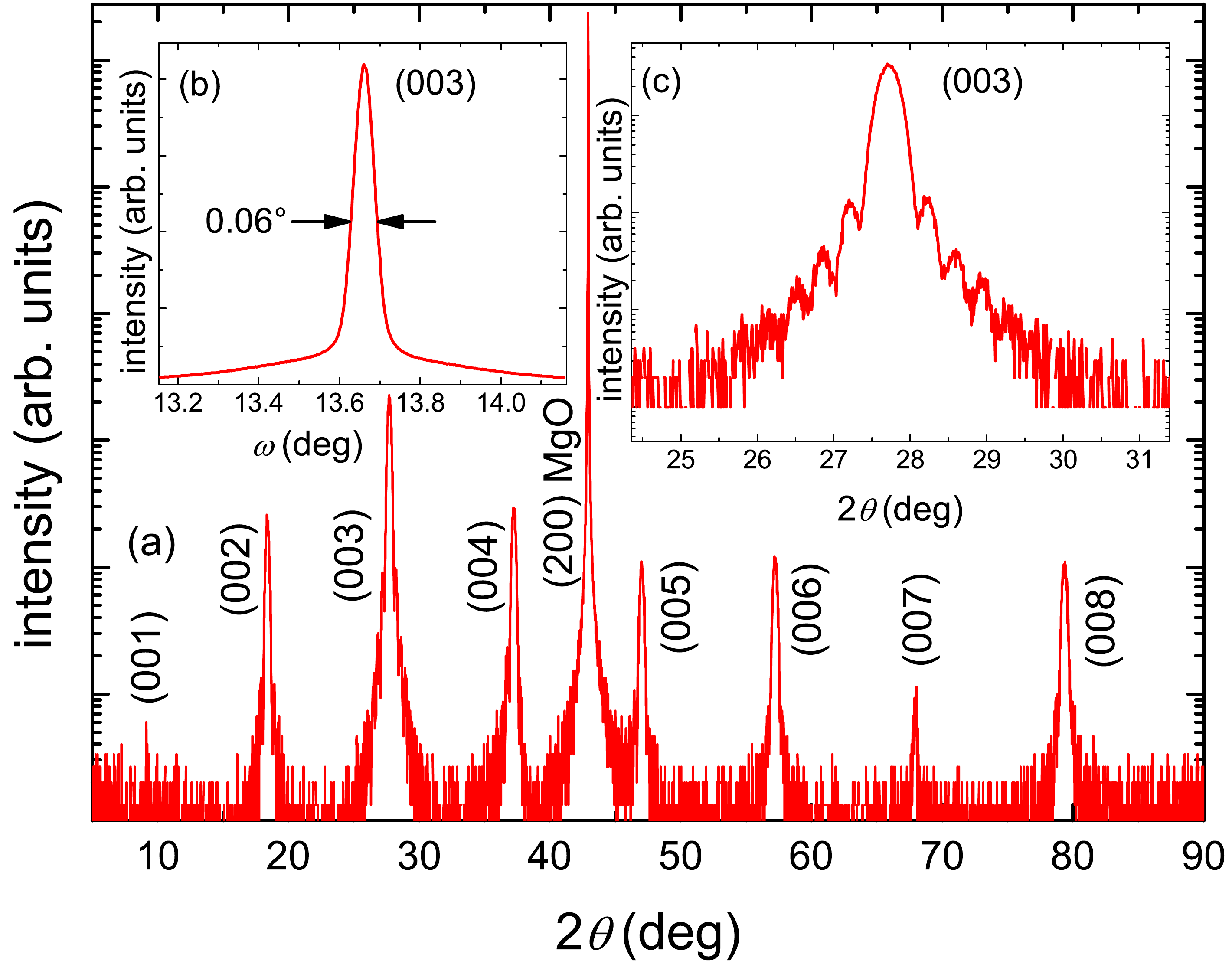}
\vspace{-0.3cm}
\caption{{\small (a) Out-of-plane (2$\theta-\omega$) XRD scans of $\lnb$ grown on a MgO(100) substrate. (b) Typical rocking curve of the (003) peak. (c) Zoom of the (003) peak with Laue fringes\cite{Buckow2013}. Reprinted with permission from Buckow $\e$\cite{Buckow2013} Copyright 2013 by Institute of Physics.}}
\vspace{-0.5cm}
\label{fig.16}
\end{center}
\end{figure}

For the case of $\smb$ and $\cmb$, Mn-atoms order antiferromagnetically\cite{Wang2011} between 270\,K and 290\,K\cite{Park2011, Wang2013} along the $c$-axis, which is different from the case of iso-structural $\emb$ where Mn atoms order at 315\,K and Eu-moments at $T_{\text{N}} = 22$\,K\cite{May2014}. A significant anisotropy enhancement was observed due to the interaction between Eu and Mn-moments that leads to a large increase in magnetoresistance $\sim 650$\% at 5\,K (12\,T) which stays unsaturated at the highest applied field of 13\,T at 5\,K\cite{May2014}. Transverse magnetoresistance behavior for $\smb, \cmb, \lab, \szs$\cite{Wang2011b, Wang2012c, Wang2012e, Wang2012f, Wang2013} goes thorough a semi-classical low field ($\sim B^2$) dependence to a linear ($\sim B$) dependence in the high-field limit around a critical field ($B^*$) as illustrated in Fig.~\ref{fig.15}. The quadratic temperature dependence of $B^*$ can be attributed to Landau level splitting of the linear energy dispersion at high field, with magnetoresistive mobility comparable to that of graphene\cite{Wang2011b}. Linear magnetic field dependence of the $MR$ supports the existence of a linear dispersion which for $\cmb$ is 105\% at 10\,T for $H\parallel c$\cite{He2012} and 120\% at 9\,T ($H\perp c$) and 2\,K\cite{Wang2012f}, for $\szs\sim  300$\% at 9\,T and 2\,K\cite{Wang2012c}. Large magneto-thermopower was measured in $\smb$ with a maximum change of 1600\% at 9\,T and 10\,K. The sign of thermopower is positive for $\smb$ and negative for $\cmb$ suggesting that hole and electron-type carriers are responsible for them respectively, although the thermal conductivity stayed independent of magnetic field\cite{Wang2012d}. The anisotropy and magnetoresistive behavior suggest the possible universal existence of Dirac fermion states in layered compounds with 2D double sized Bi square net layers\cite{Wang2013}.

\begin{figure}
\begin{center}
\includegraphics[width=8cm]{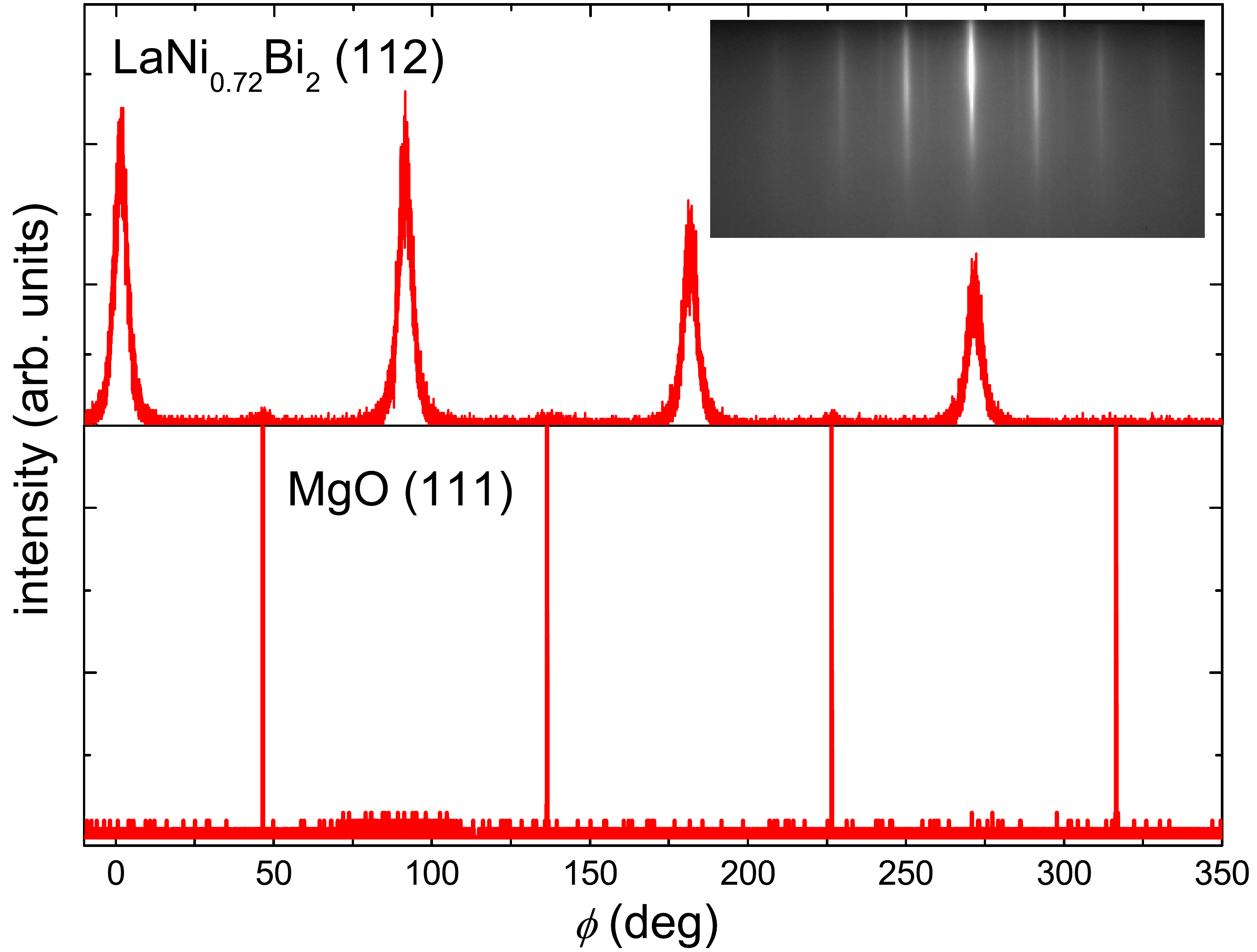}
\vspace{-0.3cm}
\caption{{\small In-plane X-ray ($\phi$) scan of $\lnb$ grown on MgO substrate and (inset) RHEED image taken along the (110) direction of MgO after deposition\cite{Buckow2013}. Reprinted with permission from Buckow $\e$\cite{Buckow2013}. Copyright 2013 by Institute of Physics.}}
\vspace{-0.5cm}
\label{fig.17}
\end{center}
\end{figure}

\section{Thin Films}
High-quality thin films of pnictide compounds will be important for applications, however, they can also do as an useful alternative to those materials where single crystals are still unavailable. Single crystalline phases of pnictide superconductors were grown using MBE and pulsed laser deposition (PLD) showing that thin film techniques are useful for stabilising pnictide compounds (including metastable materials) by using growth kinetics and substrate induced epitaxial strain which is difficult to realize in bulk form. Superconducting films of 122 and 1111-phases of pnictides were grown successfully using PLD \cite{Hiramatsu2008a, Hiramatsu2008b} and, in some cases, higher crystallinity and superior superconducting properties were observed in thin films compared to their bulk counterparts \cite{Iida2009, Katase2010}. Recent observation of high-$T_c$ ($\sim 65$\,K) in the 11-structure compound FeSe \cite{Wang2012, Liu2012, Tan2013, He2013, Sun2014} has fuelled interest in superconducting thin film systems. In situ measurements on a monolayer of FeSe grown on SrTiO$_3$ has revealed a $T_c > 100$\,K \cite{Ge2015} which is the highest (almost 10 times higher than bulk $T_c \sim 10$\,K \cite{Hsu2008}) in any pnictide system observed so far. The superconductivity in FeSe films  has been claimed to originate from interface mode coupling, thus, is most likely not a bulk effect.

\subsection{Growth by Molecular Beam Epitaxy Technique}
\label{sec.thin_film_growth}

\begin{figure}
\begin{center}
\includegraphics[width=7.5cm]{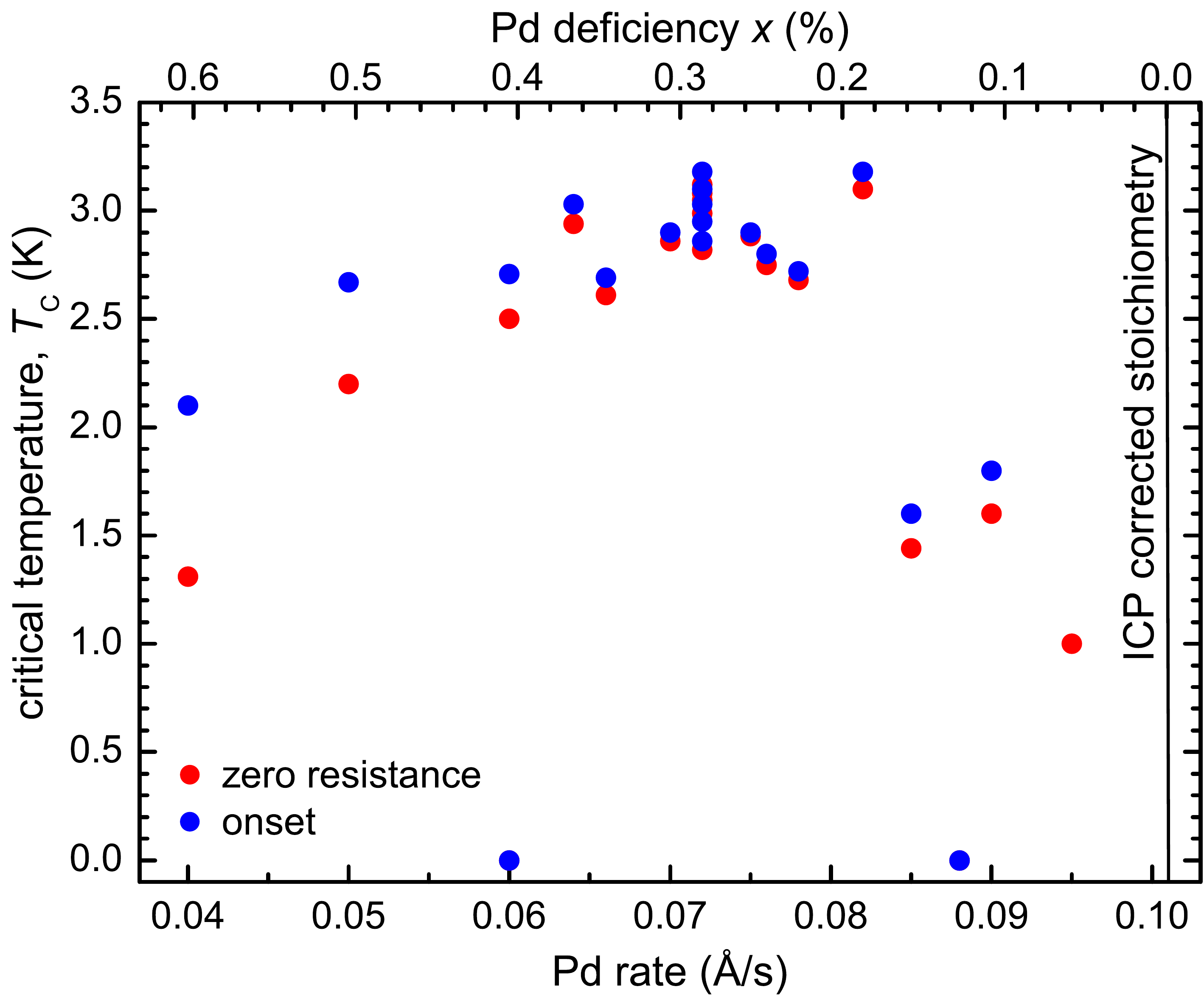}
\vspace{-0.3cm}
\caption{{\small Phase diagram of LaPd$_x$Sb$_2$ showing the variation of $\tc$ as function of the Pd content\cite{Reiner2015}. Reprinted with permission from Retzlaff $\e$\cite{Reiner2015}. Copyright 2015 by American Physics Society.}}
\vspace{-0.5cm}
\label{fig.18}
\end{center}
\end{figure}

The first single-phase superconducting thin film of the 112-phase has been LaNi$_x$Bi$_2$ grown by MBE\cite{Buckow2012}. MBE allows to grow a range of materials of various melting points using a combination of electron guns and (high-temperature) effusion cells. This is advantageous in terms of material flexibility, compositional and structural stabilisation as it allows precise doping control. The growth process was carried out in UHV atmosphere (typically at a base pressure $\sim 10^{-9}$\,mbar) and typical substrate temperatures between 300 and 600$^{\circ}$C. MgO(100) substrates were found to be suitable for epitaxial growth of most of the systems studied\cite{Buckow2012, Buckow2013, Kurian2013, Reiner2015}. The substrates were pre-annealed at 1000$^{\circ}$C in atmosphere to improve their surface quality. The growth process was monitored in situ using reflection high energy electron diffraction (RHEED).

\subsection{Thin Film Structure}

A list of samples investigated in thin film form along with their growth parameters is shown in Tab.~\ref{tab.1}. For all these materials, the out-of-plane X-ray diffraction (XRD) pattern (see Fig.~\ref{fig.16}(a)) consisted of a series of (00$l$) peaks which suggests a ZrCuSiAs-type of crystal structure with the symmetry group $P4/nmm$. The 112-phase thin films were found to be highly $c$-axis oriented and phase pure resulting in some cases in a higher crystalline quality as compared to the bulk synthesised counterpart of the same material. The high crystalline nature of the films was confirmed by the observation of Laue oscillations (see Fig.~\ref{fig.16})(c)) with very narrow FWHM typically between 0.03-0.07$^{\circ}$ (see Fig.~\ref{fig.16}(b)). The presence of streaky lines in the RHEED pattern in Fig.~\ref{fig.17} (inset) indicated the epitaxial nature of the films and smooth nature of the surface. From the X-ray $\phi$-scan in Fig.~\ref{fig.17}, the four-fold symmetry of the tetragonal structure of the films were observed. In addition to the primary 4-peaks with 90$^{\circ}$ separation, another set of 4-peaks with reduced intensities were placed symmetrically between the primary peaks suggesting the presence of two sets of growth domains rotated by 45$^{\circ}$ to each other. It was concluded that the peaks with higher intensity correspond to the majority domains which overgrew the minority domains (corresponding to low intensity peaks) which are only present at the substrate/sample interface.

\begin{figure}
\begin{center}
\includegraphics[width=7.5cm]{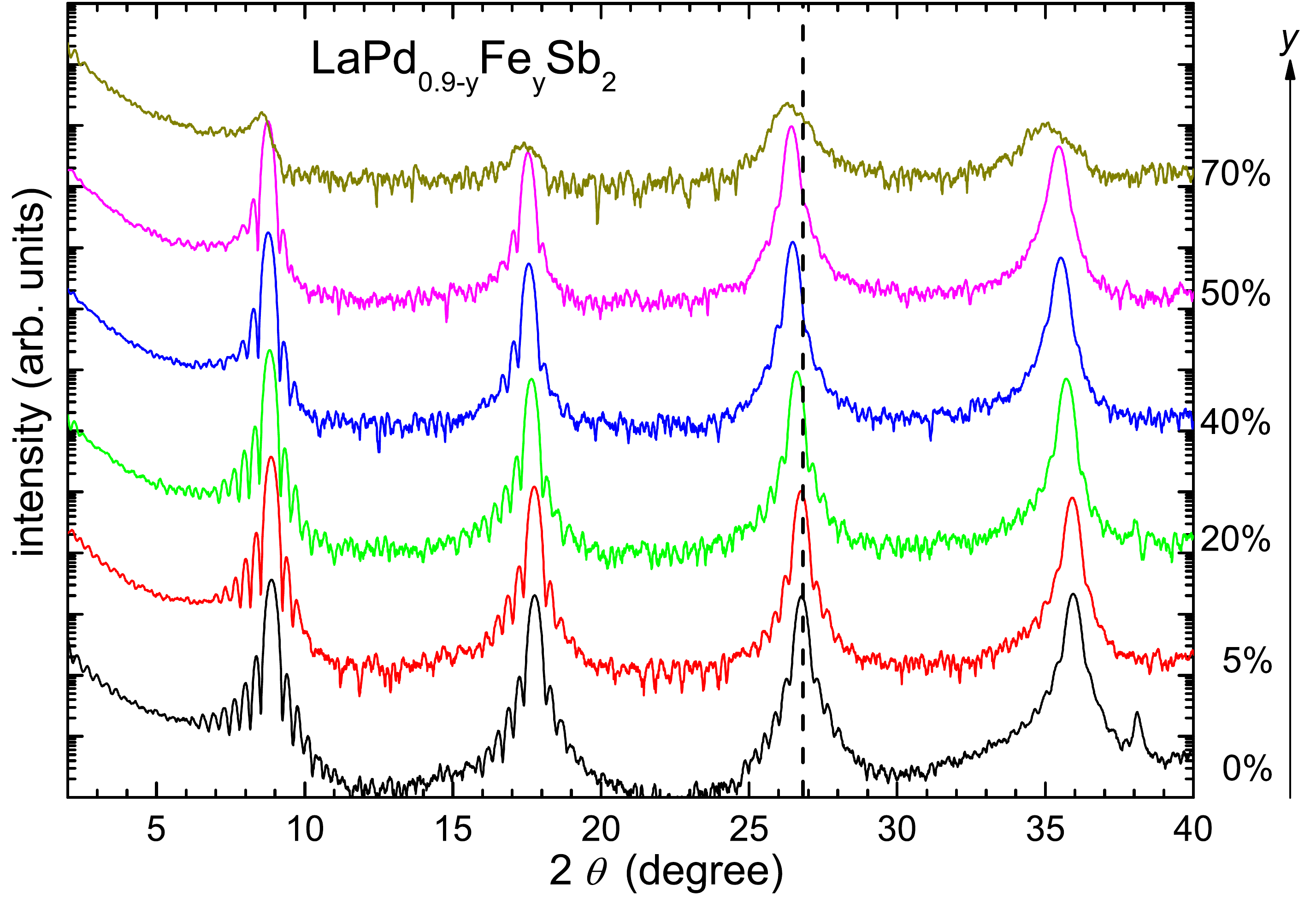}
\vspace{-0.3cm}
\caption{{\small Out-of-plane (2$\theta-\omega$) XRD scans of LaPd$_{0.9-y}$Fe$_y$Sb$_2$ for  for various Fe-doping level $0\le y\le0.7$\cite{Reiner2015}. Reprinted with permission from Retzlaff $\e$\cite{Reiner2015}. Copyright 2015 by American Physics Society.}}
\vspace{-0.5cm}
\label{fig.19}
\end{center}
\end{figure}

\subsection{Superconducting and Magnetic Properties of Thin Films}
\begin{table*}[t]
\begin{spacing}{1.3}
\begin{tabular}{l cc cc cc cc}
\hline\hline
Material && Substrate temperature && Growth rate && Lattice constant & Lattice constant & $\tc$ \\
&&  ($^{\circ}$C) && ({\AA}/s) && in-plane ({\AA}) & out-of-plane ({\AA}) &  (K) \\
\hline
LaNi$_x$Sb$_2$ \cite{Buckow2013} &&  390-450 && $\sim$ 0.5 &&  4.57$\pm$0.1 & 9.76$\pm$0.01 &  4.05 \\
CeNi$_x$Bi$_2$ \cite{Buckow2012} &&  410-440 && $\sim$ 0.5 && 4.565$\pm$0.002 & 9.64$\pm$0.01 &  4.05 \\
LaNi$_x$Bi$_2$ \cite{Kurian2013} &&  400-450 && 0.25-1.0 && - & 9.786 &  3.1 \\
LaPd$_x$Sb$_2$  \cite{Reiner2015} &&  440-520 && - && 4.52(2) & 9.88(5) &  3.27 \\
LaPd$_x$Bi$_2$ \cite{Reiner2015} &&  405-445 && - && 4.55 & 9.70(9) &  3.03 \\
\hline\hline
\end{tabular}
\caption{List of 112-type pnictide superconducting thin films and their growth parameters.}
\label{tab.1}
\end{spacing}
\vspace{-0.5cm}
\end{table*}

The $\tc$ was obtained from four probe resistivity measurements where a sharp superconducting transition was observed. At room temperature, all the films had a resistivity in the range of 100\,$\mu\Omega$cm. Between 5\,K and 30\,K, the resistivity shows metallic behavior that can be fitted using a quadratic temperature dependence $\rho(T) = \rho_0 + AT^2$, and from 30\, K tp 300\,K, the $\rho(T)$ vs.~$T$ dependence is mostly linear. The small residual resistivity ratio\cite{Reiner2015} ($\sim 1.58$) indicated a good quality and crystallinity of the samples. Magnetic measurements revealed the presence of Meissner shielding at $T<\tc$ with a significant volume of the superconducting phase. For $\lps$, the lower and upper critical fields were about 10\,G and 1.1\,T, respectively (with the field applied in out-of-plane direction).

The superconducting phase has been found to depend on the stoichiometric variation of its constituting elements. For $\lps$ within a $\pm$5\% variation of the optimal Sb concentration, $\tc$ stayed almost unaffected, while for larger variations superconductivity vanished associated with a sharp decline in crystallinity. However, for variation in Pd concentration $\tc$ showed a broad dome-shaped behaviour around the mean-value where highest $\tc$ is observed as shown in Fig.~\ref{fig.18}. The influence of Fe substitution on the structural behaviour of $\lps$ has been illustrated in Fig.~\ref{fig.19}. Replacement of Pd by Fe in LaPd$_{0.9-y}$Fe$_y$Sb$_2$ led to a sharp decline in crystallinity for $y > 0.5$ indicated by the disappearance of Laue oscillations as shown in Fig.~\ref{fig.19}, by a shift of the (003) peak position, and also by an increase in the $c$-axis lattice constant. Superconductivity disappeared instantly even on a small level ($\sim 5\%$) of Fe substitution in $\lps$, which suggests a conventional $s$-wave nature of the superconducting state as a small amount of ferromagnetic impurities lead to a strong reduction of the  BCS density of states. So far, thin films of high-$T_c$ 112-pnictides are elusive. Such films could be a new playground for investigating the symmetry of the order parameter and achieving critical temperatures above 77\,K.

\section{Conclusion and outlook}

\begin{figure*}
\begin{center}
\includegraphics[width=18cm]{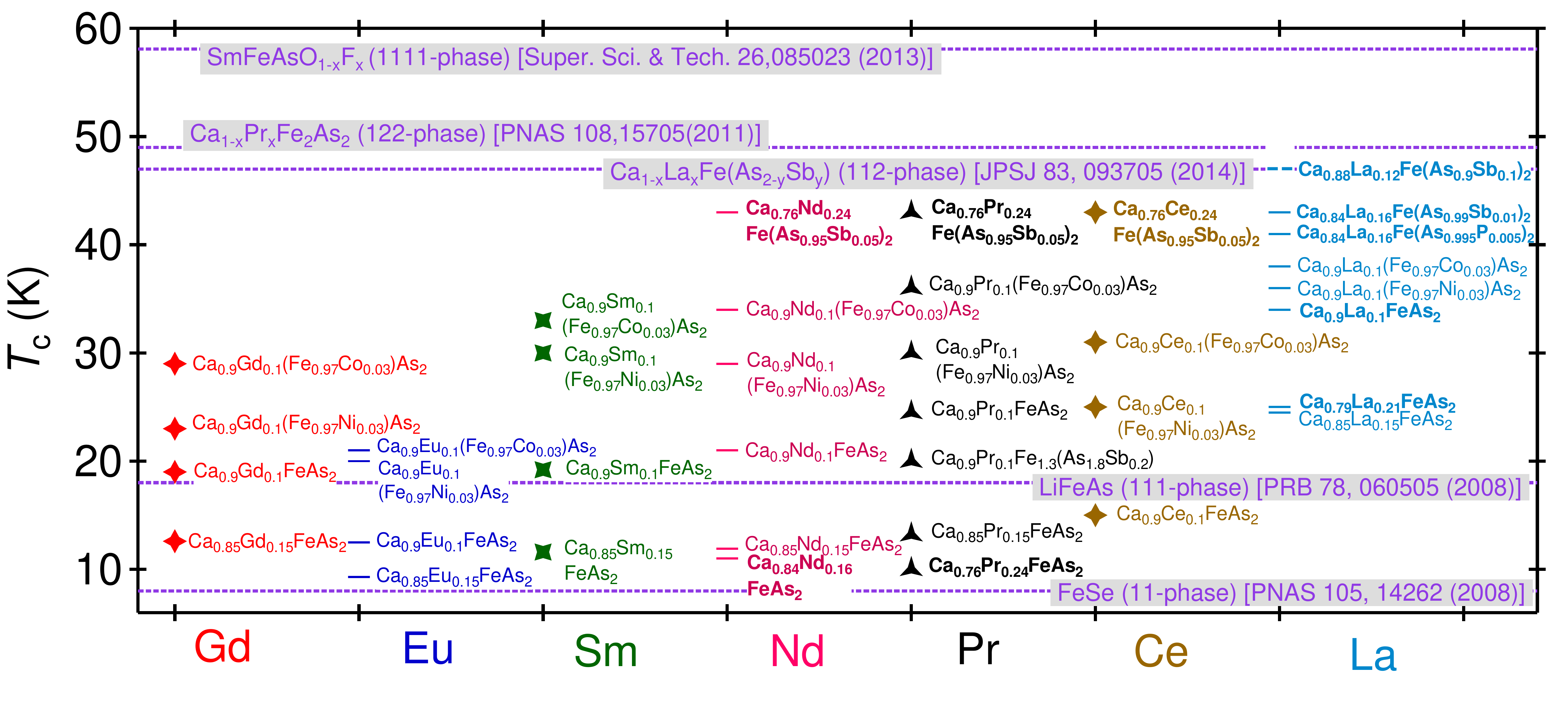}
\vspace{-0.3cm}
\caption{{\small A phase diagram based on the $\tc$ values of various (Ca,RE)FeAs$_2$ phases for various RE-elements, arranged in the order of increasing ionic radii under various doping conditions. Symbols indicate $\tc$ values reported so far for various single crystalline (labels in bold fonts) and polycrystalline phases \cite{Katayama2013, Kudo2014a, Kudo2014b, Yakita2015, Yakita2015b, Xing2015, Sala2014, Yakita2014} ; horizontal dotted lines indicate the maximum $\tc$ values reported for various types of $\p$ phases.}}
\vspace{-0.5cm}
\label{fig.20}
\end{center}
\end{figure*}

In this review, we have summarised the investigation of superconductivity and Dirac fermions in 112-type $\p$s. The existence of naturally occurring anisotropic Dirac Cones in the intermetallic 112-structure makes them interesting for possible application in nanoelectronics as an alternative to graphene while the large magnetoresistance could be of interest for spintronics. The discovery of high-$\tc$ superconductivity in $\cl$\cite{Katayama2013} has generated significant research interest in this newly discovered Fe-based superconducting system. Structurally the 112-phase is quasi-2D similar to the 1111-phase, but the presence of a metallic spacer layer and multiple-vacancies of the $Pn$-atoms in the neighboring layers are unique properties of this system.  Although the parent compound CaFeAs$_2$ has not been synthesised yet,  superconductivity can be introduced by $RE$-doping and the $\tc$ has been predicted to go beyond 50\,K for smaller levels of $RE$-content in Ca$_{1-x}$RE$_x$FeAs$_2$\cite{Kudo2014b}. Evidence of a structural and a magnetic phase transition were found both in the non-superconducting and superconducting phases\cite{Jiang2016, Kawasaki2015, Caglieris2016}. ARPES measurements revealed the presence of three hole like Fermi pockets at the zone centre and two electron like pockets at the zone corner with $3d$ character similar to the other $\p$ systems with moderate nesting between them\cite{Katayama2013, Xu2013, Wang2014, Huang2015}. In the low-$\tc$ 112-systems, $T_{\text{N}}$ lies below 10\,K and the presence of Ni-vacancies in the Ni$_x$Bi layer seems to be crucial to stabilize the crystal structure and bulk superconductivity\cite{Mizoguchi2011}. For $\cl$, $T_{\text{N}}$ can be enhanced up to 70\,K\cite{Kawasaki2015} for a high level of $RE$-doping.

In general, Sb doping helps increasing the in-plane lattice constants. A larger sized pnictogen results in a $\tc$ enhancement with increased superconducting volume fraction for various $RE$-doped Ca$_{1-x}RE_x$FeAs$_2$ with the so far highest $\tc$ of 47\,K. Preliminary measurements suggested a conventional $s$-wave type of pairing symmetry in the 112-phase, although a small amount of $TM$ doping seems to improve the superconducting properties in some cases. The moderate level of anisotropy in the 112-phase $\cl$ lies in between the 1111 and 122-type $\p$ systems, while the upper critical field is significantly higher than for other $\p$ phases with comparable critical current density. This makes it interesting from an application perspective as $J_c$ can reach $\>10^6$\,A/cm$^2$ over a large magnetic field range which sounds promising for the fabrication of superconducting tapes and Josephson junctions \cite{Zhou2014}. Also, it will be interesting to see if the substrate generated strain and interface effects can improve the $\tc$ in a thin film structure similar to the FeSe thin films. Generally, due to the lower level of $RE$-elements the 112-phase will allow significant cost reduction in producing superconducting tapes compared to the 122/1111-$\p$ phases. For this purpose, the long-time stability of the 112-compounds will be essential.  

Compared to the other $\p$ phases, the research work on the 112-system is still at a relatively younger stage and further investigation is needed to understand the potential of this $\p$ phase. A list of the $\tc$'s of various 112-(Ca,RE)FeAs$_2$ compounds studied so far has been illustrated in Fig.~\ref{fig.20}. General trend can be found that larger the size of the RE-ion, higher is the $\tc$. It is agreed that further reduction of the RE-content is essential for the further enhancement of $\tc$ in the $\cl$ system which : (a) for the bulk system needs further optimisation of the solid state synthesis techniques and (b) could be possible through various thin film growth processes, but yet to be explored for the $\cl$ system. Scattering techniques like $\mu$SR \cite{RayPRL, RaySRO} and Small Angle Neutron Scattering (SANS) could be useful to explore various parts of the phase diagram : (a) to understand the microscopic presence of magnetism and superconductivity and (b) nature of the vortex lattice symmetry and the vortex phase diagram. It can be expected that the availability of good quality and larger sized single crystals and thin films will allow diverse investigation in this system. Understanding the exact role of Sb on the enhancement of $\tc$ will help achieving higher $\tc$ in future. In an ideal case, Sb could replace As completely in superconducting 112-compounds with $\tc$ above 77\,K paving the way for a sustainable use of pnictide superconductors in applications.

\noindent\rule{8.7cm}{0.8pt}

\vspace{10pt}

\section*{Acknowledgement}
This work was supported by the Deutsche Forschungsgemeinschaft (DFG) through Grant No. SPP 1458 (LA 560/10-2). We would also like to thank A.~Buckow, P.~Komissinskiy, K.~Kupka, J.~Kurian, and R.~Retzlaff for their contributions to this project.


\vspace{10pt}

\section*{References}
\vspace{-10pt}

\bibliographystyle{apsrev}
\setlength{\bibsep}{3pt}
\bibliography{Bibliography}

\begin{thebibliography}{120}
\expandafter\ifx\csname natexlab\endcsname\relax\def\natexlab#1{#1}\fi
\expandafter\ifx\csname bibnamefont\endcsname\relax
  \def\bibnamefont#1{#1}\fi
\expandafter\ifx\csname bibfnamefont\endcsname\relax
  \def\bibfnamefont#1{#1}\fi
\expandafter\ifx\csname citenamefont\endcsname\relax
  \def\citenamefont#1{#1}\fi
\expandafter\ifx\csname url\endcsname\relax
  \def\url#1{\texttt{#1}}\fi
\expandafter\ifx\csname urlprefix\endcsname\relax\def\urlprefix{URL }\fi
\providecommand{\bibinfo}[2]{#2}
\providecommand{\eprint}[2][]{\url{#2}}

\bibitem[{\citenamefont{Kamihara et~al.}(2006)\citenamefont{Kamihara,
  Hiramatsu, Kawamura, Yanagi, Kamiya, and Hosono}}]{Kamihara2006}
\bibinfo{author}{\bibfnamefont{Y.}~\bibnamefont{Kamihara}},
  \bibinfo{author}{\bibfnamefont{M.}~\bibnamefont{Hiramatsu},
  \bibfnamefont{Hidenori�and~Hirano}},
  \bibinfo{author}{\bibfnamefont{R.}~\bibnamefont{Kawamura}},
  \bibinfo{author}{\bibfnamefont{H.}~\bibnamefont{Yanagi}},
  \bibinfo{author}{\bibfnamefont{T.}~\bibnamefont{Kamiya}}, \bibnamefont{and}
  \bibinfo{author}{\bibfnamefont{H.}~\bibnamefont{Hosono}},
  \bibinfo{journal}{Journal of the American Chemical Society}
  \textbf{\bibinfo{volume}{128}}, \bibinfo{pages}{10012}
  (\bibinfo{year}{2006}).

\bibitem[{\citenamefont{Kamihara et~al.}(2008)\citenamefont{Kamihara, Watanabe,
  Hirano, , and Hosono}}]{Kamihara2008}
\bibinfo{author}{\bibfnamefont{Y.}~\bibnamefont{Kamihara}},
  \bibinfo{author}{\bibfnamefont{T.}~\bibnamefont{Watanabe}},
  \bibinfo{author}{\bibfnamefont{M.}~\bibnamefont{Hirano}}, , \bibnamefont{and}
  \bibinfo{author}{\bibfnamefont{H.}~\bibnamefont{Hosono}},
  \bibinfo{journal}{Journal of the American Chemical Society}
  \textbf{\bibinfo{volume}{130}}, \bibinfo{pages}{3296} (\bibinfo{year}{2008}).

\bibitem[{\citenamefont{Takahashi et~al.}(2008)\citenamefont{Takahashi, Igawa,
  Arii, Kamihara, Hirano, and Hosono}}]{Takahasi2008}
\bibinfo{author}{\bibfnamefont{H.}~\bibnamefont{Takahashi}},
  \bibinfo{author}{\bibfnamefont{K.}~\bibnamefont{Igawa}},
  \bibinfo{author}{\bibfnamefont{K.}~\bibnamefont{Arii}},
  \bibinfo{author}{\bibfnamefont{Y.}~\bibnamefont{Kamihara}},
  \bibinfo{author}{\bibfnamefont{M.}~\bibnamefont{Hirano}}, \bibnamefont{and}
  \bibinfo{author}{\bibfnamefont{H.}~\bibnamefont{Hosono}},
  \bibinfo{journal}{Nature} \textbf{\bibinfo{volume}{453}},
  \bibinfo{pages}{376} (\bibinfo{year}{2008}).

\bibitem[{\citenamefont{Hosono et~al.}(2015)\citenamefont{Hosono, Tanabe,
  Takayama-Muromachi, Kageyama, Yamanaka, Kumakura, Nohara, Hiramatsu, and
  Fujitsu}}]{Hosono2015}
\bibinfo{author}{\bibfnamefont{H.}~\bibnamefont{Hosono}},
  \bibinfo{author}{\bibfnamefont{K.}~\bibnamefont{Tanabe}},
  \bibinfo{author}{\bibfnamefont{E.}~\bibnamefont{Takayama-Muromachi}},
  \bibinfo{author}{\bibfnamefont{H.}~\bibnamefont{Kageyama}},
  \bibinfo{author}{\bibfnamefont{S.}~\bibnamefont{Yamanaka}},
  \bibinfo{author}{\bibfnamefont{H.}~\bibnamefont{Kumakura}},
  \bibinfo{author}{\bibfnamefont{M.}~\bibnamefont{Nohara}},
  \bibinfo{author}{\bibfnamefont{H.}~\bibnamefont{Hiramatsu}},
  \bibnamefont{and} \bibinfo{author}{\bibfnamefont{S.}~\bibnamefont{Fujitsu}},
  \bibinfo{journal}{Science and Technology of Advanced Materials}
  \textbf{\bibinfo{volume}{16}}, \bibinfo{pages}{033503}
  (\bibinfo{year}{2015}).

\bibitem[{\citenamefont{Chen et~al.}(2008)\citenamefont{Chen, Wu, Wu, Liu,
  Chen, and Fang}}]{Chen2008}
\bibinfo{author}{\bibfnamefont{X.~H.} \bibnamefont{Chen}},
  \bibinfo{author}{\bibfnamefont{T.}~\bibnamefont{Wu}},
  \bibinfo{author}{\bibfnamefont{G.}~\bibnamefont{Wu}},
  \bibinfo{author}{\bibfnamefont{R.~H.} \bibnamefont{Liu}},
  \bibinfo{author}{\bibfnamefont{H.}~\bibnamefont{Chen}}, \bibnamefont{and}
  \bibinfo{author}{\bibfnamefont{D.~F.} \bibnamefont{Fang}},
  \bibinfo{journal}{Nature} \textbf{\bibinfo{volume}{453}},
  \bibinfo{pages}{761} (\bibinfo{year}{2008}).

\bibitem[{\citenamefont{Liu et~al.}(2008)\citenamefont{Liu, Wu, Wu, Fang, Chen,
  Li, Liu, Xie, Wang, Yang et~al.}}]{Liu2008}
\bibinfo{author}{\bibfnamefont{R.~H.} \bibnamefont{Liu}},
  \bibinfo{author}{\bibfnamefont{G.}~\bibnamefont{Wu}},
  \bibinfo{author}{\bibfnamefont{T.}~\bibnamefont{Wu}},
  \bibinfo{author}{\bibfnamefont{D.~F.} \bibnamefont{Fang}},
  \bibinfo{author}{\bibfnamefont{H.}~\bibnamefont{Chen}},
  \bibinfo{author}{\bibfnamefont{S.~Y.} \bibnamefont{Li}},
  \bibinfo{author}{\bibfnamefont{K.}~\bibnamefont{Liu}},
  \bibinfo{author}{\bibfnamefont{Y.~L.} \bibnamefont{Xie}},
  \bibinfo{author}{\bibfnamefont{X.~F.} \bibnamefont{Wang}},
  \bibinfo{author}{\bibfnamefont{R.~L.} \bibnamefont{Yang}},
  \bibnamefont{et~al.}, \bibinfo{journal}{Phys. Rev. Lett.}
  \textbf{\bibinfo{volume}{101}}, \bibinfo{pages}{087001}
  (\bibinfo{year}{2008}).

\bibitem[{\citenamefont{Wang et~al.}(2008{\natexlab{a}})\citenamefont{Wang, Li,
  Chi, Zhu, Ren, Li, Wang, Lin, Luo, Jiang et~al.}}]{Wang2008}
\bibinfo{author}{\bibfnamefont{C.}~\bibnamefont{Wang}},
  \bibinfo{author}{\bibfnamefont{L.}~\bibnamefont{Li}},
  \bibinfo{author}{\bibfnamefont{S.}~\bibnamefont{Chi}},
  \bibinfo{author}{\bibfnamefont{Z.}~\bibnamefont{Zhu}},
  \bibinfo{author}{\bibfnamefont{Z.}~\bibnamefont{Ren}},
  \bibinfo{author}{\bibfnamefont{Y.}~\bibnamefont{Li}},
  \bibinfo{author}{\bibfnamefont{Y.}~\bibnamefont{Wang}},
  \bibinfo{author}{\bibfnamefont{X.}~\bibnamefont{Lin}},
  \bibinfo{author}{\bibfnamefont{Y.}~\bibnamefont{Luo}},
  \bibinfo{author}{\bibfnamefont{S.}~\bibnamefont{Jiang}},
  \bibnamefont{et~al.}, \bibinfo{journal}{EPL (Europhysics Letters)}
  \textbf{\bibinfo{volume}{83}}, \bibinfo{pages}{67006}
  (\bibinfo{year}{2008}{\natexlab{a}}).

\bibitem[{\citenamefont{Ren et~al.}(2008)\citenamefont{Ren, Che, Dong, Yang,
  Lu, Yi, Shen, Li, Sun, Zhou et~al.}}]{Ren2008}
\bibinfo{author}{\bibfnamefont{Z.-A.} \bibnamefont{Ren}},
  \bibinfo{author}{\bibfnamefont{G.-C.} \bibnamefont{Che}},
  \bibinfo{author}{\bibfnamefont{X.-L.} \bibnamefont{Dong}},
  \bibinfo{author}{\bibfnamefont{J.}~\bibnamefont{Yang}},
  \bibinfo{author}{\bibfnamefont{W.}~\bibnamefont{Lu}},
  \bibinfo{author}{\bibfnamefont{W.}~\bibnamefont{Yi}},
  \bibinfo{author}{\bibfnamefont{X.-L.} \bibnamefont{Shen}},
  \bibinfo{author}{\bibfnamefont{Z.-C.} \bibnamefont{Li}},
  \bibinfo{author}{\bibfnamefont{L.-L.} \bibnamefont{Sun}},
  \bibinfo{author}{\bibfnamefont{F.}~\bibnamefont{Zhou}}, \bibnamefont{et~al.},
  \bibinfo{journal}{EPL (Europhysics Letters)} \textbf{\bibinfo{volume}{83}},
  \bibinfo{pages}{17002} (\bibinfo{year}{2008}).

\bibitem[{\citenamefont{Hanna et~al.}(2011)\citenamefont{Hanna, Muraba,
  Matsuishi, Igawa, Kodama, Shamoto, and Hosono}}]{Hanna2011}
\bibinfo{author}{\bibfnamefont{T.}~\bibnamefont{Hanna}},
  \bibinfo{author}{\bibfnamefont{Y.}~\bibnamefont{Muraba}},
  \bibinfo{author}{\bibfnamefont{S.}~\bibnamefont{Matsuishi}},
  \bibinfo{author}{\bibfnamefont{N.}~\bibnamefont{Igawa}},
  \bibinfo{author}{\bibfnamefont{K.}~\bibnamefont{Kodama}},
  \bibinfo{author}{\bibfnamefont{S.-i.} \bibnamefont{Shamoto}},
  \bibnamefont{and} \bibinfo{author}{\bibfnamefont{H.}~\bibnamefont{Hosono}},
  \bibinfo{journal}{Phys. Rev. B} \textbf{\bibinfo{volume}{84}},
  \bibinfo{pages}{024521} (\bibinfo{year}{2011}).

\bibitem[{\citenamefont{Fujioka et~al.}(2013)\citenamefont{Fujioka, Denholme,
  Ozaki, Okazaki, Deguchi, Demura, Hara, Watanabe, Takeya, Yamaguchi
  et~al.}}]{Fujioka2013}
\bibinfo{author}{\bibfnamefont{M.}~\bibnamefont{Fujioka}},
  \bibinfo{author}{\bibfnamefont{S.~J.} \bibnamefont{Denholme}},
  \bibinfo{author}{\bibfnamefont{T.}~\bibnamefont{Ozaki}},
  \bibinfo{author}{\bibfnamefont{H.}~\bibnamefont{Okazaki}},
  \bibinfo{author}{\bibfnamefont{K.}~\bibnamefont{Deguchi}},
  \bibinfo{author}{\bibfnamefont{S.}~\bibnamefont{Demura}},
  \bibinfo{author}{\bibfnamefont{H.}~\bibnamefont{Hara}},
  \bibinfo{author}{\bibfnamefont{T.}~\bibnamefont{Watanabe}},
  \bibinfo{author}{\bibfnamefont{H.}~\bibnamefont{Takeya}},
  \bibinfo{author}{\bibfnamefont{T.}~\bibnamefont{Yamaguchi}},
  \bibnamefont{et~al.}, \bibinfo{journal}{Superconductor Science and
  Technology} \textbf{\bibinfo{volume}{26}}, \bibinfo{pages}{085023}
  (\bibinfo{year}{2013}).

\bibitem[{\citenamefont{Rotter et~al.}(2008)\citenamefont{Rotter, Tegel, and
  Johrendt}}]{Rotter2008}
\bibinfo{author}{\bibfnamefont{M.}~\bibnamefont{Rotter}},
  \bibinfo{author}{\bibfnamefont{M.}~\bibnamefont{Tegel}}, \bibnamefont{and}
  \bibinfo{author}{\bibfnamefont{D.}~\bibnamefont{Johrendt}},
  \bibinfo{journal}{Phys. Rev. Lett.} \textbf{\bibinfo{volume}{101}},
  \bibinfo{pages}{107006} (\bibinfo{year}{2008}).

\bibitem[{\citenamefont{Hsu et~al.}(2008)\citenamefont{Hsu, Luo, Yeh, Chen,
  Huang, Wu, Lee, Huang, Chu, Yan et~al.}}]{Hsu2008}
\bibinfo{author}{\bibfnamefont{F.-C.} \bibnamefont{Hsu}},
  \bibinfo{author}{\bibfnamefont{J.-Y.} \bibnamefont{Luo}},
  \bibinfo{author}{\bibfnamefont{K.-W.} \bibnamefont{Yeh}},
  \bibinfo{author}{\bibfnamefont{T.-K.} \bibnamefont{Chen}},
  \bibinfo{author}{\bibfnamefont{T.-W.} \bibnamefont{Huang}},
  \bibinfo{author}{\bibfnamefont{P.~M.} \bibnamefont{Wu}},
  \bibinfo{author}{\bibfnamefont{Y.-C.} \bibnamefont{Lee}},
  \bibinfo{author}{\bibfnamefont{Y.-L.} \bibnamefont{Huang}},
  \bibinfo{author}{\bibfnamefont{Y.-Y.} \bibnamefont{Chu}},
  \bibinfo{author}{\bibfnamefont{D.-C.} \bibnamefont{Yan}},
  \bibnamefont{et~al.}, \bibinfo{journal}{Proceedings of the National Academy
  of Sciences} \textbf{\bibinfo{volume}{105}}, \bibinfo{pages}{14262}
  (\bibinfo{year}{2008}).

\bibitem[{\citenamefont{Tapp et~al.}(2008)\citenamefont{Tapp, Tang, Lv, Sasmal,
  Lorenz, Chu, and Guloy}}]{Tapp2008}
\bibinfo{author}{\bibfnamefont{J.~H.} \bibnamefont{Tapp}},
  \bibinfo{author}{\bibfnamefont{Z.}~\bibnamefont{Tang}},
  \bibinfo{author}{\bibfnamefont{B.}~\bibnamefont{Lv}},
  \bibinfo{author}{\bibfnamefont{K.}~\bibnamefont{Sasmal}},
  \bibinfo{author}{\bibfnamefont{B.}~\bibnamefont{Lorenz}},
  \bibinfo{author}{\bibfnamefont{P.~C.~W.} \bibnamefont{Chu}},
  \bibnamefont{and} \bibinfo{author}{\bibfnamefont{A.~M.} \bibnamefont{Guloy}},
  \bibinfo{journal}{Phys. Rev. B} \textbf{\bibinfo{volume}{78}},
  \bibinfo{pages}{060505} (\bibinfo{year}{2008}).

\bibitem[{\citenamefont{Wang et~al.}(2008{\natexlab{b}})\citenamefont{Wang,
  Liu, Lv, Gao, Yang, Yu, Li, and Jin}}]{Wang2008a}
\bibinfo{author}{\bibfnamefont{X.}~\bibnamefont{Wang}},
  \bibinfo{author}{\bibfnamefont{Q.}~\bibnamefont{Liu}},
  \bibinfo{author}{\bibfnamefont{Y.}~\bibnamefont{Lv}},
  \bibinfo{author}{\bibfnamefont{W.}~\bibnamefont{Gao}},
  \bibinfo{author}{\bibfnamefont{L.}~\bibnamefont{Yang}},
  \bibinfo{author}{\bibfnamefont{R.}~\bibnamefont{Yu}},
  \bibinfo{author}{\bibfnamefont{F.}~\bibnamefont{Li}}, \bibnamefont{and}
  \bibinfo{author}{\bibfnamefont{C.}~\bibnamefont{Jin}},
  \bibinfo{journal}{Solid State Communications} \textbf{\bibinfo{volume}{148}},
  \bibinfo{pages}{538 } (\bibinfo{year}{2008}{\natexlab{b}}), ISSN
  \bibinfo{issn}{0038-1098}.

\bibitem[{\citenamefont{Pitcher et~al.}(2008)\citenamefont{Pitcher, Parker,
  Adamson, Herkelrath, Boothroyd, Ibberson, Brunelli, and
  Clarke}}]{Pitcher2008}
\bibinfo{author}{\bibfnamefont{M.~J.} \bibnamefont{Pitcher}},
  \bibinfo{author}{\bibfnamefont{D.~R.} \bibnamefont{Parker}},
  \bibinfo{author}{\bibfnamefont{P.}~\bibnamefont{Adamson}},
  \bibinfo{author}{\bibfnamefont{S.~J.~C.} \bibnamefont{Herkelrath}},
  \bibinfo{author}{\bibfnamefont{A.~T.} \bibnamefont{Boothroyd}},
  \bibinfo{author}{\bibfnamefont{R.~M.} \bibnamefont{Ibberson}},
  \bibinfo{author}{\bibfnamefont{M.}~\bibnamefont{Brunelli}}, \bibnamefont{and}
  \bibinfo{author}{\bibfnamefont{S.~J.} \bibnamefont{Clarke}},
  \bibinfo{journal}{Chem. Commun.} pp. \bibinfo{pages}{5918--5920}
  (\bibinfo{year}{2008}).

\bibitem[{\citenamefont{Katayama et~al.}(2013)\citenamefont{Katayama, Kudo,
  Onari, Mizukami, Sugawara, Sugiyama, Kitahama, Iba, Fujimura, Nishimoto
  et~al.}}]{Katayama2013}
\bibinfo{author}{\bibfnamefont{N.}~\bibnamefont{Katayama}},
  \bibinfo{author}{\bibfnamefont{K.}~\bibnamefont{Kudo}},
  \bibinfo{author}{\bibfnamefont{S.}~\bibnamefont{Onari}},
  \bibinfo{author}{\bibfnamefont{T.}~\bibnamefont{Mizukami}},
  \bibinfo{author}{\bibfnamefont{K.}~\bibnamefont{Sugawara}},
  \bibinfo{author}{\bibfnamefont{Y.}~\bibnamefont{Sugiyama}},
  \bibinfo{author}{\bibfnamefont{Y.}~\bibnamefont{Kitahama}},
  \bibinfo{author}{\bibfnamefont{K.}~\bibnamefont{Iba}},
  \bibinfo{author}{\bibfnamefont{K.}~\bibnamefont{Fujimura}},
  \bibinfo{author}{\bibfnamefont{N.}~\bibnamefont{Nishimoto}},
  \bibnamefont{et~al.}, \bibinfo{journal}{Journal of the Physical Society of
  Japan} \textbf{\bibinfo{volume}{82}}, \bibinfo{pages}{123702}
  (\bibinfo{year}{2013}).

\bibitem[{\citenamefont{Kudo et~al.}(2014{\natexlab{a}})\citenamefont{Kudo,
  Mizukami, Kitahama, Mitsuoka, Iba, Fujimura, Nishimoto, Hiraoka, and
  Nohara}}]{Kudo2014a}
\bibinfo{author}{\bibfnamefont{K.}~\bibnamefont{Kudo}},
  \bibinfo{author}{\bibfnamefont{T.}~\bibnamefont{Mizukami}},
  \bibinfo{author}{\bibfnamefont{Y.}~\bibnamefont{Kitahama}},
  \bibinfo{author}{\bibfnamefont{D.}~\bibnamefont{Mitsuoka}},
  \bibinfo{author}{\bibfnamefont{K.}~\bibnamefont{Iba}},
  \bibinfo{author}{\bibfnamefont{K.}~\bibnamefont{Fujimura}},
  \bibinfo{author}{\bibfnamefont{N.}~\bibnamefont{Nishimoto}},
  \bibinfo{author}{\bibfnamefont{Y.}~\bibnamefont{Hiraoka}}, \bibnamefont{and}
  \bibinfo{author}{\bibfnamefont{M.}~\bibnamefont{Nohara}},
  \bibinfo{journal}{Journal of the Physical Society of Japan}
  \textbf{\bibinfo{volume}{83}}, \bibinfo{pages}{025001}
  (\bibinfo{year}{2014}{\natexlab{a}}).

\bibitem[{\citenamefont{Kudo et~al.}(2014{\natexlab{b}})\citenamefont{Kudo,
  Kitahama, Fujimura, Mizukami, Ota, and Nohara}}]{Kudo2014b}
\bibinfo{author}{\bibfnamefont{K.}~\bibnamefont{Kudo}},
  \bibinfo{author}{\bibfnamefont{Y.}~\bibnamefont{Kitahama}},
  \bibinfo{author}{\bibfnamefont{K.}~\bibnamefont{Fujimura}},
  \bibinfo{author}{\bibfnamefont{T.}~\bibnamefont{Mizukami}},
  \bibinfo{author}{\bibfnamefont{H.}~\bibnamefont{Ota}}, \bibnamefont{and}
  \bibinfo{author}{\bibfnamefont{M.}~\bibnamefont{Nohara}},
  \bibinfo{journal}{Journal of the Physical Society of Japan}
  \textbf{\bibinfo{volume}{83}}, \bibinfo{pages}{093705}
  (\bibinfo{year}{2014}{\natexlab{b}}).

\bibitem[{\citenamefont{Gurevich}(2011)}]{Gurevich2011}
\bibinfo{author}{\bibfnamefont{A.}~\bibnamefont{Gurevich}},
  \bibinfo{journal}{Reports on Progress in Physics}
  \textbf{\bibinfo{volume}{74}}, \bibinfo{pages}{124501}
  (\bibinfo{year}{2011}).

\bibitem[{\citenamefont{Moll et~al.}(2010)\citenamefont{Moll, Puzniak,
  Balakirev, Rogacki, Karpinski, Zhigadlo, and Batlogg}}]{Moll2010}
\bibinfo{author}{\bibfnamefont{P.~J.~W.} \bibnamefont{Moll}},
  \bibinfo{author}{\bibfnamefont{R.}~\bibnamefont{Puzniak}},
  \bibinfo{author}{\bibfnamefont{F.}~\bibnamefont{Balakirev}},
  \bibinfo{author}{\bibfnamefont{K.}~\bibnamefont{Rogacki}},
  \bibinfo{author}{\bibfnamefont{J.}~\bibnamefont{Karpinski}},
  \bibinfo{author}{\bibfnamefont{N.~D.} \bibnamefont{Zhigadlo}},
  \bibnamefont{and} \bibinfo{author}{\bibfnamefont{B.}~\bibnamefont{Batlogg}},
  \bibinfo{journal}{Nat Mater} \textbf{\bibinfo{volume}{9}},
  \bibinfo{pages}{628} (\bibinfo{year}{2010}).

\bibitem[{\citenamefont{Chubukov}(2012)}]{Chubukov2012}
\bibinfo{author}{\bibfnamefont{A.}~\bibnamefont{Chubukov}},
  \bibinfo{journal}{Annual Review of Condensed Matter Physics}
  \textbf{\bibinfo{volume}{3}}, \bibinfo{pages}{57} (\bibinfo{year}{2012}).

\bibitem[{\citenamefont{Johnston}(2010)}]{Johnston2010}
\bibinfo{author}{\bibfnamefont{D.~C.} \bibnamefont{Johnston}},
  \bibinfo{journal}{Advances in Physics} \textbf{\bibinfo{volume}{59}},
  \bibinfo{pages}{803} (\bibinfo{year}{2010}).

\bibitem[{\citenamefont{Lee et~al.}(2008)\citenamefont{Lee, Iyo, Eisaki, Kito,
  Fernandez-Diaz, Ito, Kihou, Matsuhata, Braden, and Yamada}}]{Lee2008}
\bibinfo{author}{\bibfnamefont{C.-H.} \bibnamefont{Lee}},
  \bibinfo{author}{\bibfnamefont{A.}~\bibnamefont{Iyo}},
  \bibinfo{author}{\bibfnamefont{H.}~\bibnamefont{Eisaki}},
  \bibinfo{author}{\bibfnamefont{H.}~\bibnamefont{Kito}},
  \bibinfo{author}{\bibfnamefont{M.~T.} \bibnamefont{Fernandez-Diaz}},
  \bibinfo{author}{\bibfnamefont{T.}~\bibnamefont{Ito}},
  \bibinfo{author}{\bibfnamefont{K.}~\bibnamefont{Kihou}},
  \bibinfo{author}{\bibfnamefont{H.}~\bibnamefont{Matsuhata}},
  \bibinfo{author}{\bibfnamefont{M.}~\bibnamefont{Braden}}, \bibnamefont{and}
  \bibinfo{author}{\bibfnamefont{K.}~\bibnamefont{Yamada}},
  \bibinfo{journal}{Journal of the Physical Society of Japan}
  \textbf{\bibinfo{volume}{77}}, \bibinfo{pages}{083704}
  (\bibinfo{year}{2008}).

\bibitem[{\citenamefont{Taylor}(2013)}]{Taylor2013}
\bibinfo{author}{\bibfnamefont{A.}~\bibnamefont{Taylor}},
  \emph{\bibinfo{title}{Magnetic Dynamics in Iron-based Superconductors Probed
  by Neutron Spectroscopy}} (\bibinfo{publisher}{University of Oxford, UK},
  \bibinfo{year}{2013}).

\bibitem[{\citenamefont{Aswathy et~al.}(2010)\citenamefont{Aswathy, Anooja,
  Sarun, and Syamaprasad}}]{Aswathy2010}
\bibinfo{author}{\bibfnamefont{P.~M.} \bibnamefont{Aswathy}},
  \bibinfo{author}{\bibfnamefont{J.~B.} \bibnamefont{Anooja}},
  \bibinfo{author}{\bibfnamefont{P.~M.} \bibnamefont{Sarun}}, \bibnamefont{and}
  \bibinfo{author}{\bibfnamefont{U.}~\bibnamefont{Syamaprasad}},
  \bibinfo{journal}{Superconductor Science and Technology}
  \textbf{\bibinfo{volume}{23}}, \bibinfo{pages}{073001}
  (\bibinfo{year}{2010}).

\bibitem[{\citenamefont{Qing-Yan et~al.}(2012)\citenamefont{Qing-Yan, Zhi,
  Wen-Hao, Zuo-Cheng, Jin-Song, Wei, Hao, Yun-Bo, Peng, Kai et~al.}}]{Wang2012}
\bibinfo{author}{\bibfnamefont{W.}~\bibnamefont{Qing-Yan}},
  \bibinfo{author}{\bibfnamefont{L.}~\bibnamefont{Zhi}},
  \bibinfo{author}{\bibfnamefont{Z.}~\bibnamefont{Wen-Hao}},
  \bibinfo{author}{\bibfnamefont{Z.}~\bibnamefont{Zuo-Cheng}},
  \bibinfo{author}{\bibfnamefont{Z.}~\bibnamefont{Jin-Song}},
  \bibinfo{author}{\bibfnamefont{L.}~\bibnamefont{Wei}},
  \bibinfo{author}{\bibfnamefont{D.}~\bibnamefont{Hao}},
  \bibinfo{author}{\bibfnamefont{O.}~\bibnamefont{Yun-Bo}},
  \bibinfo{author}{\bibfnamefont{D.}~\bibnamefont{Peng}},
  \bibinfo{author}{\bibfnamefont{C.}~\bibnamefont{Kai}}, \bibnamefont{et~al.},
  \bibinfo{journal}{Chinese Physics Letters} \textbf{\bibinfo{volume}{29}},
  \bibinfo{pages}{037402} (\bibinfo{year}{2012}).

\bibitem[{\citenamefont{Ge et~al.}(2015)\citenamefont{Ge, Liu, Liu, Gao, Qian,
  Xue, Liu, and Jia}}]{Ge2015}
\bibinfo{author}{\bibfnamefont{J.-F.} \bibnamefont{Ge}},
  \bibinfo{author}{\bibfnamefont{Z.-L.} \bibnamefont{Liu}},
  \bibinfo{author}{\bibfnamefont{C.}~\bibnamefont{Liu}},
  \bibinfo{author}{\bibfnamefont{C.-L.} \bibnamefont{Gao}},
  \bibinfo{author}{\bibfnamefont{D.}~\bibnamefont{Qian}},
  \bibinfo{author}{\bibfnamefont{Q.-K.} \bibnamefont{Xue}},
  \bibinfo{author}{\bibfnamefont{Y.}~\bibnamefont{Liu}}, \bibnamefont{and}
  \bibinfo{author}{\bibfnamefont{J.-F.} \bibnamefont{Jia}},
  \bibinfo{journal}{Nat Mater} \textbf{\bibinfo{volume}{14}},
  \bibinfo{pages}{285} (\bibinfo{year}{2015}).

\bibitem[{\citenamefont{Shim et~al.}(2009)\citenamefont{Shim, Haule, and
  Kotliar}}]{Shim2009}
\bibinfo{author}{\bibfnamefont{J.~H.} \bibnamefont{Shim}},
  \bibinfo{author}{\bibfnamefont{K.}~\bibnamefont{Haule}}, \bibnamefont{and}
  \bibinfo{author}{\bibfnamefont{G.}~\bibnamefont{Kotliar}},
  \bibinfo{journal}{Phys. Rev. B} \textbf{\bibinfo{volume}{79}},
  \bibinfo{pages}{060501} (\bibinfo{year}{2009}).

\bibitem[{\citenamefont{Wang et~al.}(2011{\natexlab{a}})\citenamefont{Wang,
  Zhao, Yin, Kotliar, Kim, Aronson, and Morosan}}]{Wang2011}
\bibinfo{author}{\bibfnamefont{J.~K.} \bibnamefont{Wang}},
  \bibinfo{author}{\bibfnamefont{L.~L.} \bibnamefont{Zhao}},
  \bibinfo{author}{\bibfnamefont{Q.}~\bibnamefont{Yin}},
  \bibinfo{author}{\bibfnamefont{G.}~\bibnamefont{Kotliar}},
  \bibinfo{author}{\bibfnamefont{M.~S.} \bibnamefont{Kim}},
  \bibinfo{author}{\bibfnamefont{M.~C.} \bibnamefont{Aronson}},
  \bibnamefont{and} \bibinfo{author}{\bibfnamefont{E.}~\bibnamefont{Morosan}},
  \bibinfo{journal}{Phys. Rev. B} \textbf{\bibinfo{volume}{84}},
  \bibinfo{pages}{064428} (\bibinfo{year}{2011}{\natexlab{a}}).

\bibitem[{\citenamefont{Wang et~al.}(2011{\natexlab{b}})\citenamefont{Wang,
  Graf, Lei, Tozer, and Petrovic}}]{Wang2011b}
\bibinfo{author}{\bibfnamefont{K.}~\bibnamefont{Wang}},
  \bibinfo{author}{\bibfnamefont{D.}~\bibnamefont{Graf}},
  \bibinfo{author}{\bibfnamefont{H.}~\bibnamefont{Lei}},
  \bibinfo{author}{\bibfnamefont{S.~W.} \bibnamefont{Tozer}}, \bibnamefont{and}
  \bibinfo{author}{\bibfnamefont{C.}~\bibnamefont{Petrovic}},
  \bibinfo{journal}{Phys. Rev. B} \textbf{\bibinfo{volume}{84}},
  \bibinfo{pages}{220401} (\bibinfo{year}{2011}{\natexlab{b}}).

\bibitem[{\citenamefont{Park et~al.}(2011)\citenamefont{Park, Lee,
  Wolff-Fabris, Koh, Eom, Kim, Farhan, Jo, Kim, Shim et~al.}}]{Park2011}
\bibinfo{author}{\bibfnamefont{J.}~\bibnamefont{Park}},
  \bibinfo{author}{\bibfnamefont{G.}~\bibnamefont{Lee}},
  \bibinfo{author}{\bibfnamefont{F.}~\bibnamefont{Wolff-Fabris}},
  \bibinfo{author}{\bibfnamefont{Y.~Y.} \bibnamefont{Koh}},
  \bibinfo{author}{\bibfnamefont{M.~J.} \bibnamefont{Eom}},
  \bibinfo{author}{\bibfnamefont{Y.~K.} \bibnamefont{Kim}},
  \bibinfo{author}{\bibfnamefont{M.~A.} \bibnamefont{Farhan}},
  \bibinfo{author}{\bibfnamefont{Y.~J.} \bibnamefont{Jo}},
  \bibinfo{author}{\bibfnamefont{C.}~\bibnamefont{Kim}},
  \bibinfo{author}{\bibfnamefont{J.~H.} \bibnamefont{Shim}},
  \bibnamefont{et~al.}, \bibinfo{journal}{Phys. Rev. Lett.}
  \textbf{\bibinfo{volume}{107}}, \bibinfo{pages}{126402}
  (\bibinfo{year}{2011}).

\bibitem[{\citenamefont{Rosa et~al.}(2015)\citenamefont{Rosa, Jesus, Adriano,
  Fisk, and Pagliuso}}]{Rosa2015b}
\bibinfo{author}{\bibfnamefont{P.~F.~S.} \bibnamefont{Rosa}},
  \bibinfo{author}{\bibfnamefont{C.~B.~R.} \bibnamefont{Jesus}},
  \bibinfo{author}{\bibfnamefont{C.}~\bibnamefont{Adriano}},
  \bibinfo{author}{\bibfnamefont{Z.}~\bibnamefont{Fisk}}, \bibnamefont{and}
  \bibinfo{author}{\bibfnamefont{P.~G.} \bibnamefont{Pagliuso}},
  \bibinfo{journal}{Journal of Physics: Conference Series}
  \textbf{\bibinfo{volume}{592}}, \bibinfo{pages}{012063}
  (\bibinfo{year}{2015}).

\bibitem[{\citenamefont{Mizoguchi et~al.}(2011)\citenamefont{Mizoguchi,
  Matsuishi, Hirano, Tachibana, Takayama-Muromachi, Kawaji, and
  Hosono}}]{Mizoguchi2011}
\bibinfo{author}{\bibfnamefont{H.}~\bibnamefont{Mizoguchi}},
  \bibinfo{author}{\bibfnamefont{S.}~\bibnamefont{Matsuishi}},
  \bibinfo{author}{\bibfnamefont{M.}~\bibnamefont{Hirano}},
  \bibinfo{author}{\bibfnamefont{M.}~\bibnamefont{Tachibana}},
  \bibinfo{author}{\bibfnamefont{E.}~\bibnamefont{Takayama-Muromachi}},
  \bibinfo{author}{\bibfnamefont{H.}~\bibnamefont{Kawaji}}, \bibnamefont{and}
  \bibinfo{author}{\bibfnamefont{H.}~\bibnamefont{Hosono}},
  \bibinfo{journal}{Phys. Rev. Lett.} \textbf{\bibinfo{volume}{106}},
  \bibinfo{pages}{057002} (\bibinfo{year}{2011}).

\bibitem[{\citenamefont{Buckow et~al.}(2012)\citenamefont{Buckow, Kupka,
  Retzlaff, Kurian, and Alff}}]{Buckow2012}
\bibinfo{author}{\bibfnamefont{A.}~\bibnamefont{Buckow}},
  \bibinfo{author}{\bibfnamefont{K.}~\bibnamefont{Kupka}},
  \bibinfo{author}{\bibfnamefont{R.}~\bibnamefont{Retzlaff}},
  \bibinfo{author}{\bibfnamefont{J.}~\bibnamefont{Kurian}}, \bibnamefont{and}
  \bibinfo{author}{\bibfnamefont{L.}~\bibnamefont{Alff}},
  \bibinfo{journal}{Applied Physics Letters} \textbf{\bibinfo{volume}{101}},
  \bibinfo{eid}{162602} (\bibinfo{year}{2012}).

\bibitem[{\citenamefont{Jung et~al.}(2002)\citenamefont{Jung, Lacerda, and
  Takabatake}}]{Jung2002}
\bibinfo{author}{\bibfnamefont{M.~H.} \bibnamefont{Jung}},
  \bibinfo{author}{\bibfnamefont{A.~H.} \bibnamefont{Lacerda}},
  \bibnamefont{and}
  \bibinfo{author}{\bibfnamefont{T.}~\bibnamefont{Takabatake}},
  \bibinfo{journal}{Phys. Rev. B} \textbf{\bibinfo{volume}{65}},
  \bibinfo{pages}{132405} (\bibinfo{year}{2002}).

\bibitem[{\citenamefont{Kodama et~al.}(2011)\citenamefont{Kodama, Wakimoto,
  Igawa, Shamoto, Mizoguchi, and Hosono}}]{Kodama2011}
\bibinfo{author}{\bibfnamefont{K.}~\bibnamefont{Kodama}},
  \bibinfo{author}{\bibfnamefont{S.}~\bibnamefont{Wakimoto}},
  \bibinfo{author}{\bibfnamefont{N.}~\bibnamefont{Igawa}},
  \bibinfo{author}{\bibfnamefont{S.}~\bibnamefont{Shamoto}},
  \bibinfo{author}{\bibfnamefont{H.}~\bibnamefont{Mizoguchi}},
  \bibnamefont{and} \bibinfo{author}{\bibfnamefont{H.}~\bibnamefont{Hosono}},
  \bibinfo{journal}{Phys. Rev. B} \textbf{\bibinfo{volume}{83}},
  \bibinfo{pages}{214512} (\bibinfo{year}{2011}).

\bibitem[{\citenamefont{Thamizhavel et~al.}(2003)\citenamefont{Thamizhavel,
  Galatanu, Yamamoto, Okubo, Yamada, Tabata, Kobayashi, Nakamura, Sugiyama,
  Kindo et~al.}}]{Thamizhavel2003}
\bibinfo{author}{\bibfnamefont{A.}~\bibnamefont{Thamizhavel}},
  \bibinfo{author}{\bibfnamefont{A.}~\bibnamefont{Galatanu}},
  \bibinfo{author}{\bibfnamefont{E.}~\bibnamefont{Yamamoto}},
  \bibinfo{author}{\bibfnamefont{T.}~\bibnamefont{Okubo}},
  \bibinfo{author}{\bibfnamefont{M.}~\bibnamefont{Yamada}},
  \bibinfo{author}{\bibfnamefont{K.}~\bibnamefont{Tabata}},
  \bibinfo{author}{\bibfnamefont{T.~C.} \bibnamefont{Kobayashi}},
  \bibinfo{author}{\bibfnamefont{N.}~\bibnamefont{Nakamura}},
  \bibinfo{author}{\bibfnamefont{K.}~\bibnamefont{Sugiyama}},
  \bibinfo{author}{\bibfnamefont{K.}~\bibnamefont{Kindo}},
  \bibnamefont{et~al.}, \bibinfo{journal}{Journal of the Physical Society of
  Japan} \textbf{\bibinfo{volume}{72}}, \bibinfo{pages}{2632}
  (\bibinfo{year}{2003}).

\bibitem[{\citenamefont{Mizoguchi et~al.}(2012)\citenamefont{Mizoguchi, Kamiya,
  and Hosono}}]{Mizoguchi2012}
\bibinfo{author}{\bibfnamefont{H.}~\bibnamefont{Mizoguchi}},
  \bibinfo{author}{\bibfnamefont{T.}~\bibnamefont{Kamiya}}, \bibnamefont{and}
  \bibinfo{author}{\bibfnamefont{H.}~\bibnamefont{Hosono}},
  \bibinfo{journal}{Solid State Communications} \textbf{\bibinfo{volume}{152}},
  \bibinfo{pages}{666 } (\bibinfo{year}{2012}), ISSN \bibinfo{issn}{0038-1098},
  \bibinfo{note}{special Issue on Iron-based Superconductors}.

\bibitem[{\citenamefont{Lin et~al.}(2013)\citenamefont{Lin, Straszheim,
  Bud’ko, and Canfield}}]{Lin2013}
\bibinfo{author}{\bibfnamefont{X.}~\bibnamefont{Lin}},
  \bibinfo{author}{\bibfnamefont{W.~E.} \bibnamefont{Straszheim}},
  \bibinfo{author}{\bibfnamefont{S.~L.} \bibnamefont{Bud’ko}},
  \bibnamefont{and} \bibinfo{author}{\bibfnamefont{P.~C.}
  \bibnamefont{Canfield}}, \bibinfo{journal}{Journal of Alloys and Compounds}
  \textbf{\bibinfo{volume}{554}}, \bibinfo{pages}{304 } (\bibinfo{year}{2013}),
  ISSN \bibinfo{issn}{0925-8388}.

\bibitem[{\citenamefont{Yakita et~al.}(2014)\citenamefont{Yakita, Ogino, Okada,
  Yamamoto, Kishio, Tohei, Ikuhara, Gotoh, Fujihisa, Kataoka
  et~al.}}]{Yakita2014}
\bibinfo{author}{\bibfnamefont{H.}~\bibnamefont{Yakita}},
  \bibinfo{author}{\bibfnamefont{H.}~\bibnamefont{Ogino}},
  \bibinfo{author}{\bibfnamefont{T.}~\bibnamefont{Okada}},
  \bibinfo{author}{\bibfnamefont{A.}~\bibnamefont{Yamamoto}},
  \bibinfo{author}{\bibfnamefont{K.}~\bibnamefont{Kishio}},
  \bibinfo{author}{\bibfnamefont{T.}~\bibnamefont{Tohei}},
  \bibinfo{author}{\bibfnamefont{Y.}~\bibnamefont{Ikuhara}},
  \bibinfo{author}{\bibfnamefont{Y.}~\bibnamefont{Gotoh}},
  \bibinfo{author}{\bibfnamefont{H.}~\bibnamefont{Fujihisa}},
  \bibinfo{author}{\bibfnamefont{K.}~\bibnamefont{Kataoka}},
  \bibnamefont{et~al.}, \bibinfo{journal}{Journal of the American Chemical
  Society} \textbf{\bibinfo{volume}{136}}, \bibinfo{pages}{846}
  (\bibinfo{year}{2014}).

\bibitem[{\citenamefont{Okada et~al.}(2014)\citenamefont{Okada, Ogino, Yakita,
  Yamamoto, Kishio, and Shimoyama}}]{Okada2014}
\bibinfo{author}{\bibfnamefont{T.}~\bibnamefont{Okada}},
  \bibinfo{author}{\bibfnamefont{H.}~\bibnamefont{Ogino}},
  \bibinfo{author}{\bibfnamefont{H.}~\bibnamefont{Yakita}},
  \bibinfo{author}{\bibfnamefont{A.}~\bibnamefont{Yamamoto}},
  \bibinfo{author}{\bibfnamefont{K.}~\bibnamefont{Kishio}}, \bibnamefont{and}
  \bibinfo{author}{\bibfnamefont{J.}~\bibnamefont{Shimoyama}},
  \bibinfo{journal}{Physica C: Superconductivity}
  \textbf{\bibinfo{volume}{505}}, \bibinfo{pages}{1 } (\bibinfo{year}{2014}).

\bibitem[{\citenamefont{Yakita et~al.}(2015{\natexlab{a}})\citenamefont{Yakita,
  Ogino, Sala, Okada, Yamamoto, Kishio, Iyo, Eisaki, and
  Shimoyama}}]{Yakita2015}
\bibinfo{author}{\bibfnamefont{H.}~\bibnamefont{Yakita}},
  \bibinfo{author}{\bibfnamefont{H.}~\bibnamefont{Ogino}},
  \bibinfo{author}{\bibfnamefont{A.}~\bibnamefont{Sala}},
  \bibinfo{author}{\bibfnamefont{T.}~\bibnamefont{Okada}},
  \bibinfo{author}{\bibfnamefont{A.}~\bibnamefont{Yamamoto}},
  \bibinfo{author}{\bibfnamefont{K.}~\bibnamefont{Kishio}},
  \bibinfo{author}{\bibfnamefont{A.}~\bibnamefont{Iyo}},
  \bibinfo{author}{\bibfnamefont{H.}~\bibnamefont{Eisaki}}, \bibnamefont{and}
  \bibinfo{author}{\bibfnamefont{J.}~\bibnamefont{Shimoyama}},
  \bibinfo{journal}{Physica C: Superconductivity and its Applications}
  \textbf{\bibinfo{volume}{518}}, \bibinfo{pages}{14 }
  (\bibinfo{year}{2015}{\natexlab{a}}), \bibinfo{note}{proceedings of the 27th
  International Symposium on Superconductivity}.

\bibitem[{\citenamefont{Katayama et~al.}(2015)\citenamefont{Katayama, Sugawara,
  Nakano, Kitou, Sugiyama, Kawaguchi, Ito, Higuchi, Fujii, and
  Sawa}}]{Katayama2015}
\bibinfo{author}{\bibfnamefont{N.}~\bibnamefont{Katayama}},
  \bibinfo{author}{\bibfnamefont{K.}~\bibnamefont{Sugawara}},
  \bibinfo{author}{\bibfnamefont{A.}~\bibnamefont{Nakano}},
  \bibinfo{author}{\bibfnamefont{S.}~\bibnamefont{Kitou}},
  \bibinfo{author}{\bibfnamefont{Y.}~\bibnamefont{Sugiyama}},
  \bibinfo{author}{\bibfnamefont{N.}~\bibnamefont{Kawaguchi}},
  \bibinfo{author}{\bibfnamefont{H.}~\bibnamefont{Ito}},
  \bibinfo{author}{\bibfnamefont{T.}~\bibnamefont{Higuchi}},
  \bibinfo{author}{\bibfnamefont{T.}~\bibnamefont{Fujii}}, \bibnamefont{and}
  \bibinfo{author}{\bibfnamefont{H.}~\bibnamefont{Sawa}},
  \bibinfo{journal}{Physica C: Superconductivity and its Applications}
  \textbf{\bibinfo{volume}{518}}, \bibinfo{pages}{10 } (\bibinfo{year}{2015}),
  \bibinfo{note}{proceedings of the 27th International Symposium on
  Superconductivity}.

\bibitem[{\citenamefont{Joseph et~al.}(2015{\natexlab{a}})\citenamefont{Joseph,
  Marini, Demitri, Capitani, Bernasconi, Zhou, Xing, and Shi}}]{Joseph2015a}
\bibinfo{author}{\bibfnamefont{B.}~\bibnamefont{Joseph}},
  \bibinfo{author}{\bibfnamefont{C.}~\bibnamefont{Marini}},
  \bibinfo{author}{\bibfnamefont{N.}~\bibnamefont{Demitri}},
  \bibinfo{author}{\bibfnamefont{F.}~\bibnamefont{Capitani}},
  \bibinfo{author}{\bibfnamefont{A.}~\bibnamefont{Bernasconi}},
  \bibinfo{author}{\bibfnamefont{W.}~\bibnamefont{Zhou}},
  \bibinfo{author}{\bibfnamefont{X.}~\bibnamefont{Xing}}, \bibnamefont{and}
  \bibinfo{author}{\bibfnamefont{Z.}~\bibnamefont{Shi}},
  \bibinfo{journal}{Superconductor Science and Technology}
  \textbf{\bibinfo{volume}{28}}, \bibinfo{pages}{092001}
  (\bibinfo{year}{2015}{\natexlab{a}}).

\bibitem[{\citenamefont{Zhou et~al.}(2015)\citenamefont{Zhou, Xing, Zhou, Xu,
  and Shi}}]{Zhou2015}
\bibinfo{author}{\bibfnamefont{W.}~\bibnamefont{Zhou}},
  \bibinfo{author}{\bibfnamefont{X.~Z.} \bibnamefont{Xing}},
  \bibinfo{author}{\bibfnamefont{X.}~\bibnamefont{Zhou}},
  \bibinfo{author}{\bibfnamefont{M.~X.} \bibnamefont{Xu}}, \bibnamefont{and}
  \bibinfo{author}{\bibfnamefont{Z.~X.} \bibnamefont{Shi}},
  \bibinfo{journal}{EPL (Europhysics Letters)} \textbf{\bibinfo{volume}{109}},
  \bibinfo{pages}{37005} (\bibinfo{year}{2015}).

\bibitem[{\citenamefont{Yakita et~al.}(2015{\natexlab{b}})\citenamefont{Yakita,
  Ogino, Sala, Okada, Yamamoto, Kishio, Iyo, Eisaki, and ichi
  Shimoyama}}]{Hiroyuki2015}
\bibinfo{author}{\bibfnamefont{H.}~\bibnamefont{Yakita}},
  \bibinfo{author}{\bibfnamefont{H.}~\bibnamefont{Ogino}},
  \bibinfo{author}{\bibfnamefont{A.}~\bibnamefont{Sala}},
  \bibinfo{author}{\bibfnamefont{T.}~\bibnamefont{Okada}},
  \bibinfo{author}{\bibfnamefont{A.}~\bibnamefont{Yamamoto}},
  \bibinfo{author}{\bibfnamefont{K.}~\bibnamefont{Kishio}},
  \bibinfo{author}{\bibfnamefont{A.}~\bibnamefont{Iyo}},
  \bibinfo{author}{\bibfnamefont{H.}~\bibnamefont{Eisaki}}, \bibnamefont{and}
  \bibinfo{author}{\bibfnamefont{J.}~\bibnamefont{ichi Shimoyama}},
  \bibinfo{journal}{Superconductor Science and Technology}
  \textbf{\bibinfo{volume}{28}}, \bibinfo{pages}{065001}
  (\bibinfo{year}{2015}{\natexlab{b}}).

\bibitem[{\citenamefont{Zhou et~al.}(2014)\citenamefont{Zhou, Zhuang, Yuan, Li,
  Xing, Sun, and Shi}}]{Zhou2014}
\bibinfo{author}{\bibfnamefont{W.}~\bibnamefont{Zhou}},
  \bibinfo{author}{\bibfnamefont{J.}~\bibnamefont{Zhuang}},
  \bibinfo{author}{\bibfnamefont{F.}~\bibnamefont{Yuan}},
  \bibinfo{author}{\bibfnamefont{X.}~\bibnamefont{Li}},
  \bibinfo{author}{\bibfnamefont{X.}~\bibnamefont{Xing}},
  \bibinfo{author}{\bibfnamefont{Y.}~\bibnamefont{Sun}}, \bibnamefont{and}
  \bibinfo{author}{\bibfnamefont{Z.}~\bibnamefont{Shi}},
  \bibinfo{journal}{Applied Physics Express} \textbf{\bibinfo{volume}{7}},
  \bibinfo{pages}{063102} (\bibinfo{year}{2014}).

\bibitem[{\citenamefont{Sala et~al.}(2014)\citenamefont{Sala, Yakita, Ogino,
  Okada, Yamamoto, Kishio, Ishida, Iyo, Eisaki, Fujioka et~al.}}]{Sala2014}
\bibinfo{author}{\bibfnamefont{A.}~\bibnamefont{Sala}},
  \bibinfo{author}{\bibfnamefont{H.}~\bibnamefont{Yakita}},
  \bibinfo{author}{\bibfnamefont{H.}~\bibnamefont{Ogino}},
  \bibinfo{author}{\bibfnamefont{T.}~\bibnamefont{Okada}},
  \bibinfo{author}{\bibfnamefont{A.}~\bibnamefont{Yamamoto}},
  \bibinfo{author}{\bibfnamefont{K.}~\bibnamefont{Kishio}},
  \bibinfo{author}{\bibfnamefont{S.}~\bibnamefont{Ishida}},
  \bibinfo{author}{\bibfnamefont{A.}~\bibnamefont{Iyo}},
  \bibinfo{author}{\bibfnamefont{H.}~\bibnamefont{Eisaki}},
  \bibinfo{author}{\bibfnamefont{M.}~\bibnamefont{Fujioka}},
  \bibnamefont{et~al.}, \bibinfo{journal}{Applied Physics Express}
  \textbf{\bibinfo{volume}{7}}, \bibinfo{pages}{073102} (\bibinfo{year}{2014}).

\bibitem[{\citenamefont{Xing et~al.}(2015)\citenamefont{Xing, Zhou, Xu, Li,
  Sun, Zhang, and Shi}}]{Xing2015}
\bibinfo{author}{\bibfnamefont{X.}~\bibnamefont{Xing}},
  \bibinfo{author}{\bibfnamefont{W.}~\bibnamefont{Zhou}},
  \bibinfo{author}{\bibfnamefont{B.}~\bibnamefont{Xu}},
  \bibinfo{author}{\bibfnamefont{N.}~\bibnamefont{Li}},
  \bibinfo{author}{\bibfnamefont{Y.}~\bibnamefont{Sun}},
  \bibinfo{author}{\bibfnamefont{Y.}~\bibnamefont{Zhang}}, \bibnamefont{and}
  \bibinfo{author}{\bibfnamefont{Z.}~\bibnamefont{Shi}},
  \bibinfo{journal}{Journal of the Physical Society of Japan}
  \textbf{\bibinfo{volume}{84}}, \bibinfo{pages}{075001}
  (\bibinfo{year}{2015}).

\bibitem[{\citenamefont{Li et~al.}(2015)\citenamefont{Li, Liu, Zhou, Yang,
  Shen, Li, Jiang, Niu, Xie, Sun et~al.}}]{Li2015}
\bibinfo{author}{\bibfnamefont{M.~Y.} \bibnamefont{Li}},
  \bibinfo{author}{\bibfnamefont{Z.~T.} \bibnamefont{Liu}},
  \bibinfo{author}{\bibfnamefont{W.}~\bibnamefont{Zhou}},
  \bibinfo{author}{\bibfnamefont{H.~F.} \bibnamefont{Yang}},
  \bibinfo{author}{\bibfnamefont{D.~W.} \bibnamefont{Shen}},
  \bibinfo{author}{\bibfnamefont{W.}~\bibnamefont{Li}},
  \bibinfo{author}{\bibfnamefont{J.}~\bibnamefont{Jiang}},
  \bibinfo{author}{\bibfnamefont{X.~H.} \bibnamefont{Niu}},
  \bibinfo{author}{\bibfnamefont{B.~P.} \bibnamefont{Xie}},
  \bibinfo{author}{\bibfnamefont{Y.}~\bibnamefont{Sun}}, \bibnamefont{et~al.},
  \bibinfo{journal}{Phys. Rev. B} \textbf{\bibinfo{volume}{91}},
  \bibinfo{pages}{045112} (\bibinfo{year}{2015}).

\bibitem[{\citenamefont{Kawasaki et~al.}(2015)\citenamefont{Kawasaki, Mabuchi,
  Maeda, Adachi, Mizukami, Kudo, Nohara, and Zheng}}]{Kawasaki2015}
\bibinfo{author}{\bibfnamefont{S.}~\bibnamefont{Kawasaki}},
  \bibinfo{author}{\bibfnamefont{T.}~\bibnamefont{Mabuchi}},
  \bibinfo{author}{\bibfnamefont{S.}~\bibnamefont{Maeda}},
  \bibinfo{author}{\bibfnamefont{T.}~\bibnamefont{Adachi}},
  \bibinfo{author}{\bibfnamefont{T.}~\bibnamefont{Mizukami}},
  \bibinfo{author}{\bibfnamefont{K.}~\bibnamefont{Kudo}},
  \bibinfo{author}{\bibfnamefont{M.}~\bibnamefont{Nohara}}, \bibnamefont{and}
  \bibinfo{author}{\bibfnamefont{G.-q.} \bibnamefont{Zheng}},
  \bibinfo{journal}{Phys. Rev. B} \textbf{\bibinfo{volume}{92}},
  \bibinfo{pages}{180508} (\bibinfo{year}{2015}).

\bibitem[{\citenamefont{Liu et~al.}(2015)\citenamefont{Liu, Kim, Sala, Ogino,
  Shimoyama, Büchner, and Borisenko}}]{Liu2015}
\bibinfo{author}{\bibfnamefont{Z.-H.} \bibnamefont{Liu}},
  \bibinfo{author}{\bibfnamefont{T.~K.} \bibnamefont{Kim}},
  \bibinfo{author}{\bibfnamefont{A.}~\bibnamefont{Sala}},
  \bibinfo{author}{\bibfnamefont{H.}~\bibnamefont{Ogino}},
  \bibinfo{author}{\bibfnamefont{J.}~\bibnamefont{Shimoyama}},
  \bibinfo{author}{\bibfnamefont{B.}~\bibnamefont{Büchner}}, \bibnamefont{and}
  \bibinfo{author}{\bibfnamefont{S.~V.} \bibnamefont{Borisenko}},
  \bibinfo{journal}{Applied Physics Letters} \textbf{\bibinfo{volume}{106}},
  \bibinfo{eid}{052602} (\bibinfo{year}{2015}).

\bibitem[{\citenamefont{Huang et~al.}(2015)\citenamefont{Huang, Yu, Liu, and
  Zou}}]{Huang2015}
\bibinfo{author}{\bibfnamefont{Y.-N.} \bibnamefont{Huang}},
  \bibinfo{author}{\bibfnamefont{X.-L.} \bibnamefont{Yu}},
  \bibinfo{author}{\bibfnamefont{D.-Y.} \bibnamefont{Liu}}, \bibnamefont{and}
  \bibinfo{author}{\bibfnamefont{L.-J.} \bibnamefont{Zou}},
  \bibinfo{journal}{Journal of Applied Physics} \textbf{\bibinfo{volume}{117}},
  \bibinfo{eid}{17E113} (\bibinfo{year}{2015}).

\bibitem[{\citenamefont{Jiang et~al.}(2016)\citenamefont{Jiang, Liu, Cao,
  Birol, Allred, Tian, Liu, Cho, Krogstad, Ma et~al.}}]{Jiang2016}
\bibinfo{author}{\bibfnamefont{S.}~\bibnamefont{Jiang}},
  \bibinfo{author}{\bibfnamefont{C.}~\bibnamefont{Liu}},
  \bibinfo{author}{\bibfnamefont{H.}~\bibnamefont{Cao}},
  \bibinfo{author}{\bibfnamefont{T.}~\bibnamefont{Birol}},
  \bibinfo{author}{\bibfnamefont{J.~M.} \bibnamefont{Allred}},
  \bibinfo{author}{\bibfnamefont{W.}~\bibnamefont{Tian}},
  \bibinfo{author}{\bibfnamefont{L.}~\bibnamefont{Liu}},
  \bibinfo{author}{\bibfnamefont{K.}~\bibnamefont{Cho}},
  \bibinfo{author}{\bibfnamefont{M.~J.} \bibnamefont{Krogstad}},
  \bibinfo{author}{\bibfnamefont{J.}~\bibnamefont{Ma}}, \bibnamefont{et~al.},
  \bibinfo{journal}{Phys. Rev. B} \textbf{\bibinfo{volume}{93}},
  \bibinfo{pages}{054522} (\bibinfo{year}{2016}).

\bibitem[{\citenamefont{Retzlaff et~al.}(2015)\citenamefont{Retzlaff, Buckow,
  Komissinskiy, Ray, Schmidt, M\"uhlig, Schmidl, Seidel, Kurian, and
  Alff}}]{Reiner2015}
\bibinfo{author}{\bibfnamefont{R.}~\bibnamefont{Retzlaff}},
  \bibinfo{author}{\bibfnamefont{A.}~\bibnamefont{Buckow}},
  \bibinfo{author}{\bibfnamefont{P.}~\bibnamefont{Komissinskiy}},
  \bibinfo{author}{\bibfnamefont{S.}~\bibnamefont{Ray}},
  \bibinfo{author}{\bibfnamefont{S.}~\bibnamefont{Schmidt}},
  \bibinfo{author}{\bibfnamefont{H.}~\bibnamefont{M\"uhlig}},
  \bibinfo{author}{\bibfnamefont{F.}~\bibnamefont{Schmidl}},
  \bibinfo{author}{\bibfnamefont{P.}~\bibnamefont{Seidel}},
  \bibinfo{author}{\bibfnamefont{J.}~\bibnamefont{Kurian}}, \bibnamefont{and}
  \bibinfo{author}{\bibfnamefont{L.}~\bibnamefont{Alff}},
  \bibinfo{journal}{Phys. Rev. B} \textbf{\bibinfo{volume}{91}},
  \bibinfo{pages}{104519} (\bibinfo{year}{2015}).

\bibitem[{\citenamefont{Han et~al.}(2013)\citenamefont{Han, Malliakas,
  Stoumpos, Sturza, Claus, Chung, and Kanatzidis}}]{Han2013}
\bibinfo{author}{\bibfnamefont{F.}~\bibnamefont{Han}},
  \bibinfo{author}{\bibfnamefont{C.~D.} \bibnamefont{Malliakas}},
  \bibinfo{author}{\bibfnamefont{C.~C.} \bibnamefont{Stoumpos}},
  \bibinfo{author}{\bibfnamefont{M.}~\bibnamefont{Sturza}},
  \bibinfo{author}{\bibfnamefont{H.}~\bibnamefont{Claus}},
  \bibinfo{author}{\bibfnamefont{D.~Y.} \bibnamefont{Chung}}, \bibnamefont{and}
  \bibinfo{author}{\bibfnamefont{M.~G.} \bibnamefont{Kanatzidis}},
  \bibinfo{journal}{Phys. Rev. B} \textbf{\bibinfo{volume}{88}},
  \bibinfo{pages}{144511} (\bibinfo{year}{2013}).

\bibitem[{\citenamefont{Buckow et~al.}(2013)\citenamefont{Buckow, Retzlaff,
  Kurian, and Alff}}]{Buckow2013}
\bibinfo{author}{\bibfnamefont{A.}~\bibnamefont{Buckow}},
  \bibinfo{author}{\bibfnamefont{R.}~\bibnamefont{Retzlaff}},
  \bibinfo{author}{\bibfnamefont{J.}~\bibnamefont{Kurian}}, \bibnamefont{and}
  \bibinfo{author}{\bibfnamefont{L.}~\bibnamefont{Alff}},
  \bibinfo{journal}{Superconductor Science and Technology}
  \textbf{\bibinfo{volume}{26}}, \bibinfo{pages}{015014}
  (\bibinfo{year}{2013}).

\bibitem[{\citenamefont{Kurian et~al.}(2013)\citenamefont{Kurian, Buckow,
  Retzlaff, and Alff}}]{Kurian2013}
\bibinfo{author}{\bibfnamefont{J.}~\bibnamefont{Kurian}},
  \bibinfo{author}{\bibfnamefont{A.}~\bibnamefont{Buckow}},
  \bibinfo{author}{\bibfnamefont{R.}~\bibnamefont{Retzlaff}}, \bibnamefont{and}
  \bibinfo{author}{\bibfnamefont{L.}~\bibnamefont{Alff}},
  \bibinfo{journal}{Physica C: Superconductivity}
  \textbf{\bibinfo{volume}{484}}, \bibinfo{pages}{171 } (\bibinfo{year}{2013}).

\bibitem[{\citenamefont{Jo et~al.}(2014)\citenamefont{Jo, Park, Lee, Eom, Choi,
  Shim, Kang, and Kim}}]{Jo2014}
\bibinfo{author}{\bibfnamefont{Y.~J.} \bibnamefont{Jo}},
  \bibinfo{author}{\bibfnamefont{J.}~\bibnamefont{Park}},
  \bibinfo{author}{\bibfnamefont{G.}~\bibnamefont{Lee}},
  \bibinfo{author}{\bibfnamefont{M.~J.} \bibnamefont{Eom}},
  \bibinfo{author}{\bibfnamefont{E.~S.} \bibnamefont{Choi}},
  \bibinfo{author}{\bibfnamefont{J.~H.} \bibnamefont{Shim}},
  \bibinfo{author}{\bibfnamefont{W.}~\bibnamefont{Kang}}, \bibnamefont{and}
  \bibinfo{author}{\bibfnamefont{J.~S.} \bibnamefont{Kim}},
  \bibinfo{journal}{Phys. Rev. Lett.} \textbf{\bibinfo{volume}{113}},
  \bibinfo{pages}{156602} (\bibinfo{year}{2014}).

\bibitem[{\citenamefont{Wang et~al.}(2012{\natexlab{a}})\citenamefont{Wang,
  Wang, and Petrovic}}]{Wang2012d}
\bibinfo{author}{\bibfnamefont{K.}~\bibnamefont{Wang}},
  \bibinfo{author}{\bibfnamefont{L.}~\bibnamefont{Wang}}, \bibnamefont{and}
  \bibinfo{author}{\bibfnamefont{C.}~\bibnamefont{Petrovic}},
  \bibinfo{journal}{Applied Physics Letters} \textbf{\bibinfo{volume}{100}},
  \bibinfo{eid}{112111} (\bibinfo{year}{2012}{\natexlab{a}}).

\bibitem[{\citenamefont{Lee et~al.}(2013)\citenamefont{Lee, Farhan, Kim, and
  Shim}}]{Lee2013}
\bibinfo{author}{\bibfnamefont{G.}~\bibnamefont{Lee}},
  \bibinfo{author}{\bibfnamefont{M.~A.} \bibnamefont{Farhan}},
  \bibinfo{author}{\bibfnamefont{J.~S.} \bibnamefont{Kim}}, \bibnamefont{and}
  \bibinfo{author}{\bibfnamefont{J.~H.} \bibnamefont{Shim}},
  \bibinfo{journal}{Phys. Rev. B} \textbf{\bibinfo{volume}{87}},
  \bibinfo{pages}{245104} (\bibinfo{year}{2013}).

\bibitem[{\citenamefont{Jia et~al.}(2014)\citenamefont{Jia, Liu, Cai, Qian,
  Wang, Miao, Richard, Zhao, Li, Wang et~al.}}]{Jia2014}
\bibinfo{author}{\bibfnamefont{L.-L.} \bibnamefont{Jia}},
  \bibinfo{author}{\bibfnamefont{Z.-H.} \bibnamefont{Liu}},
  \bibinfo{author}{\bibfnamefont{Y.-P.} \bibnamefont{Cai}},
  \bibinfo{author}{\bibfnamefont{T.}~\bibnamefont{Qian}},
  \bibinfo{author}{\bibfnamefont{X.-P.} \bibnamefont{Wang}},
  \bibinfo{author}{\bibfnamefont{H.}~\bibnamefont{Miao}},
  \bibinfo{author}{\bibfnamefont{P.}~\bibnamefont{Richard}},
  \bibinfo{author}{\bibfnamefont{Y.-G.} \bibnamefont{Zhao}},
  \bibinfo{author}{\bibfnamefont{Y.}~\bibnamefont{Li}},
  \bibinfo{author}{\bibfnamefont{D.-M.} \bibnamefont{Wang}},
  \bibnamefont{et~al.}, \bibinfo{journal}{Phys. Rev. B}
  \textbf{\bibinfo{volume}{90}}, \bibinfo{pages}{035133}
  (\bibinfo{year}{2014}).

\bibitem[{\citenamefont{Guo et~al.}(2014)\citenamefont{Guo, Princep, Zhang,
  Manuel, Khalyavin, Mazin, Shi, and Boothroyd}}]{Guo2014}
\bibinfo{author}{\bibfnamefont{Y.~F.} \bibnamefont{Guo}},
  \bibinfo{author}{\bibfnamefont{A.~J.} \bibnamefont{Princep}},
  \bibinfo{author}{\bibfnamefont{X.}~\bibnamefont{Zhang}},
  \bibinfo{author}{\bibfnamefont{P.}~\bibnamefont{Manuel}},
  \bibinfo{author}{\bibfnamefont{D.}~\bibnamefont{Khalyavin}},
  \bibinfo{author}{\bibfnamefont{I.~I.} \bibnamefont{Mazin}},
  \bibinfo{author}{\bibfnamefont{Y.~G.} \bibnamefont{Shi}}, \bibnamefont{and}
  \bibinfo{author}{\bibfnamefont{A.~T.} \bibnamefont{Boothroyd}},
  \bibinfo{journal}{Phys. Rev. B} \textbf{\bibinfo{volume}{90}},
  \bibinfo{pages}{075120} (\bibinfo{year}{2014}).

\bibitem[{\citenamefont{Feng et~al.}(2014)\citenamefont{Feng, Wang, Chen, Shi,
  Xie, Yi, Liang, He, He, Peng et~al.}}]{Feng2014}
\bibinfo{author}{\bibfnamefont{Y.}~\bibnamefont{Feng}},
  \bibinfo{author}{\bibfnamefont{Z.}~\bibnamefont{Wang}},
  \bibinfo{author}{\bibfnamefont{C.}~\bibnamefont{Chen}},
  \bibinfo{author}{\bibfnamefont{Y.}~\bibnamefont{Shi}},
  \bibinfo{author}{\bibfnamefont{Z.}~\bibnamefont{Xie}},
  \bibinfo{author}{\bibfnamefont{H.}~\bibnamefont{Yi}},
  \bibinfo{author}{\bibfnamefont{A.}~\bibnamefont{Liang}},
  \bibinfo{author}{\bibfnamefont{S.}~\bibnamefont{He}},
  \bibinfo{author}{\bibfnamefont{J.}~\bibnamefont{He}},
  \bibinfo{author}{\bibfnamefont{Y.}~\bibnamefont{Peng}}, \bibnamefont{et~al.},
  \bibinfo{journal}{Scientific Reports} \textbf{\bibinfo{volume}{4}},
  \bibinfo{pages}{5385 EP } (\bibinfo{year}{2014}).

\bibitem[{\citenamefont{Wang et~al.}(2012{\natexlab{b}})\citenamefont{Wang,
  Graf, Wang, Lei, Tozer, and Petrovic}}]{Wang2012f}
\bibinfo{author}{\bibfnamefont{K.}~\bibnamefont{Wang}},
  \bibinfo{author}{\bibfnamefont{D.}~\bibnamefont{Graf}},
  \bibinfo{author}{\bibfnamefont{L.}~\bibnamefont{Wang}},
  \bibinfo{author}{\bibfnamefont{H.}~\bibnamefont{Lei}},
  \bibinfo{author}{\bibfnamefont{S.~W.} \bibnamefont{Tozer}}, \bibnamefont{and}
  \bibinfo{author}{\bibfnamefont{C.}~\bibnamefont{Petrovic}},
  \bibinfo{journal}{Phys. Rev. B} \textbf{\bibinfo{volume}{85}},
  \bibinfo{pages}{041101} (\bibinfo{year}{2012}{\natexlab{b}}).

\bibitem[{\citenamefont{He et~al.}(2012)\citenamefont{He, Wang, and
  Chen}}]{He2012}
\bibinfo{author}{\bibfnamefont{J.~B.} \bibnamefont{He}},
  \bibinfo{author}{\bibfnamefont{D.~M.} \bibnamefont{Wang}}, \bibnamefont{and}
  \bibinfo{author}{\bibfnamefont{G.~F.} \bibnamefont{Chen}},
  \bibinfo{journal}{Applied Physics Letters} \textbf{\bibinfo{volume}{100}},
  \bibinfo{eid}{112405} (\bibinfo{year}{2012}).

\bibitem[{\citenamefont{Wang and Petrovic}(2012{\natexlab{a}})}]{Wang2012e}
\bibinfo{author}{\bibfnamefont{K.}~\bibnamefont{Wang}} \bibnamefont{and}
  \bibinfo{author}{\bibfnamefont{C.}~\bibnamefont{Petrovic}},
  \bibinfo{journal}{Phys. Rev. B} \textbf{\bibinfo{volume}{86}},
  \bibinfo{pages}{155213} (\bibinfo{year}{2012}{\natexlab{a}}).

\bibitem[{\citenamefont{Wang et~al.}(2013)\citenamefont{Wang, Graf, and
  Petrovic}}]{Wang2013}
\bibinfo{author}{\bibfnamefont{K.}~\bibnamefont{Wang}},
  \bibinfo{author}{\bibfnamefont{D.}~\bibnamefont{Graf}}, \bibnamefont{and}
  \bibinfo{author}{\bibfnamefont{C.}~\bibnamefont{Petrovic}},
  \bibinfo{journal}{Phys. Rev. B} \textbf{\bibinfo{volume}{87}},
  \bibinfo{pages}{235101} (\bibinfo{year}{2013}).

\bibitem[{\citenamefont{Shi et~al.}(2015)\citenamefont{Shi, Richard, Wang, Liu,
  Matt, Xu, Dhaka, Ristic, Qian, Yang et~al.}}]{Shi2015}
\bibinfo{author}{\bibfnamefont{X.}~\bibnamefont{Shi}},
  \bibinfo{author}{\bibfnamefont{P.}~\bibnamefont{Richard}},
  \bibinfo{author}{\bibfnamefont{K.}~\bibnamefont{Wang}},
  \bibinfo{author}{\bibfnamefont{M.}~\bibnamefont{Liu}},
  \bibinfo{author}{\bibfnamefont{C.~E.} \bibnamefont{Matt}},
  \bibinfo{author}{\bibfnamefont{N.}~\bibnamefont{Xu}},
  \bibinfo{author}{\bibfnamefont{R.~S.} \bibnamefont{Dhaka}},
  \bibinfo{author}{\bibfnamefont{Z.}~\bibnamefont{Ristic}},
  \bibinfo{author}{\bibfnamefont{T.}~\bibnamefont{Qian}},
  \bibinfo{author}{\bibfnamefont{Y.-F.} \bibnamefont{Yang}},
  \bibnamefont{et~al.}, \bibinfo{howpublished}{Arxiv: 1512.04187}
  (\bibinfo{year}{2015}).

\bibitem[{\citenamefont{May et~al.}(2014)\citenamefont{May, McGuire, and
  Sales}}]{May2014}
\bibinfo{author}{\bibfnamefont{A.~F.} \bibnamefont{May}},
  \bibinfo{author}{\bibfnamefont{M.~A.} \bibnamefont{McGuire}},
  \bibnamefont{and} \bibinfo{author}{\bibfnamefont{B.~C.} \bibnamefont{Sales}},
  \bibinfo{journal}{Phys. Rev. B} \textbf{\bibinfo{volume}{90}},
  \bibinfo{pages}{075109} (\bibinfo{year}{2014}).

\bibitem[{\citenamefont{Wang and Petrovic}(2012{\natexlab{b}})}]{Wang2012c}
\bibinfo{author}{\bibfnamefont{K.}~\bibnamefont{Wang}} \bibnamefont{and}
  \bibinfo{author}{\bibfnamefont{C.}~\bibnamefont{Petrovic}},
  \bibinfo{journal}{Applied Physics Letters} \textbf{\bibinfo{volume}{101}},
  \bibinfo{eid}{152102} (\bibinfo{year}{2012}{\natexlab{b}}).

\bibitem[{\citenamefont{Yakita et~al.}(2015{\natexlab{c}})\citenamefont{Yakita,
  Ogino, Sala, Okada, Yamamoto, Kishio, Iyo, Eisaki, and ichi
  Shimoyama}}]{Yakita2015b}
\bibinfo{author}{\bibfnamefont{H.}~\bibnamefont{Yakita}},
  \bibinfo{author}{\bibfnamefont{H.}~\bibnamefont{Ogino}},
  \bibinfo{author}{\bibfnamefont{A.}~\bibnamefont{Sala}},
  \bibinfo{author}{\bibfnamefont{T.}~\bibnamefont{Okada}},
  \bibinfo{author}{\bibfnamefont{A.}~\bibnamefont{Yamamoto}},
  \bibinfo{author}{\bibfnamefont{K.}~\bibnamefont{Kishio}},
  \bibinfo{author}{\bibfnamefont{A.}~\bibnamefont{Iyo}},
  \bibinfo{author}{\bibfnamefont{H.}~\bibnamefont{Eisaki}}, \bibnamefont{and}
  \bibinfo{author}{\bibfnamefont{J.}~\bibnamefont{ichi Shimoyama}},
  \bibinfo{journal}{Superconductor Science and Technology}
  \textbf{\bibinfo{volume}{28}}, \bibinfo{pages}{065001}
  (\bibinfo{year}{2015}{\natexlab{c}}).

\bibitem[{\citenamefont{Han et~al.}(2015)\citenamefont{Han, Wan, Phelan,
  Stoumpos, Sturza, Malliakas, Li, Han, Zhao, Chung et~al.}}]{Han2015}
\bibinfo{author}{\bibfnamefont{F.}~\bibnamefont{Han}},
  \bibinfo{author}{\bibfnamefont{X.}~\bibnamefont{Wan}},
  \bibinfo{author}{\bibfnamefont{D.}~\bibnamefont{Phelan}},
  \bibinfo{author}{\bibfnamefont{C.~C.} \bibnamefont{Stoumpos}},
  \bibinfo{author}{\bibfnamefont{M.}~\bibnamefont{Sturza}},
  \bibinfo{author}{\bibfnamefont{C.~D.} \bibnamefont{Malliakas}},
  \bibinfo{author}{\bibfnamefont{Q.}~\bibnamefont{Li}},
  \bibinfo{author}{\bibfnamefont{T.-H.} \bibnamefont{Han}},
  \bibinfo{author}{\bibfnamefont{Q.}~\bibnamefont{Zhao}},
  \bibinfo{author}{\bibfnamefont{D.~Y.} \bibnamefont{Chung}},
  \bibnamefont{et~al.}, \bibinfo{journal}{Phys. Rev. B}
  \textbf{\bibinfo{volume}{92}}, \bibinfo{pages}{045112}
  (\bibinfo{year}{2015}).

\bibitem[{\citenamefont{Harter et~al.}(2016)\citenamefont{Harter, Chu, Jiang,
  Ni, and Hsieh}}]{Harter2016}
\bibinfo{author}{\bibfnamefont{J.~W.} \bibnamefont{Harter}},
  \bibinfo{author}{\bibfnamefont{H.}~\bibnamefont{Chu}},
  \bibinfo{author}{\bibfnamefont{S.}~\bibnamefont{Jiang}},
  \bibinfo{author}{\bibfnamefont{N.}~\bibnamefont{Ni}}, \bibnamefont{and}
  \bibinfo{author}{\bibfnamefont{D.}~\bibnamefont{Hsieh}},
  \bibinfo{journal}{Phys. Rev. B} \textbf{\bibinfo{volume}{93}},
  \bibinfo{pages}{104506} (\bibinfo{year}{2016}).

\bibitem[{\citenamefont{Matsuishi et~al.}(2008)\citenamefont{Matsuishi, Inoue,
  Nomura, Yanagi, Hirano, and Hosono}}]{Matsuishi2008}
\bibinfo{author}{\bibfnamefont{S.}~\bibnamefont{Matsuishi}},
  \bibinfo{author}{\bibfnamefont{Y.}~\bibnamefont{Inoue}},
  \bibinfo{author}{\bibfnamefont{T.}~\bibnamefont{Nomura}},
  \bibinfo{author}{\bibfnamefont{H.}~\bibnamefont{Yanagi}},
  \bibinfo{author}{\bibfnamefont{M.}~\bibnamefont{Hirano}}, \bibnamefont{and}
  \bibinfo{author}{\bibfnamefont{H.}~\bibnamefont{Hosono}},
  \bibinfo{journal}{Journal of the American Chemical Society}
  \textbf{\bibinfo{volume}{130}}, \bibinfo{pages}{14428}
  (\bibinfo{year}{2008}).

\bibitem[{\citenamefont{Joseph et~al.}(2015{\natexlab{b}})\citenamefont{Joseph,
  Iadecola, Bernasconi, Rispoli, Demitri, Xing, Zhou, and Shi}}]{Joseph2015b}
\bibinfo{author}{\bibfnamefont{B.}~\bibnamefont{Joseph}},
  \bibinfo{author}{\bibfnamefont{A.}~\bibnamefont{Iadecola}},
  \bibinfo{author}{\bibfnamefont{A.}~\bibnamefont{Bernasconi}},
  \bibinfo{author}{\bibfnamefont{P.}~\bibnamefont{Rispoli}},
  \bibinfo{author}{\bibfnamefont{N.}~\bibnamefont{Demitri}},
  \bibinfo{author}{\bibfnamefont{X.}~\bibnamefont{Xing}},
  \bibinfo{author}{\bibfnamefont{W.}~\bibnamefont{Zhou}}, \bibnamefont{and}
  \bibinfo{author}{\bibfnamefont{Z.}~\bibnamefont{Shi}},
  \bibinfo{journal}{Journal of Physics and Chemistry of Solids}
  \textbf{\bibinfo{volume}{84}}, \bibinfo{pages}{24 }
  (\bibinfo{year}{2015}{\natexlab{b}}).

\bibitem[{\citenamefont{Caglieris et~al.}(2016)\citenamefont{Caglieris, Sala,
  Fujioka, Hummel, Pallecchi, Lamura, Johrendt, Takano, Ishida, Iyo
  et~al.}}]{Caglieris2016}
\bibinfo{author}{\bibfnamefont{F.}~\bibnamefont{Caglieris}},
  \bibinfo{author}{\bibfnamefont{A.}~\bibnamefont{Sala}},
  \bibinfo{author}{\bibfnamefont{M.}~\bibnamefont{Fujioka}},
  \bibinfo{author}{\bibfnamefont{F.}~\bibnamefont{Hummel}},
  \bibinfo{author}{\bibfnamefont{I.}~\bibnamefont{Pallecchi}},
  \bibinfo{author}{\bibfnamefont{G.}~\bibnamefont{Lamura}},
  \bibinfo{author}{\bibfnamefont{D.}~\bibnamefont{Johrendt}},
  \bibinfo{author}{\bibfnamefont{Y.}~\bibnamefont{Takano}},
  \bibinfo{author}{\bibfnamefont{S.}~\bibnamefont{Ishida}},
  \bibinfo{author}{\bibfnamefont{A.}~\bibnamefont{Iyo}}, \bibnamefont{et~al.},
  \bibinfo{journal}{APL Mater.} \textbf{\bibinfo{volume}{4}},
  \bibinfo{eid}{020702} (\bibinfo{year}{2016}).

\bibitem[{\citenamefont{Xu et~al.}(2013)\citenamefont{Xu, De-Fa, Lin, Qi,
  Qing-Ge, Dong-Yun, Bing, He-Mian, Jian-Wei, Jun-Feng et~al.}}]{Xu2013}
\bibinfo{author}{\bibfnamefont{L.}~\bibnamefont{Xu}},
  \bibinfo{author}{\bibfnamefont{L.}~\bibnamefont{De-Fa}},
  \bibinfo{author}{\bibfnamefont{Z.}~\bibnamefont{Lin}},
  \bibinfo{author}{\bibfnamefont{G.}~\bibnamefont{Qi}},
  \bibinfo{author}{\bibfnamefont{M.}~\bibnamefont{Qing-Ge}},
  \bibinfo{author}{\bibfnamefont{C.}~\bibnamefont{Dong-Yun}},
  \bibinfo{author}{\bibfnamefont{S.}~\bibnamefont{Bing}},
  \bibinfo{author}{\bibfnamefont{Y.}~\bibnamefont{He-Mian}},
  \bibinfo{author}{\bibfnamefont{H.}~\bibnamefont{Jian-Wei}},
  \bibinfo{author}{\bibfnamefont{H.}~\bibnamefont{Jun-Feng}},
  \bibnamefont{et~al.}, \bibinfo{journal}{Chinese Physics Letters}
  \textbf{\bibinfo{volume}{30}}, \bibinfo{pages}{127402}
  (\bibinfo{year}{2013}).

\bibitem[{\citenamefont{Wang et~al.}(2014)\citenamefont{Wang, Shi, Zhang, and
  Yi}}]{Wang2014}
\bibinfo{author}{\bibfnamefont{G.}~\bibnamefont{Wang}},
  \bibinfo{author}{\bibfnamefont{X.}~\bibnamefont{Shi}},
  \bibinfo{author}{\bibfnamefont{L.}~\bibnamefont{Zhang}}, \bibnamefont{and}
  \bibinfo{author}{\bibfnamefont{X.}~\bibnamefont{Yi}}, \bibinfo{journal}{Solid
  State Communications} \textbf{\bibinfo{volume}{200}}, \bibinfo{pages}{61 }
  (\bibinfo{year}{2014}).

\bibitem[{\citenamefont{Mazin et~al.}(2008)\citenamefont{Mazin, Singh,
  Johannes, and Du}}]{Mazin2008}
\bibinfo{author}{\bibfnamefont{I.~I.} \bibnamefont{Mazin}},
  \bibinfo{author}{\bibfnamefont{D.~J.} \bibnamefont{Singh}},
  \bibinfo{author}{\bibfnamefont{M.~D.} \bibnamefont{Johannes}},
  \bibnamefont{and} \bibinfo{author}{\bibfnamefont{M.~H.} \bibnamefont{Du}},
  \bibinfo{journal}{Phys. Rev. Lett.} \textbf{\bibinfo{volume}{101}},
  \bibinfo{pages}{057003} (\bibinfo{year}{2008}).

\bibitem[{\citenamefont{Singh and Du}(2008)}]{Singh2008}
\bibinfo{author}{\bibfnamefont{D.~J.} \bibnamefont{Singh}} \bibnamefont{and}
  \bibinfo{author}{\bibfnamefont{M.-H.} \bibnamefont{Du}},
  \bibinfo{journal}{Phys. Rev. Lett.} \textbf{\bibinfo{volume}{100}},
  \bibinfo{pages}{237003} (\bibinfo{year}{2008}).

\bibitem[{\citenamefont{Ding et~al.}(2008)\citenamefont{Ding, Richard,
  Nakayama, Sugawara, Arakane, Sekiba, Takayama, Souma, Sato, Takahashi
  et~al.}}]{Ding2008}
\bibinfo{author}{\bibfnamefont{H.}~\bibnamefont{Ding}},
  \bibinfo{author}{\bibfnamefont{P.}~\bibnamefont{Richard}},
  \bibinfo{author}{\bibfnamefont{K.}~\bibnamefont{Nakayama}},
  \bibinfo{author}{\bibfnamefont{K.}~\bibnamefont{Sugawara}},
  \bibinfo{author}{\bibfnamefont{T.}~\bibnamefont{Arakane}},
  \bibinfo{author}{\bibfnamefont{Y.}~\bibnamefont{Sekiba}},
  \bibinfo{author}{\bibfnamefont{A.}~\bibnamefont{Takayama}},
  \bibinfo{author}{\bibfnamefont{S.}~\bibnamefont{Souma}},
  \bibinfo{author}{\bibfnamefont{T.}~\bibnamefont{Sato}},
  \bibinfo{author}{\bibfnamefont{T.}~\bibnamefont{Takahashi}},
  \bibnamefont{et~al.}, \bibinfo{journal}{EPL (Europhysics Letters)}
  \textbf{\bibinfo{volume}{83}}, \bibinfo{pages}{47001} (\bibinfo{year}{2008}).

\bibitem[{\citenamefont{Wu et~al.}(2014)\citenamefont{Wu, Le, Liang, Qin, Fan,
  and Hu}}]{Wu2014}
\bibinfo{author}{\bibfnamefont{X.}~\bibnamefont{Wu}},
  \bibinfo{author}{\bibfnamefont{C.}~\bibnamefont{Le}},
  \bibinfo{author}{\bibfnamefont{Y.}~\bibnamefont{Liang}},
  \bibinfo{author}{\bibfnamefont{S.}~\bibnamefont{Qin}},
  \bibinfo{author}{\bibfnamefont{H.}~\bibnamefont{Fan}}, \bibnamefont{and}
  \bibinfo{author}{\bibfnamefont{J.}~\bibnamefont{Hu}}, \bibinfo{journal}{Phys.
  Rev. B} \textbf{\bibinfo{volume}{89}}, \bibinfo{pages}{205102}
  (\bibinfo{year}{2014}).

\bibitem[{\citenamefont{Wu et~al.}(2015)\citenamefont{Wu, Qin, Liang, Le, Fan,
  and Hu}}]{Wu2015b}
\bibinfo{author}{\bibfnamefont{X.}~\bibnamefont{Wu}},
  \bibinfo{author}{\bibfnamefont{S.}~\bibnamefont{Qin}},
  \bibinfo{author}{\bibfnamefont{Y.}~\bibnamefont{Liang}},
  \bibinfo{author}{\bibfnamefont{C.}~\bibnamefont{Le}},
  \bibinfo{author}{\bibfnamefont{H.}~\bibnamefont{Fan}}, \bibnamefont{and}
  \bibinfo{author}{\bibfnamefont{J.}~\bibnamefont{Hu}}, \bibinfo{journal}{Phys.
  Rev. B} \textbf{\bibinfo{volume}{91}}, \bibinfo{pages}{081111}
  (\bibinfo{year}{2015}).

\bibitem[{\citenamefont{Nagai et~al.}(2015)\citenamefont{Nagai, Nakamura,
  Machida, and Kuroki}}]{Nagai2015}
\bibinfo{author}{\bibfnamefont{Y.}~\bibnamefont{Nagai}},
  \bibinfo{author}{\bibfnamefont{H.}~\bibnamefont{Nakamura}},
  \bibinfo{author}{\bibfnamefont{M.}~\bibnamefont{Machida}}, \bibnamefont{and}
  \bibinfo{author}{\bibfnamefont{K.}~\bibnamefont{Kuroki}},
  \bibinfo{journal}{Journal of the Physical Society of Japan}
  \textbf{\bibinfo{volume}{84}}, \bibinfo{pages}{093702}
  (\bibinfo{year}{2015}).

\bibitem[{\citenamefont{Thirupathaiah et~al.}(2013)\citenamefont{Thirupathaiah,
  St\"urzer, Zabolotnyy, Johrendt, B\"uchner, and
  Borisenko}}]{Thirupathaiah2013}
\bibinfo{author}{\bibfnamefont{S.}~\bibnamefont{Thirupathaiah}},
  \bibinfo{author}{\bibfnamefont{T.}~\bibnamefont{St\"urzer}},
  \bibinfo{author}{\bibfnamefont{V.~B.} \bibnamefont{Zabolotnyy}},
  \bibinfo{author}{\bibfnamefont{D.}~\bibnamefont{Johrendt}},
  \bibinfo{author}{\bibfnamefont{B.}~\bibnamefont{B\"uchner}},
  \bibnamefont{and} \bibinfo{author}{\bibfnamefont{S.~V.}
  \bibnamefont{Borisenko}}, \bibinfo{journal}{Phys. Rev. B}
  \textbf{\bibinfo{volume}{88}}, \bibinfo{pages}{140505}
  (\bibinfo{year}{2013}).

\bibitem[{\citenamefont{Charnukha et~al.}(2015)\citenamefont{Charnukha,
  Thirupathaiah, Zabolotnyy, B{\"u}chner, Zhigadlo, Batlogg, Yaresko, and
  Borisenko}}]{Charnukha2015}
\bibinfo{author}{\bibfnamefont{A.}~\bibnamefont{Charnukha}},
  \bibinfo{author}{\bibfnamefont{S.}~\bibnamefont{Thirupathaiah}},
  \bibinfo{author}{\bibfnamefont{V.~B.} \bibnamefont{Zabolotnyy}},
  \bibinfo{author}{\bibfnamefont{B.}~\bibnamefont{B{\"u}chner}},
  \bibinfo{author}{\bibfnamefont{N.~D.} \bibnamefont{Zhigadlo}},
  \bibinfo{author}{\bibfnamefont{B.}~\bibnamefont{Batlogg}},
  \bibinfo{author}{\bibfnamefont{A.~N.} \bibnamefont{Yaresko}},
  \bibnamefont{and} \bibinfo{author}{\bibfnamefont{S.~V.}
  \bibnamefont{Borisenko}}, \bibinfo{journal}{Scientific Reports}
  \textbf{\bibinfo{volume}{5}}, \bibinfo{pages}{10392 EP }
  (\bibinfo{year}{2015}).

\bibitem[{\citenamefont{Steglich et~al.}(1979)\citenamefont{Steglich, Aarts,
  Bredl, Lieke, Meschede, Franz, and Sch\"afer}}]{Steglich1979}
\bibinfo{author}{\bibfnamefont{F.}~\bibnamefont{Steglich}},
  \bibinfo{author}{\bibfnamefont{J.}~\bibnamefont{Aarts}},
  \bibinfo{author}{\bibfnamefont{C.~D.} \bibnamefont{Bredl}},
  \bibinfo{author}{\bibfnamefont{W.}~\bibnamefont{Lieke}},
  \bibinfo{author}{\bibfnamefont{D.}~\bibnamefont{Meschede}},
  \bibinfo{author}{\bibfnamefont{W.}~\bibnamefont{Franz}}, \bibnamefont{and}
  \bibinfo{author}{\bibfnamefont{H.}~\bibnamefont{Sch\"afer}},
  \bibinfo{journal}{Phys. Rev. Lett.} \textbf{\bibinfo{volume}{43}},
  \bibinfo{pages}{1892} (\bibinfo{year}{1979}).

\bibitem[{\citenamefont{Kito et~al.}(2008)\citenamefont{Kito, Eisaki, and
  Iyo}}]{Kito2008}
\bibinfo{author}{\bibfnamefont{H.}~\bibnamefont{Kito}},
  \bibinfo{author}{\bibfnamefont{H.}~\bibnamefont{Eisaki}}, \bibnamefont{and}
  \bibinfo{author}{\bibfnamefont{A.}~\bibnamefont{Iyo}},
  \bibinfo{journal}{Journal of the Physical Society of Japan}
  \textbf{\bibinfo{volume}{77}}, \bibinfo{pages}{063707}
  (\bibinfo{year}{2008}).

\bibitem[{\citenamefont{Nicklas et~al.}(2007)\citenamefont{Nicklas, Stockert,
  Park, Habicht, Kiefer, Pham, Thompson, Fisk, and Steglich}}]{Nicklas2007}
\bibinfo{author}{\bibfnamefont{M.}~\bibnamefont{Nicklas}},
  \bibinfo{author}{\bibfnamefont{O.}~\bibnamefont{Stockert}},
  \bibinfo{author}{\bibfnamefont{T.}~\bibnamefont{Park}},
  \bibinfo{author}{\bibfnamefont{K.}~\bibnamefont{Habicht}},
  \bibinfo{author}{\bibfnamefont{K.}~\bibnamefont{Kiefer}},
  \bibinfo{author}{\bibfnamefont{L.~D.} \bibnamefont{Pham}},
  \bibinfo{author}{\bibfnamefont{J.~D.} \bibnamefont{Thompson}},
  \bibinfo{author}{\bibfnamefont{Z.}~\bibnamefont{Fisk}}, \bibnamefont{and}
  \bibinfo{author}{\bibfnamefont{F.}~\bibnamefont{Steglich}},
  \bibinfo{journal}{Phys. Rev. B} \textbf{\bibinfo{volume}{76}},
  \bibinfo{pages}{052401} (\bibinfo{year}{2007}).

\bibitem[{\citenamefont{Fujiwara et~al.}(2013)\citenamefont{Fujiwara, Tsutsumi,
  Iimura, Matsuishi, Hosono, Yamakawa, and Kontani}}]{Fujiwara2013}
\bibinfo{author}{\bibfnamefont{N.}~\bibnamefont{Fujiwara}},
  \bibinfo{author}{\bibfnamefont{S.}~\bibnamefont{Tsutsumi}},
  \bibinfo{author}{\bibfnamefont{S.}~\bibnamefont{Iimura}},
  \bibinfo{author}{\bibfnamefont{S.}~\bibnamefont{Matsuishi}},
  \bibinfo{author}{\bibfnamefont{H.}~\bibnamefont{Hosono}},
  \bibinfo{author}{\bibfnamefont{Y.}~\bibnamefont{Yamakawa}}, \bibnamefont{and}
  \bibinfo{author}{\bibfnamefont{H.}~\bibnamefont{Kontani}},
  \bibinfo{journal}{Phys. Rev. Lett.} \textbf{\bibinfo{volume}{111}},
  \bibinfo{pages}{097002} (\bibinfo{year}{2013}).

\bibitem[{\citenamefont{Hiraishi et~al.}(2014)\citenamefont{Hiraishi, Iimura,
  Kojima, Yamaura, Hiraka, Ikeda, Miao, Ishikawa, Torii, Miyazaki
  et~al.}}]{Hiraishi2014}
\bibinfo{author}{\bibfnamefont{M.}~\bibnamefont{Hiraishi}},
  \bibinfo{author}{\bibfnamefont{S.}~\bibnamefont{Iimura}},
  \bibinfo{author}{\bibfnamefont{K.~M.} \bibnamefont{Kojima}},
  \bibinfo{author}{\bibfnamefont{J.}~\bibnamefont{Yamaura}},
  \bibinfo{author}{\bibfnamefont{H.}~\bibnamefont{Hiraka}},
  \bibinfo{author}{\bibfnamefont{K.}~\bibnamefont{Ikeda}},
  \bibinfo{author}{\bibfnamefont{P.}~\bibnamefont{Miao}},
  \bibinfo{author}{\bibfnamefont{Y.}~\bibnamefont{Ishikawa}},
  \bibinfo{author}{\bibfnamefont{S.}~\bibnamefont{Torii}},
  \bibinfo{author}{\bibfnamefont{M.}~\bibnamefont{Miyazaki}},
  \bibnamefont{et~al.}, \bibinfo{journal}{Nat Phys}
  \textbf{\bibinfo{volume}{10}}, \bibinfo{pages}{300} (\bibinfo{year}{2014}).

\bibitem[{\citenamefont{Sakurai et~al.}(2015)\citenamefont{Sakurai, Fujiwara,
  Kawaguchi, Yamakawa, Kontani, Iimura, Matsuishi, and Hosono}}]{Sakurai2015}
\bibinfo{author}{\bibfnamefont{R.}~\bibnamefont{Sakurai}},
  \bibinfo{author}{\bibfnamefont{N.}~\bibnamefont{Fujiwara}},
  \bibinfo{author}{\bibfnamefont{N.}~\bibnamefont{Kawaguchi}},
  \bibinfo{author}{\bibfnamefont{Y.}~\bibnamefont{Yamakawa}},
  \bibinfo{author}{\bibfnamefont{H.}~\bibnamefont{Kontani}},
  \bibinfo{author}{\bibfnamefont{S.}~\bibnamefont{Iimura}},
  \bibinfo{author}{\bibfnamefont{S.}~\bibnamefont{Matsuishi}},
  \bibnamefont{and} \bibinfo{author}{\bibfnamefont{H.}~\bibnamefont{Hosono}},
  \bibinfo{journal}{Phys. Rev. B} \textbf{\bibinfo{volume}{91}},
  \bibinfo{pages}{064509} (\bibinfo{year}{2015}).

\bibitem[{\citenamefont{Saha et~al.}(2012)\citenamefont{Saha, Butch, Drye,
  Magill, Ziemak, Kirshenbaum, Zavalij, Lynn, and Paglione}}]{Saha2012}
\bibinfo{author}{\bibfnamefont{S.~R.} \bibnamefont{Saha}},
  \bibinfo{author}{\bibfnamefont{N.~P.} \bibnamefont{Butch}},
  \bibinfo{author}{\bibfnamefont{T.}~\bibnamefont{Drye}},
  \bibinfo{author}{\bibfnamefont{J.}~\bibnamefont{Magill}},
  \bibinfo{author}{\bibfnamefont{S.}~\bibnamefont{Ziemak}},
  \bibinfo{author}{\bibfnamefont{K.}~\bibnamefont{Kirshenbaum}},
  \bibinfo{author}{\bibfnamefont{P.~Y.} \bibnamefont{Zavalij}},
  \bibinfo{author}{\bibfnamefont{J.~W.} \bibnamefont{Lynn}}, \bibnamefont{and}
  \bibinfo{author}{\bibfnamefont{J.}~\bibnamefont{Paglione}},
  \bibinfo{journal}{Phys. Rev. B} \textbf{\bibinfo{volume}{85}},
  \bibinfo{pages}{024525} (\bibinfo{year}{2012}).

\bibitem[{\citenamefont{Kudo et~al.}(2013)\citenamefont{Kudo, Iba, Takasuga,
  Kitahama, Matsumura, Danura, Nogami, and Nohara}}]{Kudo2013}
\bibinfo{author}{\bibfnamefont{K.}~\bibnamefont{Kudo}},
  \bibinfo{author}{\bibfnamefont{K.}~\bibnamefont{Iba}},
  \bibinfo{author}{\bibfnamefont{M.}~\bibnamefont{Takasuga}},
  \bibinfo{author}{\bibfnamefont{Y.}~\bibnamefont{Kitahama}},
  \bibinfo{author}{\bibfnamefont{J.-i.} \bibnamefont{Matsumura}},
  \bibinfo{author}{\bibfnamefont{M.}~\bibnamefont{Danura}},
  \bibinfo{author}{\bibfnamefont{Y.}~\bibnamefont{Nogami}}, \bibnamefont{and}
  \bibinfo{author}{\bibfnamefont{M.}~\bibnamefont{Nohara}},
  \bibinfo{journal}{Scientific Reports} \textbf{\bibinfo{volume}{3}},
  \bibinfo{pages}{1478 EP } (\bibinfo{year}{2013}).

\bibitem[{\citenamefont{Lv et~al.}(2011)\citenamefont{Lv, Deng, Gooch, Wei,
  Sun, Meen, Xue, Lorenz, and Chu}}]{Lv2011}
\bibinfo{author}{\bibfnamefont{B.}~\bibnamefont{Lv}},
  \bibinfo{author}{\bibfnamefont{L.}~\bibnamefont{Deng}},
  \bibinfo{author}{\bibfnamefont{M.}~\bibnamefont{Gooch}},
  \bibinfo{author}{\bibfnamefont{F.}~\bibnamefont{Wei}},
  \bibinfo{author}{\bibfnamefont{Y.}~\bibnamefont{Sun}},
  \bibinfo{author}{\bibfnamefont{J.~K.} \bibnamefont{Meen}},
  \bibinfo{author}{\bibfnamefont{Y.-Y.} \bibnamefont{Xue}},
  \bibinfo{author}{\bibfnamefont{B.}~\bibnamefont{Lorenz}}, \bibnamefont{and}
  \bibinfo{author}{\bibfnamefont{C.-W.} \bibnamefont{Chu}},
  \bibinfo{journal}{Proceedings of the National Academy of Sciences}
  \textbf{\bibinfo{volume}{108}}, \bibinfo{pages}{15705}
  (\bibinfo{year}{2011}).

\bibitem[{\citenamefont{Sefat et~al.}(2008{\natexlab{a}})\citenamefont{Sefat,
  Jin, McGuire, Sales, Singh, and Mandrus}}]{Sefat2008}
\bibinfo{author}{\bibfnamefont{A.~S.} \bibnamefont{Sefat}},
  \bibinfo{author}{\bibfnamefont{R.}~\bibnamefont{Jin}},
  \bibinfo{author}{\bibfnamefont{M.~A.} \bibnamefont{McGuire}},
  \bibinfo{author}{\bibfnamefont{B.~C.} \bibnamefont{Sales}},
  \bibinfo{author}{\bibfnamefont{D.~J.} \bibnamefont{Singh}}, \bibnamefont{and}
  \bibinfo{author}{\bibfnamefont{D.}~\bibnamefont{Mandrus}},
  \bibinfo{journal}{Phys. Rev. Lett.} \textbf{\bibinfo{volume}{101}},
  \bibinfo{pages}{117004} (\bibinfo{year}{2008}{\natexlab{a}}).

\bibitem[{\citenamefont{Sefat et~al.}(2008{\natexlab{b}})\citenamefont{Sefat,
  Huq, McGuire, Jin, Sales, Mandrus, Cranswick, Stephens, and
  Stone}}]{Sefat2008b}
\bibinfo{author}{\bibfnamefont{A.~S.} \bibnamefont{Sefat}},
  \bibinfo{author}{\bibfnamefont{A.}~\bibnamefont{Huq}},
  \bibinfo{author}{\bibfnamefont{M.~A.} \bibnamefont{McGuire}},
  \bibinfo{author}{\bibfnamefont{R.}~\bibnamefont{Jin}},
  \bibinfo{author}{\bibfnamefont{B.~C.} \bibnamefont{Sales}},
  \bibinfo{author}{\bibfnamefont{D.}~\bibnamefont{Mandrus}},
  \bibinfo{author}{\bibfnamefont{L.~M.~D.} \bibnamefont{Cranswick}},
  \bibinfo{author}{\bibfnamefont{P.~W.} \bibnamefont{Stephens}},
  \bibnamefont{and} \bibinfo{author}{\bibfnamefont{K.~H.} \bibnamefont{Stone}},
  \bibinfo{journal}{Phys. Rev. B} \textbf{\bibinfo{volume}{78}},
  \bibinfo{pages}{104505} (\bibinfo{year}{2008}{\natexlab{b}}).

\bibitem[{\citenamefont{Jaroszynski et~al.}(2008)\citenamefont{Jaroszynski,
  Hunte, Balicas, Jo, Rai\ifmmode \check{c}\else
  \v{c}\fi{}evi\ifmmode~\acute{c}\else \'{c}\fi{}, Gurevich, Larbalestier,
  Balakirev, Fang, Cheng et~al.}}]{Jaroszynski2008}
\bibinfo{author}{\bibfnamefont{J.}~\bibnamefont{Jaroszynski}},
  \bibinfo{author}{\bibfnamefont{F.}~\bibnamefont{Hunte}},
  \bibinfo{author}{\bibfnamefont{L.}~\bibnamefont{Balicas}},
  \bibinfo{author}{\bibfnamefont{Y.-j.} \bibnamefont{Jo}},
  \bibinfo{author}{\bibfnamefont{I.}~\bibnamefont{Rai\ifmmode \check{c}\else
  \v{c}\fi{}evi\ifmmode~\acute{c}\else \'{c}\fi{}}},
  \bibinfo{author}{\bibfnamefont{A.}~\bibnamefont{Gurevich}},
  \bibinfo{author}{\bibfnamefont{D.~C.} \bibnamefont{Larbalestier}},
  \bibinfo{author}{\bibfnamefont{F.~F.} \bibnamefont{Balakirev}},
  \bibinfo{author}{\bibfnamefont{L.}~\bibnamefont{Fang}},
  \bibinfo{author}{\bibfnamefont{P.}~\bibnamefont{Cheng}},
  \bibnamefont{et~al.}, \bibinfo{journal}{Phys. Rev. B}
  \textbf{\bibinfo{volume}{78}}, \bibinfo{pages}{174523}
  (\bibinfo{year}{2008}).

\bibitem[{\citenamefont{Yuan et~al.}(2009)\citenamefont{Yuan, Singleton,
  Balakirev, Baily, Chen, Luo, and Wang}}]{Yuan2009}
\bibinfo{author}{\bibfnamefont{H.~Q.} \bibnamefont{Yuan}},
  \bibinfo{author}{\bibfnamefont{J.}~\bibnamefont{Singleton}},
  \bibinfo{author}{\bibfnamefont{F.~F.} \bibnamefont{Balakirev}},
  \bibinfo{author}{\bibfnamefont{S.~A.} \bibnamefont{Baily}},
  \bibinfo{author}{\bibfnamefont{G.~F.} \bibnamefont{Chen}},
  \bibinfo{author}{\bibfnamefont{J.~L.} \bibnamefont{Luo}}, \bibnamefont{and}
  \bibinfo{author}{\bibfnamefont{N.~L.} \bibnamefont{Wang}},
  \bibinfo{journal}{Nature} \textbf{\bibinfo{volume}{457}},
  \bibinfo{pages}{565} (\bibinfo{year}{2009}).

\bibitem[{\citenamefont{Park et~al.}(2016)\citenamefont{Park, Mine, Yamada,
  Ohtake, Akiyama, Sun, Pyon, Tamegai, Kitahama, Mizukami et~al.}}]{Park2016}
\bibinfo{author}{\bibfnamefont{A.}~\bibnamefont{Park}},
  \bibinfo{author}{\bibfnamefont{A.}~\bibnamefont{Mine}},
  \bibinfo{author}{\bibfnamefont{T.}~\bibnamefont{Yamada}},
  \bibinfo{author}{\bibfnamefont{F.}~\bibnamefont{Ohtake}},
  \bibinfo{author}{\bibfnamefont{H.}~\bibnamefont{Akiyama}},
  \bibinfo{author}{\bibfnamefont{Y.}~\bibnamefont{Sun}},
  \bibinfo{author}{\bibfnamefont{S.}~\bibnamefont{Pyon}},
  \bibinfo{author}{\bibfnamefont{T.}~\bibnamefont{Tamegai}},
  \bibinfo{author}{\bibfnamefont{Y.}~\bibnamefont{Kitahama}},
  \bibinfo{author}{\bibfnamefont{T.}~\bibnamefont{Mizukami}},
  \bibnamefont{et~al.}, \bibinfo{journal}{Superconductor Science and
  Technology} \textbf{\bibinfo{volume}{29}}, \bibinfo{pages}{055006}
  (\bibinfo{year}{2016}).

\bibitem[{\citenamefont{Sun et~al.}(2013)\citenamefont{Sun, Taen, Tsuchiya,
  Ding, Pyon, Shi, and Tamegai}}]{Sun2013}
\bibinfo{author}{\bibfnamefont{Y.}~\bibnamefont{Sun}},
  \bibinfo{author}{\bibfnamefont{T.}~\bibnamefont{Taen}},
  \bibinfo{author}{\bibfnamefont{Y.}~\bibnamefont{Tsuchiya}},
  \bibinfo{author}{\bibfnamefont{Q.}~\bibnamefont{Ding}},
  \bibinfo{author}{\bibfnamefont{S.}~\bibnamefont{Pyon}},
  \bibinfo{author}{\bibfnamefont{Z.}~\bibnamefont{Shi}}, \bibnamefont{and}
  \bibinfo{author}{\bibfnamefont{T.}~\bibnamefont{Tamegai}},
  \bibinfo{journal}{Applied Physics Express} \textbf{\bibinfo{volume}{6}},
  \bibinfo{pages}{043101} (\bibinfo{year}{2013}).

\bibitem[{\citenamefont{Ahilan et~al.}(2009)\citenamefont{Ahilan, Ning, Imai,
  Sefat, McGuire, Sales, and Mandrus}}]{Ahilan2009}
\bibinfo{author}{\bibfnamefont{K.}~\bibnamefont{Ahilan}},
  \bibinfo{author}{\bibfnamefont{F.~L.} \bibnamefont{Ning}},
  \bibinfo{author}{\bibfnamefont{T.}~\bibnamefont{Imai}},
  \bibinfo{author}{\bibfnamefont{A.~S.} \bibnamefont{Sefat}},
  \bibinfo{author}{\bibfnamefont{M.~A.} \bibnamefont{McGuire}},
  \bibinfo{author}{\bibfnamefont{B.~C.} \bibnamefont{Sales}}, \bibnamefont{and}
  \bibinfo{author}{\bibfnamefont{D.}~\bibnamefont{Mandrus}},
  \bibinfo{journal}{Phys. Rev. B} \textbf{\bibinfo{volume}{79}},
  \bibinfo{pages}{214520} (\bibinfo{year}{2009}).

\bibitem[{\citenamefont{Wang et~al.}(2012{\natexlab{c}})\citenamefont{Wang,
  Xiang, Ying, Yan, Cheng, Ye, Luo, and Chen}}]{Wang2012b}
\bibinfo{author}{\bibfnamefont{A.~F.} \bibnamefont{Wang}},
  \bibinfo{author}{\bibfnamefont{Z.~J.} \bibnamefont{Xiang}},
  \bibinfo{author}{\bibfnamefont{J.~J.} \bibnamefont{Ying}},
  \bibinfo{author}{\bibfnamefont{Y.~J.} \bibnamefont{Yan}},
  \bibinfo{author}{\bibfnamefont{P.}~\bibnamefont{Cheng}},
  \bibinfo{author}{\bibfnamefont{G.~J.} \bibnamefont{Ye}},
  \bibinfo{author}{\bibfnamefont{X.~G.} \bibnamefont{Luo}}, \bibnamefont{and}
  \bibinfo{author}{\bibfnamefont{X.~H.} \bibnamefont{Chen}},
  \bibinfo{journal}{New Journal of Physics} \textbf{\bibinfo{volume}{14}},
  \bibinfo{pages}{113043} (\bibinfo{year}{2012}{\natexlab{c}}).

\bibitem[{\citenamefont{Zhang et~al.}(2005)\citenamefont{Zhang, Tan, Stormer,
  and Kim}}]{Zhang2005}
\bibinfo{author}{\bibfnamefont{Y.}~\bibnamefont{Zhang}},
  \bibinfo{author}{\bibfnamefont{Y.-W.} \bibnamefont{Tan}},
  \bibinfo{author}{\bibfnamefont{H.~L.} \bibnamefont{Stormer}},
  \bibnamefont{and} \bibinfo{author}{\bibfnamefont{P.}~\bibnamefont{Kim}},
  \bibinfo{journal}{Nature} \textbf{\bibinfo{volume}{438}},
  \bibinfo{pages}{201} (\bibinfo{year}{2005}).

\bibitem[{\citenamefont{Hasan and Kane}(2010)}]{Hasan2010}
\bibinfo{author}{\bibfnamefont{M.~Z.} \bibnamefont{Hasan}} \bibnamefont{and}
  \bibinfo{author}{\bibfnamefont{C.~L.} \bibnamefont{Kane}},
  \bibinfo{journal}{Rev. Mod. Phys.} \textbf{\bibinfo{volume}{82}},
  \bibinfo{pages}{3045} (\bibinfo{year}{2010}).

\bibitem[{\citenamefont{Pardo and Pickett}(2009)}]{Pardo2009}
\bibinfo{author}{\bibfnamefont{V.}~\bibnamefont{Pardo}} \bibnamefont{and}
  \bibinfo{author}{\bibfnamefont{W.~E.} \bibnamefont{Pickett}},
  \bibinfo{journal}{Phys. Rev. Lett.} \textbf{\bibinfo{volume}{102}},
  \bibinfo{pages}{166803} (\bibinfo{year}{2009}).

\bibitem[{\citenamefont{Choi et~al.}(2010)\citenamefont{Choi, Jhi, and
  Son}}]{Choi2010}
\bibinfo{author}{\bibfnamefont{S.-M.} \bibnamefont{Choi}},
  \bibinfo{author}{\bibfnamefont{S.-H.} \bibnamefont{Jhi}}, \bibnamefont{and}
  \bibinfo{author}{\bibfnamefont{Y.-W.} \bibnamefont{Son}},
  \bibinfo{journal}{Phys. Rev. B} \textbf{\bibinfo{volume}{81}},
  \bibinfo{pages}{081407} (\bibinfo{year}{2010}).

\bibitem[{\citenamefont{Min et~al.}(2006)\citenamefont{Min, Hill, Sinitsyn,
  Sahu, Kleinman, and MacDonald}}]{Min2006}
\bibinfo{author}{\bibfnamefont{H.}~\bibnamefont{Min}},
  \bibinfo{author}{\bibfnamefont{J.~E.} \bibnamefont{Hill}},
  \bibinfo{author}{\bibfnamefont{N.~A.} \bibnamefont{Sinitsyn}},
  \bibinfo{author}{\bibfnamefont{B.~R.} \bibnamefont{Sahu}},
  \bibinfo{author}{\bibfnamefont{L.}~\bibnamefont{Kleinman}}, \bibnamefont{and}
  \bibinfo{author}{\bibfnamefont{A.~H.} \bibnamefont{MacDonald}},
  \bibinfo{journal}{Phys. Rev. B} \textbf{\bibinfo{volume}{74}},
  \bibinfo{pages}{165310} (\bibinfo{year}{2006}).

\bibitem[{\citenamefont{Farhan et~al.}(2014)\citenamefont{Farhan, Lee, and
  Shim}}]{Farhan2014}
\bibinfo{author}{\bibfnamefont{M.~A.} \bibnamefont{Farhan}},
  \bibinfo{author}{\bibfnamefont{G.}~\bibnamefont{Lee}}, \bibnamefont{and}
  \bibinfo{author}{\bibfnamefont{J.~H.} \bibnamefont{Shim}},
  \bibinfo{journal}{Journal of Physics: Condensed Matter}
  \textbf{\bibinfo{volume}{26}}, \bibinfo{pages}{042201}
  (\bibinfo{year}{2014}).

\bibitem[{\citenamefont{Hiramatsu
  et~al.}(2008{\natexlab{a}})\citenamefont{Hiramatsu, Katase, Kamiya, Hirano,
  and Hosono}}]{Hiramatsu2008a}
\bibinfo{author}{\bibfnamefont{H.}~\bibnamefont{Hiramatsu}},
  \bibinfo{author}{\bibfnamefont{T.}~\bibnamefont{Katase}},
  \bibinfo{author}{\bibfnamefont{T.}~\bibnamefont{Kamiya}},
  \bibinfo{author}{\bibfnamefont{M.}~\bibnamefont{Hirano}}, \bibnamefont{and}
  \bibinfo{author}{\bibfnamefont{H.}~\bibnamefont{Hosono}},
  \bibinfo{journal}{Applied Physics Express} \textbf{\bibinfo{volume}{1}},
  \bibinfo{pages}{101702} (\bibinfo{year}{2008}{\natexlab{a}}).

\bibitem[{\citenamefont{Hiramatsu
  et~al.}(2008{\natexlab{b}})\citenamefont{Hiramatsu, Katase, Kamiya, Hirano,
  and Hosono}}]{Hiramatsu2008b}
\bibinfo{author}{\bibfnamefont{H.}~\bibnamefont{Hiramatsu}},
  \bibinfo{author}{\bibfnamefont{T.}~\bibnamefont{Katase}},
  \bibinfo{author}{\bibfnamefont{T.}~\bibnamefont{Kamiya}},
  \bibinfo{author}{\bibfnamefont{M.}~\bibnamefont{Hirano}}, \bibnamefont{and}
  \bibinfo{author}{\bibfnamefont{H.}~\bibnamefont{Hosono}},
  \bibinfo{journal}{Applied Physics Express} \textbf{\bibinfo{volume}{1}},
  \bibinfo{pages}{101702} (\bibinfo{year}{2008}{\natexlab{b}}).

\bibitem[{\citenamefont{Iida et~al.}(2009)\citenamefont{Iida, Hänisch, Hühne,
  Kurth, Kidszun, Haindl, Werner, Schultz, and Holzapfel}}]{Iida2009}
\bibinfo{author}{\bibfnamefont{K.}~\bibnamefont{Iida}},
  \bibinfo{author}{\bibfnamefont{J.}~\bibnamefont{Hänisch}},
  \bibinfo{author}{\bibfnamefont{R.}~\bibnamefont{Hühne}},
  \bibinfo{author}{\bibfnamefont{F.}~\bibnamefont{Kurth}},
  \bibinfo{author}{\bibfnamefont{M.}~\bibnamefont{Kidszun}},
  \bibinfo{author}{\bibfnamefont{S.}~\bibnamefont{Haindl}},
  \bibinfo{author}{\bibfnamefont{J.}~\bibnamefont{Werner}},
  \bibinfo{author}{\bibfnamefont{L.}~\bibnamefont{Schultz}}, \bibnamefont{and}
  \bibinfo{author}{\bibfnamefont{B.}~\bibnamefont{Holzapfel}},
  \bibinfo{journal}{Applied Physics Letters} \textbf{\bibinfo{volume}{95}},
  \bibinfo{eid}{192501} (\bibinfo{year}{2009}).

\bibitem[{\citenamefont{Katase et~al.}(2010)\citenamefont{Katase, Hiramatsu,
  Kamiya, and Hosono}}]{Katase2010}
\bibinfo{author}{\bibfnamefont{T.}~\bibnamefont{Katase}},
  \bibinfo{author}{\bibfnamefont{H.}~\bibnamefont{Hiramatsu}},
  \bibinfo{author}{\bibfnamefont{T.}~\bibnamefont{Kamiya}}, \bibnamefont{and}
  \bibinfo{author}{\bibfnamefont{H.}~\bibnamefont{Hosono}},
  \bibinfo{journal}{Applied Physics Express} \textbf{\bibinfo{volume}{3}},
  \bibinfo{pages}{063101} (\bibinfo{year}{2010}).

\bibitem[{\citenamefont{Liu et~al.}(2012)\citenamefont{Liu, Zhang, Mou, He, Ou,
  Wang, Li, Wang, Zhao, He et~al.}}]{Liu2012}
\bibinfo{author}{\bibfnamefont{D.}~\bibnamefont{Liu}},
  \bibinfo{author}{\bibfnamefont{W.}~\bibnamefont{Zhang}},
  \bibinfo{author}{\bibfnamefont{D.}~\bibnamefont{Mou}},
  \bibinfo{author}{\bibfnamefont{J.}~\bibnamefont{He}},
  \bibinfo{author}{\bibfnamefont{Y.-B.} \bibnamefont{Ou}},
  \bibinfo{author}{\bibfnamefont{Q.-Y.} \bibnamefont{Wang}},
  \bibinfo{author}{\bibfnamefont{Z.}~\bibnamefont{Li}},
  \bibinfo{author}{\bibfnamefont{L.}~\bibnamefont{Wang}},
  \bibinfo{author}{\bibfnamefont{L.}~\bibnamefont{Zhao}},
  \bibinfo{author}{\bibfnamefont{S.}~\bibnamefont{He}}, \bibnamefont{et~al.},
  \bibinfo{journal}{Nat Commun} \textbf{\bibinfo{volume}{3}},
  \bibinfo{pages}{931} (\bibinfo{year}{2012}).

\bibitem[{\citenamefont{Tan et~al.}(2013)\citenamefont{Tan, Zhang, Xia, Ye,
  Chen, Xie, Peng, Xu, Fan, Xu et~al.}}]{Tan2013}
\bibinfo{author}{\bibfnamefont{S.}~\bibnamefont{Tan}},
  \bibinfo{author}{\bibfnamefont{Y.}~\bibnamefont{Zhang}},
  \bibinfo{author}{\bibfnamefont{M.}~\bibnamefont{Xia}},
  \bibinfo{author}{\bibfnamefont{Z.}~\bibnamefont{Ye}},
  \bibinfo{author}{\bibfnamefont{F.}~\bibnamefont{Chen}},
  \bibinfo{author}{\bibfnamefont{X.}~\bibnamefont{Xie}},
  \bibinfo{author}{\bibfnamefont{R.}~\bibnamefont{Peng}},
  \bibinfo{author}{\bibfnamefont{D.}~\bibnamefont{Xu}},
  \bibinfo{author}{\bibfnamefont{Q.}~\bibnamefont{Fan}},
  \bibinfo{author}{\bibfnamefont{H.}~\bibnamefont{Xu}}, \bibnamefont{et~al.},
  \bibinfo{journal}{Nat Mater} \textbf{\bibinfo{volume}{12}},
  \bibinfo{pages}{634} (\bibinfo{year}{2013}).

\bibitem[{\citenamefont{He et~al.}(2013)\citenamefont{He, He, Zhang, Zhao, Liu,
  Liu, Mou, Ou, Wang, Li et~al.}}]{He2013}
\bibinfo{author}{\bibfnamefont{S.}~\bibnamefont{He}},
  \bibinfo{author}{\bibfnamefont{J.}~\bibnamefont{He}},
  \bibinfo{author}{\bibfnamefont{W.}~\bibnamefont{Zhang}},
  \bibinfo{author}{\bibfnamefont{L.}~\bibnamefont{Zhao}},
  \bibinfo{author}{\bibfnamefont{D.}~\bibnamefont{Liu}},
  \bibinfo{author}{\bibfnamefont{X.}~\bibnamefont{Liu}},
  \bibinfo{author}{\bibfnamefont{D.}~\bibnamefont{Mou}},
  \bibinfo{author}{\bibfnamefont{Y.-B.} \bibnamefont{Ou}},
  \bibinfo{author}{\bibfnamefont{Q.-Y.} \bibnamefont{Wang}},
  \bibinfo{author}{\bibfnamefont{Z.}~\bibnamefont{Li}}, \bibnamefont{et~al.},
  \bibinfo{journal}{Nat Mater} \textbf{\bibinfo{volume}{12}},
  \bibinfo{pages}{605} (\bibinfo{year}{2013}).

\bibitem[{\citenamefont{Sun et~al.}(2014)\citenamefont{Sun, Zhang, Xing, Li,
  Zhao, Xia, Wang, Ma, Xue, and Wang}}]{Sun2014}
\bibinfo{author}{\bibfnamefont{Y.}~\bibnamefont{Sun}},
  \bibinfo{author}{\bibfnamefont{W.}~\bibnamefont{Zhang}},
  \bibinfo{author}{\bibfnamefont{Y.}~\bibnamefont{Xing}},
  \bibinfo{author}{\bibfnamefont{F.}~\bibnamefont{Li}},
  \bibinfo{author}{\bibfnamefont{Y.}~\bibnamefont{Zhao}},
  \bibinfo{author}{\bibfnamefont{Z.}~\bibnamefont{Xia}},
  \bibinfo{author}{\bibfnamefont{L.}~\bibnamefont{Wang}},
  \bibinfo{author}{\bibfnamefont{X.}~\bibnamefont{Ma}},
  \bibinfo{author}{\bibfnamefont{Q.-K.} \bibnamefont{Xue}}, \bibnamefont{and}
  \bibinfo{author}{\bibfnamefont{J.}~\bibnamefont{Wang}},
  \bibinfo{journal}{Scientific Reports} \textbf{\bibinfo{volume}{4}},
  \bibinfo{pages}{6040 EP } (\bibinfo{year}{2014}).

\bibitem[{\citenamefont{Heron et~al.}(2013)\citenamefont{Heron, Ray, Lister,
  Aegerter, Keller, Kes, Menon, and Lee}}]{RayPRL}
\bibinfo{author}{\bibfnamefont{D.~O.~G.} \bibnamefont{Heron}},
  \bibinfo{author}{\bibfnamefont{S.~J.} \bibnamefont{Ray}},
  \bibinfo{author}{\bibfnamefont{S.~J.} \bibnamefont{Lister}},
  \bibinfo{author}{\bibfnamefont{C.~M.} \bibnamefont{Aegerter}},
  \bibinfo{author}{\bibfnamefont{H.}~\bibnamefont{Keller}},
  \bibinfo{author}{\bibfnamefont{P.~H.} \bibnamefont{Kes}},
  \bibinfo{author}{\bibfnamefont{G.~I.} \bibnamefont{Menon}}, \bibnamefont{and}
  \bibinfo{author}{\bibfnamefont{S.~L.} \bibnamefont{Lee}},
  \bibinfo{journal}{Phys. Rev. Lett.} \textbf{\bibinfo{volume}{110}},
  \bibinfo{pages}{107004} (\bibinfo{year}{2013}).

\bibitem[{\citenamefont{Ray et~al.}(2014)\citenamefont{Ray, Gibbs, Bending,
  Curran, Babaev, Baines, Mackenzie, and Lee}}]{RaySRO}
\bibinfo{author}{\bibfnamefont{S.~J.} \bibnamefont{Ray}},
  \bibinfo{author}{\bibfnamefont{A.~S.} \bibnamefont{Gibbs}},
  \bibinfo{author}{\bibfnamefont{S.~J.} \bibnamefont{Bending}},
  \bibinfo{author}{\bibfnamefont{P.~J.} \bibnamefont{Curran}},
  \bibinfo{author}{\bibfnamefont{E.}~\bibnamefont{Babaev}},
  \bibinfo{author}{\bibfnamefont{C.}~\bibnamefont{Baines}},
  \bibinfo{author}{\bibfnamefont{A.~P.} \bibnamefont{Mackenzie}},
  \bibnamefont{and} \bibinfo{author}{\bibfnamefont{S.~L.} \bibnamefont{Lee}},
  \bibinfo{journal}{Phys. Rev. B} \textbf{\bibinfo{volume}{89}},
  \bibinfo{pages}{094504} (\bibinfo{year}{2014}).

\end{thebibliography}

\end{document}